\documentclass[prd, amsfonts, twocolumn, nofootinbib, showpacs]{revtex4-1}

\usepackage{graphicx, epsfig}
\usepackage{color}
\usepackage{amsmath}
\usepackage{mathtools}
\newcommand{\be}{\begin{equation}}
\newcommand{\ee}{\end{equation}}
\newcommand{\bea}{\begin{eqnarray}}
\newcommand{\eea}{\end{eqnarray}}

\newcommand{\gapp}{\mathrel{\raise.3ex\hbox{$>$}\mkern-14mu \lower0.6ex\hbox{$\sim$}}}
\newcommand{\lapp}{\mathrel{\raise.3ex\hbox{$<$}\mkern-14mu \lower0.6ex\hbox{$\sim$}}}
\newcommand{\LSM}{L$\Sigma$M}

\newcommand{\GMLfull}{Gell-Mann-L\'evy~}

\newcommand{\HVEV}{\langle H\rangle}
\newcommand{\SVEV}{\langle S\rangle}
\newcommand{\half}{\frac{1}{2}}
\newcommand{\mpisq}{m_\pi^2}

\newcommand{\GlnuDSM}{{\nu_D}{\mathrm {SM}}^{\rm G}_{tb\tau\nu_\tau}}
\newcommand{\Dim}{d}
\newcommand{\Extended}{E-AHM}
\def\bbox{{\,\lower0.9pt\vbox{\hrule \hbox{\vrule height 0.2 cm
\hskip 0.2 cm \vrule  height 0.2 cm}\hrule}\,}}

\begin{document} 


\title{
Global $U(1)_Y \otimes$BRST symmetry and the LSS theorem:
Ward-Takahashi identities 
\\governing 
Green's functions, on-shell T-Matrix elements, and the effective potential, 
in the scalar-sector
of the spontaneously broken extended Abelian Higgs model
}

\author{Bryan W. Lynn$^{1,2}$, Glenn D. Starkman$^{1}$
}
\affiliation{$^1$ ISO/CERCA/Department of Physics, Case Western Reserve University, 
Cleveland, OH 44106-7079}
\affiliation{$^2$ Department of Physics and Astronomy, University College London, London WC1E 6BT, UK}
\email{bryan.lynn@cern.ch,  gds6@case.edu
}

\begin{abstract}
The  weak-scale $U(1)_Y$ Abelian Higgs Model (AHM) 
is the  simplest spontaneous symmetry breaking (SSB) gauge theory: a scalar 
$\phi = \frac{1}{\sqrt 2}(H+i\pi) \equiv \frac{1}{\sqrt 2}{\tilde H}e^{i{\tilde \pi}/\HVEV}$ and a vector $A^\mu$.
The extended AHM (E-AHM) adds 
certain heavy ($M_{\Phi}^2,M_{\psi}^2\sim M_{Heavy}^2 \gg \HVEV^2 \sim m_{Weak}^2$) 
spin $S=0$ scalars $\Phi$ and $S=\half$ 
fermions $\psi$. 
In Lorenz gauge, $\partial_\mu A^\mu =0$, 
the SSB AHM (and E-AHM) has 
a global $U(1)_Y$ conserved physical current, but no conserved charge.
As shown by T.W.B. Kibble, the Goldstone theorem applies, so 
$\tilde \pi$ is a massless derivatively coupled Nambu-Goldstone boson (NGB).

Proof of all-loop-orders renormalizability and unitarity for the SSB case is tricky because 
the BRST-invariant Lagrangian is not $U(1)_Y$ symmetric. 
Nevertheless, Slavnov-Taylor identities 
guarantee that 
on-shell T-matrix elements of  physical states
$A^\mu$,$\phi$, $\Phi$, $\psi$ (but not ghosts $\omega, {\bar \eta}$)
are independent of anomaly-free local $U(1)_Y$ gauge transformations.
We observe here that they are therefore also independent of the usual 
anomaly-free 
$U(1)_Y$ global/rigid transformations.  
It follows that the associated global current, 
which is classically conserved only up to gauge-fixing terms, 
is exactly conserved for amplitudes of physical states
in the AHM and E-AHM.
We identify
corresponding 
``un-deformed" (i.e. with full global $U(1)_Y$ symmetry) 
Ward-Takahashi identities (WTI).
The proof of renormalizability and unitarity, which relies on BRST invariance, is undisturbed.

In Lorenz gauge, two towers of ``1-soft-pion" SSB global WTI govern the $\phi$-sector,
and represent a new global $U(1)_Y\otimes$BRST symmetry not of the Lagrangian but of the physics.  
The first gives relations among off-shell Green's functions, 
yielding powerful constraints on the all-loop-orders $\phi$-sector SSB E-AHM low-energy effective Lagrangian and 
an additional global shift symmetry for the NGB: 
${\tilde \pi}\to {\tilde \pi}+\HVEV \theta$. 
A second tower, governing on-shell T-matrix elements, 
replaces the old Adler self-consistency conditions 
with those  for gauge theories,
further severely constrains the effective potential, and
guarantees infra-red finiteness for zero NGB ($ {\tilde \pi}$) mass. 
The on-shell WTI include a Lee-Stora-Symanzik (LSS) theorem, also for gauge theories. 
This enforces the strong condition $\mpisq = 0$ on the pseudoscalar $\pi$ 
(not just the much weaker condition $m_{\tilde \pi}^2 = 0$ on the NGB $\tilde\pi$),
and causes all relevant-operator contributions to the effective Lagrangian
to vanish exactly.

In consequence,
certain heavy $CP$-conserving $\Phi,\psi$ matter decouple completely in the $M_{Heavy}^2/m_{Weak}^2 \to \infty$ limit: we prove 4 new low-energy heavy-particle decoupling theorems which, more powerful than the usual Appelquist-Carazzone decoupling theorem, including all virtual $\Phi,\psi$ loop-contributions to {\it relevant operators}, which {\it vanish exactly due to the exact $U(1)_Y$ symmetry} of 1-soft-$\pi$ Adler-self-consistency relations governing {\it on-shell} T-Matrix elements.

Underlying our results is that global $U(1)_Y$ transformations $\delta_{U(1)_Y}$, and nilpotent $s^2=0$ BRST transformations, commute: we prove $\Big[ \delta_{U(1)_Y}, s \Big]$
in G. 't Hooft's $R_\xi$ gauges. 
With its on-shell T-Matrix constraints, SSB E-AHM physics therefore has more symmetry
than does its BRST-invariant Lagrangian $L_{E-AHM}^{R_\xi}$: i.e. global $U(1)_Y \otimes$BRST symmetry.  

The NGB ${\tilde \pi}$ decouples from the observable particle spectrum
$B^\mu$,$\tilde h$, $\tilde \Phi$, $\tilde \psi$ in the usual way, 
when the observable vector
$B_\mu \equiv A_\mu+\frac{1}{e\HVEV}\partial_\mu{\tilde \pi}$ absorbs it, 
as if it were a gauge transformation, 
hiding 
both towers of $U(1)_Y$ WTI from observable particle physics. 

\end{abstract}

\maketitle

\section{Introduction}
\label{Introduction}

What are the symmetries driving spontaneously broken Abelian Higgs Model (AHM) physics \cite{LSS-3Proof}?
Although the symmetries of the $U(1)_Y$ AHM Lagrangian are well known \cite{Ramond2004}, local gauge-invariance  is lost in the AHM Lagrangian, broken by gauge-fixing terms, and replaced with global BRST invariance \cite{BecchiRouetStora,Tyutin1975,Tyutin1976}. 

In their seminal work,  Elisabeth Kraus and Klaus Sibold 
\cite{KrausSiboldAHM} 
showed important new practicalities of the renormalizability and unitarity (to all-loop-orders) of the spontaneous symmetry breaking (SSB) Abelian Higgs model (AHM). They did this by deriving rigid invariance from BRST invariance.
The SSB case is tricky because the globally BRST-invariant Lagrangian is not $U(1)_Y$ symmetric. But they identified a set of ``deformed" ({i.e. with no remnant of the original $U(1)_Y$ group symmetry}) {\it rigid/global} AHM transformations which, after inclusion of well-defined $U(1)_Y$-breaking by quantum loops (e.g. in scalar wavefunction renormalization beyond the classical AHM), {\it are} compatible with BRST symmetry. 

Kraus and Sibold then constructed deformed Ward-Takahashi identities (WTI) 
for quantum AHM Green's functions, showing them
(with appropriate normalization conditions) 
to obey all-loop-orders renormalizability and unitarity.
Because their renormalization relies only on  deformed  WTI, 
Kraus and Sibold's results are independent of regularization scheme, 
for any acceptable scheme (i.e. if one exists).
They did not construct WTI for on-shell T-Matrix elements.

Nevertheless, Slavnov-Taylor identities \cite{JCTaylor1976} prove that the
on-shell S-Matrix elements of ``physical states" 
$A^\mu$,$\phi$, $\Phi$, $\psi$, 
(i.e. spin $S=0$ scalars $h,\pi, \Phi$, $S=\half$ {(CP-conserving) } fermions $\psi$, 
and $S=1$ gauge bosons $A_\mu$,
but not fermionic ghosts $\omega$ or anti-ghosts ${\bar \eta}$)
are independent, in the AHM, 
of the usual undeformed anomaly-free $U(1)_Y$ local/gauge transformations, 
even though these break the Lagrangian's BRST symmetry. 
We observe here that they are therefore also 
independent of  anomaly-free undeformed $U(1)_Y$ global/rigid transformations, 
resulting in ``new" global/rigid currents 
and appropriate un-deformed $U(1)_Y$ Ward-Takahashi Identities. 

We here distinguish carefully between off-shell Green's function WTI, 
which constrain the (un-observable) effective 
Lagrangian and action, 
and on-shell T-Matrix WTI,
which further severely constrain observable physics.
We show here that, in the SSB Abelian Higgs Model,
a tower of Ward-Takahashi Identities (WTI) relates all relevant-operator
contributions to AHM physical-scalar-sector physical observables to one another. 
An on-shell T-Matrix WTI, i.e. the equivalent of an Adler Self-Consistency relation but for this gauge theory,
then causes all such contributions  to vanish.
It does so through its insistence that the scalar mass-squared  vanish exactly 
\begin{equation}
\label{LSSTheoremPrime}
\mpisq=0
\end{equation}
in spontaneously broken ($\HVEV\neq0$) theories, 
which we term the Lee-Stora-Symanzik (LSS)
\footnote{
Note from BWL and GDS: Raymond Stora, who was an active participant in this research until his death, would never have named anything after himself, but we judge that, given the stature of B.W. Lee, R. Stora and K. Symanzik (now all deceased) in the history of the relevant physics, the community would refer to that result as the ``LSS theorem" anyway. 
}
theorem 
after the three physicists who recognized its central role 
in the renormalization of global Linear Sigma Models, 
and the one who was central to our understanding of its role 
in the renormalization
of  gauge theories.%
\footnote{
	As first noted by Kibble \cite{Kibble1967}, 
	in Lorenz gauge a relation similar in appearance to (\ref{LSSTheoremPrime}), 
	$m_{\tilde\pi}^2=0$,
	enforces the masslessness of a Nambu Goldstone Boson (NGB) $\tilde\pi$,
	i.e. is a Goldstone Theorem \cite{Nambu1959,Goldstone1961,Goldstone1962}
	for this gauge theory.  
	This is regardless of the fact that the NGB is not a physical degree of freedom,
	but is absorbed ("eaten") by the gauge boson.
	However, as we describe in greater detail below (cf. equation (\ref{Kibble}),
	${\tilde\pi}$ is the angular degree of freedom 
	in the Kibble representation of the complex scalar field, 
	while $\pi$ is the pseudoscalar degree of freedom in the linear representation.
	In global Linear Sigma Models $(L\Sigma M)$, 
	the masslessness of the NGB and the LSS condition (\ref{LSSTheorem}) are  equivalent. 
	Indeed, B. Lee \cite{Lee1970}, K. Symanzik \cite{Symanzik1970a,Symanzik1970b},
	A. Vassiliev \cite{Vassiliev1970} and classic texts \cite{ItzyksonZuber} advocate 
	that the spontaneously broken (``Goldstone") mode of a $U(1)$ global $L\Sigma M$ 
	is to be understood as the zero-explicit-breaking limit (i.e. $\mpisq \to 0$) 
	of the explicit $U(1)$-breaking Partially Conserved Axial-vector Current (PCAC) term,
	$L_{PCAC}=\HVEV\mpisq H$, 
	included in the $U(1)$ version 
	of the Gell-Mann and L${\acute e}$vy $L\Sigma M$ \cite{GellMannLevy1960}.
	The existence and masslessness of the purely derivatively coupled NGB
	is a result of and requires 
	the vanishing of the explicit-symmetry-breaking pseudo-scalar mass-squared.
	\newline\indent
	{\bf In the ${\mathbf {U(1)_Y}}$ AHM gauge theory, 
	the Goldstone theorem and the LSS theorem are not equivalent}. 
	To see this (or to at least suspect it) the reader should remember 
	that one cannot incorporate 
	explicit PCAC breaking of the local $U(1)_Y$ symmetry 
	into the AHM  gauge theory \cite{JCTaylor1976},
	without spoiling unitarity.
}
In addition to constraining the parameters of the theory, 
LSS permits us to employ pion-pole-dominance to compute the WTI.

The crucial advance over \cite{LSS-2}, 
which considered the  global $SU(2)_L\times U(1)_Y$ Linear Sigma Model,
is a proof that the WTI remain in place in a SSB gauge theory,
with the LSS theorem playing 
the same protective role as did the Goldstone Theorem in the global theory \cite{LSS-2}.

Our  new rigid $U(1)_Y$ WTIs 
govern the scalar-sector of the AHM
and of the extensions we consider in Section \ref{E-AHM}. 
They are therefore independent of regularization-scheme (assuming one exists).
Although not a gauge-independent procedure, 
it may help the reader to imagine that loop integrals 
are cut off at a short-distance  finite Euclidean UV scale, $\Lambda$, 
never taking the $\Lambda^2 \to \infty$ limit. 
Although that cut-off can be imagined to be near the Planck scale $\Lambda\simeq M_{Pl}$, 
quantum gravitational loops are not  included.

The structure of this paper is as follows:

Section \ref{AbelianHiggsModelSymmetry} introduces $U(1)_Y\otimes$BRST symmetry for the AHM and E-AHM in a general 't Hooft $R_\xi$ gauge, and explains why physical quantities obey that new symmetry.

Section \ref{AbelianHiggsModel} 
concerns the correct renormalization of the spontaneously broken AHM in Lorenz gauge.
We treat the AHM in isolation, 
as a stand-alone flat-space weak-scale quantum field theory,
not embedded or integrated into any higher-scale ``Beyond-AHM"  physics. 


Section \ref{E-AHM} extends our AHM
results to include the all-loop-orders virtual contributions of certain $M_{Heavy}^2 \gg m_{Weak}^2$ heavy $U(1)_Y$ matter representations (which might arise in certain Beyond-AHM models). 


Section \ref{ParticlePhysicsAHM} reminds the reader \cite{Guralnik1964} how the NGB $\tilde \pi$  disappears from the {\it observable} particle spectrum of the E-AHM. 
 

Section \ref{RigueurMathematiqueExigeante} 
discusses the exacting mathematical rigor that might have fully satisfied Raymond Stora, and persuaded him to take up the co-authorship of this paper he deserved,
had he not passed away.

Section \ref{Conclusions} reminds us that historically (with an important exception) 
the  decoupling of heavy particles is the usual experience of physics.

Appendix \ref{DerivationWTIAHM} gives a complete and pedagogical derivation
of the $U(1)_Y$ WTIs governing the $\phi$-sector of the AHM. 
Our renormalized WTIs  include all contributions from virtual transverse gauge bosons,
$\phi$-scalars, and ghosts --
$A^\mu$, $h$ and $\pi$, and ${\bar \eta}$ and $\omega$ respectively.

Appendix \ref{DerivationWTIE-AHM} gives a complete and pedagogical derivation
of $U(1)_Y$ $(h,\pi)$-sector WTIs in the E-AHM, 
which now  include the all-loop-orders contributions 
of certain additional $U(1)_Y$ matter representations: 
spin $S=0$ scalars $\Phi$, 
and $S=\half$ anomaly-cancelling {(CP-conserving)} fermions $\psi$. 
They  include all contributions from virtual transverse gauge bosons, 
ghosts, scalars, and fermions -- $A^\mu;h,\pi;{\bar \eta},\omega;\Phi;\psi$. 

\section{$U(1)_Y\otimes$BRST symmetry in 't Hooft 
$R_\xi$ 
gauges}
\label{AbelianHiggsModelSymmetry}

The BRST-invariant \cite{BecchiRouetStora,Tyutin1975,Tyutin1976} Lagrangian 
of the $U(1)_Y$ AHM gauge theory 
may be written, 
in a general 't Hooft $R_\xi$ gauge, 
in terms of  
a transverse vector $A_\mu$, 
a complex scalar $\phi$, 
a ghost $\omega$, 
and an anti-ghost $\bar \eta$:
\be
	\label{LagrangianAHMRxiGauge}
	L_{AHM}^{R_\xi} =  L_{AHM}^{GaugeInvariant} 
	+ L_{AHM}^{GaugeFix;R_\xi}  
	+L_{AHM}^{Ghost;R_\xi} 
\ee
where
\be
\label{LagrangianAHM-GI}
	L_{AHM}^{GaugeInvariant}=
	\vert D_\mu \phi \vert ^2 - \frac{1}{4}A_{\mu \nu}A^{\mu \nu} -V_{AHM}(\phi^\dagger \phi) 
\ee
with
\bea
D_\mu \phi &=&(\partial_\mu -ieY_\phi A_\mu)\phi \nonumber \\
A_{\mu \nu}&=&\partial_\mu A_\nu - \partial_\nu A_\mu \nonumber \\
V_{AHM}&=&\mu_\phi^2 \Big( \phi^\dagger \phi \Big)+ \lambda_\phi ^2 \Big(\phi^\dagger \phi\Big)^2 
\eea
and
\be
\label{LinearScalarRep}
\phi = \frac{1}{{\sqrt 2}}(H+i\pi), \quad H=\HVEV + h \quad {\mathrm {and} }\quad  Y_\phi =-1.
\ee

In G. 't Hooft's $R_\xi$ gauges, gauge fixing and DeWitt-Fadeev-Popov ghost terms 
\cite{DeWitt1967,Fadeev1967}
are written in terms of a Nakanishi-Lautrup field $b$
\cite{Nakanashi1966,Lautrup1967}, 
and the SSB vector mass $m_A=eY_\phi\HVEV =-e\HVEV >0$.
\bea
\label{tHooftGaugeFixing}
&&L_{AHM}^{GaugeFix;R_\xi}+ L_{AHM}^{Ghost;R_\xi} \nonumber \\
&&\qquad =\frac{1}{2}\xi b^2 +b\Big(\partial_\mu A^\mu +\xi m_A \pi \Big)
- {\bar \eta} \Big(\partial^2 +\xi \frac{m_A^2}{\HVEV}  H \Big) \omega \nonumber \\
&&\qquad =s\Big[{\bar \eta} \Big(F_A + \half \xi b \Big)\Big] \nonumber \\
&&F_A =\partial_\mu A^\mu +\xi m_A \pi 
\end{eqnarray}
with global BRST transformations \cite{BecchiRouetStora,Tyutin1975,Tyutin1976,Nakanashi1966,Lautrup1967,Weinberg1995} $s$
\bea
\label{BRSTTransformations}
sA_\mu &=& \partial_\mu \omega, \qquad s{\bar \eta} = b; \nonumber \\
sH &=& -e\pi  \omega, \qquad sb = 0; \nonumber \\
s\pi &=&e H  \omega, \qquad s\omega = 0; 
\eea
so that the Lagrangian (\ref{LagrangianAHMRxiGauge}) is BRST invariant
\bea
\label{BRSTTransformationLagrangian}
s L_{AHM}^{R_\xi} &=& 0 \,.
\eea

The {\it classical} Eq. of motion for the ghost is
\bea
\label{GhostEqOfM}
sF_A&=& \Big( \partial^2 +\xi \frac{m_A^2}{Y_\phi \HVEV} H \Big) \omega =0
\eea

Now define the properties of the various fields under the usual anomaly-free un-deformed rigid/global $U(1)_Y$ transformation by a constant $\Omega$:
\bea
\label{U(1)Transformations}
\delta_{U(1)_Y}A_\mu &=& 0, \qquad \delta_{U(1)_Y}{\bar \eta} = 0; \nonumber \\
\delta_{U(1)_Y}H &=& -e\pi\Omega, \qquad \delta_{U(1)_Y}b = 0; \nonumber \\
\delta_{U(1)_Y}\pi &=& eH\Omega, \qquad \delta_{U(1)_Y}\omega = 0\,.
\eea
We discover that the $R_\xi$-gauge Lagrangian (\ref{LagrangianAHMRxiGauge}) is not invariant 
under such $U(1)_Y$ transformations
\bea
\label{U(1)TransformationLagrangian}
\delta_{U(1)_Y} L_{AHM}^{R_\xi} &=& \delta_{U(1)_Y}  \Big(s\Big[{\bar \eta} \Big(F_A + \half \xi b \Big)\Big] \Big) \nonumber \\
&=& \xi em_A \Big( bH + e{\bar \eta}\pi \omega \Big)\Omega \nonumber \\
&=& s\Big(\delta_{U(1)_Y} \Big[{\bar \eta} \Big(F_A + \half \xi b \Big)\Big] \Big) \nonumber \\
&\neq& 0
\eea

{
Still, 
the actions of the {\bf BRST transformations (\ref{BRSTTransformations}) 
and the $U(1)_Y$ transformation (\ref{U(1)Transformations}) commute} on all fields. 
\bea
\label{SU2BRSTFieldCommutators}
\Big[ \delta_{U(1)_Y}, s \Big] A^\mu&=& 0;\quad \Big[ \delta_{U(1)_Y}, s \Big] \omega=0; \nonumber \\
\Big[ \delta_{U(1)_Y}, s \Big] H&=& 0;\quad \Big[ \delta_{U(1)_Y}, s \Big] {\bar \eta}=0; \nonumber \\
\Big[ \delta_{U(1)_Y}, s \Big] \pi&=& 0;\quad \Big[ \delta_{U(1)_Y}, s \Big] b=0; 
\eea
Thus, with the nilpotent property $s^2=0$ applied in (\ref{U(1)TransformationLagrangian})
\bea
\label{U(1)BRSTCommutatorAHM}
\Big[ \delta_{U(1)_Y}, s \Big] L_{AHM}^{R_\xi} &=& 0\,,
\eea 
and the two separate global symmetries can therefore co-exist in AHM  physics.}

Now add to (\ref{LagrangianAHMRxiGauge}) 
{ ANY}
$U(1)_Y$ local/gauge invariant, { and therefore BRST invariant},
Lagrangian
$L_{BeyondAHM}^{GaugeInvariant}(A_\mu,\phi;\Phi,\psi)$  
{ involving new bosonic spin-zero fields $\Phi$ 
and new anomaly-cancelling fermionic spin-$\half$ fields $\psi$}
so as to form the extended Abelian Higgs Model (E-AHM).  Then
\bea
\label{U(1)BRSTCommutatorE-AHM}
s L_{E-AHM}^{R_\xi} &=& 0  \nonumber \\
\delta_{U(1)_Y} L_{E-AHM}^{R_\xi} &=&  s \Big( \xi {em_A}  {\bar \eta} H \Omega \Big)\neq 0  \\
\Big[ \delta_{U(1)_Y}, s \Big] L_{E-AHM}^{R_\xi} &=& 0\,. \nonumber
\eea

We will show in this paper that, 
due to (\ref{BRSTTransformations},\ref{U(1)Transformations},\ref{U(1)BRSTCommutatorAHM},
		\ref{U(1)BRSTCommutatorE-AHM}), 
{\it the AHM, and the E-AHM, simultaneously obey 
both  the usual BRST symmetry 
and a global $U(1)_Y$ symmetry that controls Green's functions and on-shell T-Matrix elements.}
We will also show that our effective potential can be made gauge-independent. 

We reason as follows:
\begin{itemize}
\item All aspects of the SSB AHM and E-AHM obey BRST symmetry. 
%
\item In both the special $\xi\to 0$ case of Landau gauge and in the closely-related Lorenz gauge,
\bea
\label{LandauLorenzGauges}
L_{AHM}^{Landau}&=&L_{AHM}^{GaugeInvariant} \nonumber \\
&-&\lim_{\xi\to 0}\frac{1}{2\xi}\Big(\partial_\mu A^\mu +\xi m_A \pi \Big) ^2 - {\bar \eta} \partial^2  \omega\nonumber \\
L_{AHM}^{Lorenz}&=&L_{AHM}^{GaugeInvariant} \nonumber \\
&-&\lim_{\xi\to 0}\frac{1}{2\xi}\Big(\partial_\mu A^\mu \Big) ^2 - {\bar \eta} \partial^2  \omega\,,
\eea 
global $U(1)_Y$ symmetry 
and the larger global $U(1)_Y\otimes$BRST symmetry are preserved: 
{
\bea
\label{LandauLorenzConserveU(1)}
&&\delta_{U(1)_Y} L_{AHM}^{Lorenz} = 0 \nonumber \\
&&\delta_{U(1)_Y} L_{AHM}^{R_\xi} {\buildrel  {\xi \to 0} \over {=\joinrel=\joinrel\Longrightarrow}} \delta_{U(1)_Y} L_{AHM}^{Landau} = 0\, \\
&&s L_{AHM}^{Lorenz} = 0 \nonumber\\
&&s L_{AHM}^{Landau} = 0 \nonumber\,.
\eea  
Similarly for $L_{E-AHM}^{Lorenz}$ and $L_{E-AHM}^{Landau}$.
}
\item Physical states and time-ordered amplitudes of the exact renormalized scalar $\phi = \frac{1}{{\sqrt 2}}(H+i\pi)$ and vector $A_\mu$ obey G. 't Hooft's gauge condition
\cite{tHooft1971} 
\begin{eqnarray}
\label{GaugeConditionsPrimePrime}
\quad &&0=\big< 0\vert T\Big[ \Big( \partial_{\mu}{A}^{\mu}(z)\Big)  \\
&&\quad \quad \times h(x_1)...h(x_N)\pi(y_1)...\pi(y_M)\Big]\vert 0\big>_{\rm connected} \nonumber 
\end{eqnarray}
in Landau or Lorenz gauges.
Here we have N external renormalized scalars $h=H-\HVEV$ (coordinates $x_i$), 
and M external ($CP=-1$) renormalized pseudo-scalars ${ \pi}$ (coordinates $y_i$). 

\item We  prove in Appendix \ref{DerivationWTIAHM} for the AHM 
and in Appendix \ref{DerivationWTIE-AHM} for the E-AHM that,
in Lorenz gauge $\partial_\mu A^\mu =0$, 
scalar-sector connected amputated {\it on-shell} T-Matrix elements
obey (\ref{GaugeConditionsPrimePrime}) and the $U(1)_Y$ symmetry.
Such on-shell WTI are gauge-independent,
(i.e. true for general $R_\xi$ gauges) 
even though (\ref{U(1)TransformationLagrangian}) and (\ref{U(1)BRSTCommutatorE-AHM}) 
show that the BRST-invariant AHM (and E-AHM) Lagrangian  
is not invariant under the $U(1)_Y$ symmetry.

\item We  prove in Appendix \ref{DerivationWTIAHM} for the AHM 
and in Appendix \ref{DerivationWTIE-AHM} for the E-AHM that,
in Lorenz gauge $\partial_\mu A^\mu =0$, 
scalar-sector connected amputated gauge-dependent Green's functions
also obey (\ref{GaugeConditionsPrimePrime}) and the $U(1)_Y$ symmetry.

\item We show that our AHM and E-AHM effective potentials can be made physical (i.e.  gauge-independent) 
in sub-subsection \ref{GaugeIndependence},
thus generalizing them to 't Hooft $R_\xi$ (and all other well-behaved) gauges.
\end{itemize}

\section{The Abelian Higgs model in Lorenz gauge}
\label{AbelianHiggsModel}

\subsection{The Abelian Higgs model in Lorenz gauge}
\label{DefineAHM}

We form the AHM Lagrangian in Lorenz gauge
\begin{eqnarray}
	\label{LagrangianAHM}
	L_{AHM}^{Lorenz}&=&L_{AHM}^{GaugeInvariant} \\
	&+&L_{AHM}^{GaugeFix;Lorenz}  
	+L_{AHM}^{Ghost;Lorenz} \nonumber
\eea
with (\ref{LagrangianAHM-GI}), by writing the gauge-fixing and ghost terms:
\bea
L_{AHM}^{GaugeFix;Lorenz}&=& 
-\lim_{\xi \to0}\frac{1}{2\xi}\Big(\partial_\mu A^\mu  \Big) ^2   \nonumber \\
L_{AHM}^{Ghost;Lorenz}&=&- {\bar \eta} \partial^2  \omega.
\end{eqnarray}
The complex scalar $\phi$ is manifestly renormalizable  
in the linear representation (\ref{LinearScalarRep}).
After SSB, $m_A^2=e^2Y_\phi^2\HVEV^2$.

This paper distinguishes carefully between 
the local BRST-invariant $U(1)_Y$ Lagrangian (\ref{LagrangianAHM}) 
and its three physical modes \cite{Lee1970,Symanzik1970a,Symanzik1970b,Vassiliev1970,ItzyksonZuber}: 
symmetric Wigner mode, the classically scale-invariant  point and physical Goldstone mode.
\smallskip

{\bf 1) Symmetric Wigner mode} $\HVEV=0, m_A^2=0,\mpisq =m_{BEH}^2 = \mu^2_\phi \neq 0$:

This is QED with massless photons and massive charged scalars. 
Thankfully, Nature is not in Wigner mode! 
Further analysis and renormalization of the Wigner mode 
lies outside the scope of this paper. 

{\bf 2) Classically scale-invariant point} $\HVEV=0, m_A^2=0,\mpisq =m_{BEH}^2 = 0$:

Analysis of the scale-invariant point is also outside the scope of this paper.

{\bf 3) Spontaneously broken Goldstone mode} 
$\HVEV\neq0, m_A^2=e^2 \HVEV ^2 \neq0,\mpisq =0, m_{BEH}^2 \neq 0$:

 The famous Abelian Higgs model, 
 with its Nambu-Goldstone boson (NGB) ``eaten" by the Brout-Englert-Higgs mechanism 
(and, as we shall see, WTI governed by the (Goldstone-like) LSS theorem) 
is  the SSB ``Goldstone mode" 
of the BRST-invariant local Lagrangian (\ref{LagrangianAHM}), 
and is the subject of this paper.
We work in  Lorenz gauge for many reasons: 
\begin{itemize}
\item The $U(1)_Y$  ghosts $({\bar \eta},\omega)$  
decouple from the quantum loop dynamics, 
and can (and will) be benevolently ignored going forward.

\item After a subtlety concerning their mixing,
$\pi$ and $A^\mu$ are orthonormal species.
A term $\sim A_\mu \partial^\mu\pi$ 
arises from $\vert D_\mu \phi\vert^2$ after SSB in (\ref{LagrangianAHM});
a term $\sim \pi\partial^\mu A_\mu$ is shown to vanish for physical states in 
(\ref{GaugeConditions},\ref{GaugeConditionExtended}).
The resultant surface term $\partial^\mu \big( \pi A_\mu \big)$ 
vanishes (for physical states) because $A_\mu$ is massive.

\item Only in the SSB Goldstone mode of the BRST-invariant Lagrangian (\ref{LagrangianAHM}), 
and only after first renomalizing in the linear $\phi$ representation, 
does the renormalized Kibble $\phi$ unitary representation
\begin{eqnarray}
\label{Kibble}
\phi = \frac{1}{{\sqrt 2}}\big(H+i\pi\big) &\equiv&  \frac{1}{{\sqrt 2}} {\tilde H}e^{-i Y_\phi {\tilde \pi}/\HVEV} \nonumber \\
{H}=\HVEV + { h}; \quad {\tilde H} &=&\HVEV + {\tilde h} \\
{\tilde \pi} &\equiv&  \HVEV \vartheta  \nonumber
\end{eqnarray}
make sense. Here the $\phi$-hypercharge $Y_\phi =-1$.

\item We will prove to  all-loop-orders  the AHM Lee-Stora-Symanzik theorem (\ref{TMatrixGoldstoneTheoremPrime},
\ref{TMatrixGoldstoneTheorem}),
a gauge theory analogue of an old theorem for global $L\Sigma M$ \cite{Lee1970},
which forces the $\pi$ mass-squared $\mpisq=0$.

\item We use   ``pion-pole dominance" (i.e. $\mpisq =0)$ arguments to derive $U(1)_Y$ SSB WTIs
 (\ref{AdlerSelfConsistencyPrime},\ref{AdlerSelfConsistency},\ref{InternalTMatrix}).

\item 
We prove  with $U(1)_Y$ WTI
that, {\bf in SSB Goldstone mode,  ${\tilde \pi}$ in (\ref{Kibble}) is a Nambu-Goldstone boson} (NGB), and that the resultant SSB gauge theory 
has a  ``shift symmetry" ${\tilde \pi} \to {\tilde \pi} +\HVEV \theta$ for constant $\theta$. 
\end{itemize}

Analysis is done in terms of the exact renormalized interacting fields, 
which asymptotically become the in/out states, i.e. free fields for physical S-Matrix elements.

An important issue is the classification and disposal of relevant operators, 
in this case the ${ \pi}$, $h$ and $A_\mu$ inverse propagators (together with  tadpoles).
 Define the exact renormalized pseudo-scalar propagator 
in terms of a massless $ \pi$, the K$\ddot a$ll$\acute e$n-Lehmann \cite{Bjorken1965,Lee1970} spectral density $\rho^{\pi}_{AHM}$, and  wavefunction renormalization $Z_{AHM}^\phi$. In Lorenz gauge: 
\begin{eqnarray}
\label{pNGBPropagator}
&&\Delta^{\pi}_{AHM}(q^2) = -i(2\pi)^2\langle 0\vert T\left[ \pi(y)\pi(0)\right]\vert 0\rangle\vert^{Fourier}_{Transform} \nonumber \\
&&\quad \quad = \frac{1}{q^2
+ i\epsilon} + \int dm^2 \frac{\rho^{\pi}_{AHM}(m^2)}{q^2-m^2 + i\epsilon}  \\
&&\Big[Z^{\phi}_{AHM}\Big]^{-1} = 1+ \int dm^2 \rho^{\pi}_{AHM}(m^2)\,. \nonumber
\end{eqnarray} 

Define also the BEH scalar propagator in terms of a BEH scalar pole and the (subtracted) spectral density $\rho_{BEH}$, and the {\it same} wavefunction renormalization. We assume $h$ decays weakly, and resembles a resonance:
\begin{eqnarray}
\label{BEHPropagator}
&&\Delta^{BEH}_{AHM}(q^2) = -i (2\pi)^2\langle 0\vert T\left[ h(x) h(0)\right]\vert 0\rangle\vert^{Fourier}_{Transform} \nonumber \\
&& \quad \quad =\frac{1}{q^2-m_{BEH;Pole}^2 + i\epsilon}+ \int dm^2 \frac{\rho^{BEH}_{AHM}(m^2)}{q^2-m^2 + i\epsilon} \nonumber \\
&&\Big[ Z^{\phi}_{AHM}\Big]^{-1} = 1+ \int dm^2 \rho^{BEH}_{AHM}(m^2)  \nonumber \\
&& \int dm^2 \rho^{\pi}_{AHM}(m^2) = \int dm^2 \rho^{BEH}_{AHM}(m^2) 
\end{eqnarray}
The spectral density parts of the propagators are
\begin{eqnarray}
\label{SpectralDensityPropagators}
 \Delta^{\pi ;Spectral}_{AHM}(q^2) &\equiv& \int dm^2 \frac{\rho^{\pi}_{AHM}(m^2)}{q^2-m^2 + i\epsilon}  \nonumber \\ 
\Delta^{BEH; Spectral}_{AHM}(q^2) &\equiv& \int dm^2 \frac{\rho^{BEH}_{AHM}(m^2)}{q^2-m^2 + i\epsilon} \nonumber
\end{eqnarray}
Dimensional analysis of the wavefunction renormalizations (\ref{pNGBPropagator},\ref{BEHPropagator}), 
shows that the contribution of a state  of mass/energy $\sim M_{Heavy}$ 
to the spectral densities $\rho_{AHM}^{\pi}(M_{Heavy}^2)$ and  
$\rho_{AHM}^{BEH}(M_{Heavy}^2)\sim \frac{1}{M_{Heavy}^2}$, 
and similarly its contribution to $\Delta_{AHM}^{\pi ;Spectral}$ and $\Delta_{AHM}^{BEH;Spectral}$ 
includes only irrelevant terms $\sim \frac{1}{M_{Heavy}^2}$. 
The finite Euclidean cut-off contributes only irrelevant terms $\sim \frac{1}{\Lambda^2}$.

\subsection{Rigid/global $U(1)_Y$ WTI and  conserved rigid/global current, for the {\em physical states} of the SSB AHM, in 
Lorenz gauge. Rigid/global $U(1)_Y$ Charge is {\it not} conserved!}
\label{GlobalCurrents}

{ In their seminal work,  E. Kraus and K. Sibold 
\cite{KrausSiboldAHM} 
identified, in the Abelian Higgs model, 
``rigid and current Ward identity ({\it sic}) in accordance with ... BRS[T] invariance.''  
They are called ``deformed" because they have no remnant 
of the original anomaly-free $U(1)_Y$ symmetry. }

The SSB case is tricky because gauge-fixing terms explicitly break both local and global $U(1)_Y$ symmetry in the BRST-invariant Lagrangian.
Still, Kraus and Sibold's construction 
allowed them to demonstrate
(with appropriate normalization conditions) 
proof of all-loop-orders renormalizability and unitarity for the SSB Abelian Higgs model. 
Because their renormalization relies only on  deformed  WTI, 
Kraus and Sibold's results are independent of regularization scheme, 
for any acceptable scheme (i.e. if one exists).%
\footnote{
	E. Kraus and K. Sibold  also constructed, 
	in terms of deformed WTI, 
	all-loop-orders renormalized QED, QCD, 
	and the electro-weak Standard Model 
	\cite{KrausSiboldSM1996,KrausSiboldSM1997,KrausSM1997},
	independent of regularization  scheme. 
	From this grew the powerful technology of ``Algebraic Renormalization", 
	used by them, W. Hollik and others \cite{Hollik2002b}, 
	to renormalize SUSY QED, SUSY QCD, and the MSSM.
}

Nevertheless, Slavnov-Taylor identities \cite{JCTaylor1976} prove that the
on-shell S-Matrix elements of ``physical particles'' 
(i.e. spin $S=0$ scalars $h,\pi$, and $S=1$ transverse gauge bosons $A_\mu$, 
but not fermionic ghosts $({\bar \eta},\omega)$)
are independent of the usual 
(undeformed) 
anomaly-free 
$U(1)_Y$ local/gauge transformations, 
even though these break the Lagrangian's BRST symmetry. 

We observe here that SSB S-Matrix elements are therefore 
also independent of anomaly-free undeformed
$U(1)_Y$ global/rigid transformations, 
resulting in a ``new" global/rigid current 
and appropriate un-deformed $U(1)_Y$ Ward-Takahashi identities.
All this is done without reference to the unbroken Wigner mode and scale-invariant point. 

We are interested in rigid-symmetric relations among 
1-$(h,\pi)$-Irreducible (1-$\phi$-I) connected amputated Green's functions $\Gamma_{N,M}$, 
and among 1-$(h,\pi)$-Reducible (1-$\phi$-R) 
connected amputated transition-matrix (T-Matrix) elements $T_{N,M}$,   
with external $\phi$ scalars. 
Because these are 1-$A_\mu$-R in the AHM, and also 1-$\Phi$-R in the E-AHM  (i.e. reducible by cutting an $A_\mu$ or $\Phi$ line), it is convenient to use the powerful old tools (e.g. canonical quantization) 
from Vintage Quantum Field Theory (Vintage-QFT), a name coined by Ergin Sezgin.  

We focus on the rigid/global AHM current%
\footnote{
	This is related to the rigid/global hypercharge current of the third-generation Global Dirac Neutrino Standard Model ($\GlnuDSM$) explored in \cite{LSS-2}: replace
	$ \pi \to \pi_3,\pi^2 \to {\vec \pi}^2$; un-gauge $A_\mu$;
	add a charged pion current $\pi_2 \partial^{\mu}\pi_1-\pi_1 \partial^{\mu}\pi_2$;
	add the third generation of SM quarks (3 colors, 2 flavors) and leptons (1 charged flavor)
	add one $\nu_R$ with SSB Dirac mass $m_{\nu}$;
	change the overall sign $ {J}^{\mu;SoModified}_{AHM} \to - {J}^{\mu}_{Y;\nu_D SM}$.
	}
constructed with (\ref{U(1)Transformations}),
\begin{eqnarray}
\label{AHMCurrentPrime}
 {J}^{\mu}_{AHM}&=& \pi \partial^\mu H-H\partial^\mu \pi-eA^\mu\Big(\pi^2 + H^2 \Big)\,. \quad \quad
\end{eqnarray}

Rigid/global transformations of the fields arise, as usual, 
from the equal-time commutators (\ref{EqTimeCommAHM}):
\begin{eqnarray}
\label{TransformationsAHMFields}
 \delta_{U(1)_Y} H(t,{\vec y})&=&-i \int d^3 z\left[ {J}^0_{AHM}(t,{\vec z}),H(t,{\vec y})\right] e\Omega \nonumber \\
&=& - \int d^3 z \pi(t,{\vec z})\delta^3({\vec z}-{\vec y})e\Omega \nonumber \\
&=& - \pi (t,{\vec y}) e\Omega \\
 \delta_{U(1)_Y} \pi(t,{\vec y})&=&-i \int d^3 z\left[ {J}^0_{AHM}(t,{\vec z}),\pi(t,{\vec y})\right] e\Omega \nonumber \\
&=&  \int d^3 z H(t,{\vec z})\delta^3({\vec z}-{\vec y}) e\Omega\nonumber \\
&=&  H (t,{\vec y}) e\Omega \nonumber
\end{eqnarray} 
so ${J}^\mu_{AHM}(t,{\vec z})$ serves as a ``proper" local current for commutator purposes. 

In contrast, we show below that, in  Lorenz gauge, $U(1)_Y$ AHM 
(and therefore also $U(1)_Y$ E-AHM) 
has no associated proper global charge $Q$
because $\frac{d}{dt}Q(t)\neq0$. 
(See Eqn. (\ref{ChargeAHMTimeOrderedProductsNotConserved}) below.)

The classical equations of motion reveal a crucial fact: due to gauge-fixing terms in the BRST-invariant Lagrangian (\ref{LagrangianAHM}), the 
classical current 
(\ref{AHMCurrentPrime}) is 
not conserved. In Lorenz gauge  
\begin{eqnarray}
\label{DivergenceAHMCurrentPrime}
\partial_{\mu} {J}^{\mu}_{AHM}&=& H m_A F_A \,,
\end{eqnarray}
with
\begin{eqnarray}
m_A &=& e\HVEV
\end{eqnarray}
and $F_A$ the gauge fixing condition,
\begin{eqnarray}
F_A &\equiv& \partial_{\beta}{A}^{\beta}\,.
\end{eqnarray}
 

The global $U(1)_Y$ current (\ref{AHMCurrentPrime}) 
is, however, conserved by the physical states, 
and therefore still qualifies as a ``real" current for commutator purposes (\ref{TransformationsAHMFields}). 
Strict quantum constraints must be imposed
to force the relativistically-covariant theory of gauge bosons 
to propagate {\it only} its true number of quantum spin $S=1$ degrees of freedom. 
These constraints are 
implemented, in the modern literature, 
by use of spin $S=0$ fermionic Fadeev-Popov ghosts
$({\bar \eta},\omega)$.  
The physical states and their time-ordered products, 
but not the BRST-invariant Lagrangian  (\ref{LagrangianAHM}),  
then obey G. 't Hooft's \cite{tHooft1971} Lorenz-gauge  
gauge-fixing condition (\ref{GaugeConditionsPrimePrime}).

Eqs. (\ref{GaugeConditionsPrimePrime},\ref{GaugeConditions}) restore conservation 
of the rigid/global $U(1)_Y$ current 
for $\phi$-sector connected time-ordered products
\begin{eqnarray}
\label{PhysicalAHMCurrentConservation}
&&\Big< 0\vert T\Big[ \Big( \partial_{\mu}{J}^{\mu}_{AHM}(z) \Big) \\
&&\quad \quad \times h(x_1)...h(x_N) \pi(y_1)...\pi(y_M)\Big]\vert 0\Big>_{\rm connected} =0 \nonumber
\end{eqnarray}
{
It is in this ``physical" connected-time-ordered-product sense 
that the rigid global $U(1)_Y$ ``physical current" is conserved: 
the current-conservation equation 
(\ref{PhysicalAHMCurrentConservation}) is obeyed
only when the divergence of the current is projected in this way
on  the physical states. 
Current conservation is not a property
of the abstract Noether-current operator 
derived from the BRST-invariant Lagrangian (\ref{LagrangianAHM}).}

Appendix A derives 2 towers of quantum $U(1)_Y$ WTIs that 
exhaust the information content of (\ref{PhysicalAHMCurrentConservation});
severely constrain the dynamics (i.e. the connected time-ordered products) 
of the $\phi$-sector physical states of the SSB AHM
and realize the new $U(1)_Y\otimes$BRST symmetry of Section \ref{AbelianHiggsModelSymmetry}. 

{ We might have hoped to also build a charge
\bea
\label{AHMCharge}
Q_{AHM}(t)=\int d^3z J^0_{AHM}(t,{\vec z}) 
\eea
which would be conserved 
when similarly restricted to physical connected time-ordered products:}
\begin{eqnarray}
\label{ChargeAHMTimeOrderedProducts}
&&\Big< 0\vert T\Big[ \Big(  \frac{d}{dt}Q_{AHM}(t) \Big) \nonumber \\
&&\quad \quad \times h(x_1)...h(x_N) \pi(y_1)...\pi(y_M)\Big]\vert 0\Big>_{\rm connected} \nonumber \\
&&\quad =\int d^3 z \Big< 0\vert T\Big[ \Big(  {\vec \nabla} \cdot {\vec J}_{AHM}(t,{\vec z})\Big) \nonumber \\
&&\quad \quad \times h(x_1)...h(x_N) \pi(y_1)...\pi(y_M)\Big]\vert 0\Big>_{\rm connected} \\
&& \quad =\int_{\rm 2-surface} \!\!\!\!\!\!\!\!\!\!\!\!\!\!\!\!\!\! d^2z \quad 
	{\widehat {z}}^{\rm ~2-surface} \cdot \Big< 0\vert 
	T\Big[ \Big(  {\vec J}_{AHM}(t,{\vec z})\Big) \nonumber \\
&&\quad \quad \times h(x_1)...h(x_N) \pi(y_1)...\pi(y_M)\Big]\vert 0\Big>_{\rm connected}\,, \nonumber 
\end{eqnarray}
where we have used Stokes' theorem, 
and ${\widehat {z}_\mu}^{\rm 2-surface}$ is a unit vector normal to the $2$-surface. 
The time-ordered-product constrains the $2$-surface to lie on-or-inside the light-cone.

At a given point on the surface of a large enough 3-volume $\int d^3z$ 
(eg. the volume of all space) 
that lies on or inside the light cone, 
all fields on the $z^{\rm 2-surface}$: 
are asymptotic in-states and out-states; 
are properly quantized as free fields, 
with each field species orthogonal to the others;
and are evaluated at equal times, so that time-ordering is unnecessary.

Nevertheless, the time derivative of this charge
does not vanish even in this restricted physical sense, because, 
{with the symmetry spontaneously broken, }
a specific term in {the surface integral of the right hand side of} 
(\ref{AHMCurrentPrime}) does not vanish:
\begin{eqnarray}
\label{NGBSurfaceIntegralPrime}
\int_{\rm light cone\to\infty} \!\!\!\!\!\!\!\!\!\!\!\!\!\!\!\!\!\!  dz && \quad 
{\widehat {z}}^{\rm light cone} \cdot \Big< 0\vert T\Big[  
\Big(-\HVEV {\vec \nabla} \pi (z) \Big)   \\
&&
\times h(x_1)...h(x_N) \pi_(y_1)...\pi(y_M)\Big]\vert 0\Big> \nonumber 
\neq 0\,.
\end{eqnarray} 
In the SSB AHM, 
$\pi$ is massless (in 
Lorenz gauge), and so
capable of carrying (along the light cone) long-ranged pseudo-scalar forces 
out to the  very ends of the light cone $(z^{\rm light cone}\to \infty)$.

Eqns. (\ref{ChargeAHMTimeOrderedProducts},\ref{NGBSurfaceIntegralPrime}) then show 
that the {\it spontaneously broken $U(1)_Y$ AHM charge is not conserved}, 
even for connected time-ordered products, in Lorenz gauge
\begin{eqnarray}
\label{ChargeAHMTimeOrderedProductsNotConserved}
&&\Big< 0\vert T\Big[ \Big(  \frac{d}{dt}Q_{AHM}(t) \Big)  \\
&&\times  h(x_1)...h(x_N) \pi_{t_1}(y_1)...\pi_{t_M}(y_M)\Big]\vert 0\Big>_{\rm connected} 
\neq 0 \,,\nonumber 
\end{eqnarray}
dashing, { at least for the authors,} all further hope of a conserved charge.

The classic 
proof of the Goldstone theorem \cite{Goldstone1961,Goldstone1962,Kibble1967} 
requires a conserved charge $\frac{d}{dt}Q=0$, 
so that proof fails for spontaneously broken gauge theories.
This is a very famous result \cite{Higgs1964,Englert1964,Guralnik1964,Kibble1967}, 
and allows the spontaneously broken AHM to generate a mass-gap $m_A$ for the vector $A^\mu$ 
and to avoid massless particles in its observable physical spectrum. 
This is true even in 
Lorenz gauge, where there {\em is} a Goldstone theorem,
and consequently  
$\tilde \pi$ is a derivatively coupled (hence massless) NGB \cite{Guralnik1964,Kibble1967},
and where there is an LSS theorem, so $\pi$ is massless.

{\bf Massless $\pi$ (not $\tilde\pi$) is the basis  
of our pion-pole-dominance-based $U(1)_Y$ WTIs}, derived in Appendix A, which give: 
relations among 1-$\phi$-I connected amputated 
$\phi$-sector Greens functions $\Gamma_{N,M}$ (\ref{GreensWTIPrime}, \ref{GreensFWTI}); 
1-soft-pion theorems  (\ref{AdlerSelfConsistencyPrime}, 
	\ref{AdlerSelfConsistency}, \ref{InternalTMatrix}); 
infra-red finiteness for $\mpisq =0$ (\ref{AdlerSelfConsistencyPrime}, \ref{AdlerSelfConsistency});
an LSS
(and Goldstone) theorem (\ref{TMatrixGoldstoneTheoremPrime}, \ref{TMatrixGoldstoneTheorem}); 
vanishing 1-$\phi$-R connected amputated
on-shell $\phi$-sector T-Matrix elements $T_{N,M}$ 
(\ref{AdlerSelfConsistencyPrime},  \ref{InternalTMatrix}) that 
realize the full $U(1)_Y \otimes$BRST symmetry of Section \ref{AbelianHiggsModelSymmetry}.

\subsection{Construction of the scalar-sector effective Lagrangian 
from those $U(1)_Y$ WTIs that  govern connected amputated 1-$\phi$-I Greens functions}
\label{GreensFunctionsAHM}

In Appendix \ref{DerivationWTIAHM} we derive $U(1)_Y$ ``pion-pole-dominance"  1-$\phi$-R
connected amputated T-Matrix  WTI (\ref{InternalTMatrix})for the SSB AHM. 
Their solution is a tower of recursive $U(1)_Y$ WTI (\ref{GreensFWTI}) 
that govern 1-$\phi$-I  $\phi$-sector connected amputated Greens functions $\Gamma_{N,M}$.
For $ \pi$ with $CP=-1$, the result
\begin{eqnarray}
	\label{GreensWTIPrime}
	&&\HVEV\Gamma_{N,M+1}(p_1 ...p_N;0q_1...q_M) \nonumber  \nonumber \\
	&&\quad \quad =\sum ^M_{m=1} \Gamma_{N+1,M-1}(q_mp_1...p_N;q_1...{\widehat {q_m}}...q_M)  \\
	&&\quad \quad -\sum ^N_{n=1}\Gamma_{N-1,M+1}(p_1 ...{\widehat {p_n}}...p_N;p_nq_1...q_M)\nonumber
\end{eqnarray} 
is valid for $N,M \ge0$. 
On the left-hand-side of (\ref{GreensWTIPrime}) there are 
N renormalized $h$ external legs (coordinates x, momenta p), 
M renormalized ($CP=-1$) ${ \pi}$ external legs (coordinates y, momenta q), 
and 1 renormalized soft external $\pi(k_\mu=0)$  (coordinates z, momenta k).
``Hatted" fields with momenta $({\widehat {p_n}},{\widehat {q_m}})$ are omitted.

The rigid $U(1)_Y$ WTI 1-soft-pion theorems (\ref{GreensWTIPrime})
relate a 1-$\phi$-I Green's function with $(N+M+1)$ external fields 
(which include a zero-momentum ${\pi}$), 
to two 1-$\phi$-I  Green's functions with $(N+M)$ external fields.%
\footnote{
	The rigid  $U(1)_Y$ WTI (\ref{GreensWTIPrime})  for the $U(1)_Y$ AHM {\em gauge theory} 
	are a generalization of the classic work of B.W. Lee \cite{Lee1970},  
	who constructed two all-loop-orders renormalized towers of WTI's 
	for the global $SU(2)_L\times SU(2)_R$ Gell-Mann L${\acute e}$vy (GML) model \cite{GellMannLevy1960} 
	with Partially Conserved Axial-vector Currents (PCAC).
	We replace GML's strongly-interacting Linear Sigma Model (\LSM) 
	with a weakly-interacting BEH \LSM, with explicit PCAC breaking. 
	Replace $\sigma \to H$, ${\vec\pi}\to{\pi}$,$m_{\sigma}\to m_{BEH}$ and $f_{\pi} \to \HVEV$, 
	and add local gauge group $U(1)_Y$. 
	This generates a set of  global $U(1)_Y$ WTI 
	governing relations among weak-interaction 1-$\phi$-R T-Matrix elements $T_{N,M}$. 
	A solution-set of those $U(1)_Y$ WTI 
	then govern relations among $U(1)_Y$ 1-$\phi$-I Green's functions $\Gamma_{N,M}$. 
	\hfil\break
	As observed by Lee for GML, 
	one of those on-shell T-Matrix WTI is equivalent to the Goldstone theorem. 
	This equivalence relies on the ability to incorporate a PCAC term into the global theory,
	and then retrieve the spontaneously broken theory in the 
	appropriate zero-explicit-breaking limit, namely $\mpisq\to0$.
	In the gauge theory, although  explicit-breaking terms are allowed by power-counting,
	they violate the BRST symmetry and spoil unitarity \cite{Storaprivate}.
	Yet, the T-matrix WTI persists and forces $\mpisq=0$ in Lorenz gauge, 
	which is now  the new LSS theorem.
	The Goldstone theorem also persists in Lorenz gauge, and forces $m_{\tilde\pi}^2=0$.
	\hfil\break
	Appendix \ref{DerivationWTIAHM} includes, in Table 1,  a translation between the WTI proofs in this paper (a gauge theory) and in  B.W. Lee (a global theory). 
}
The Green's functions $\Gamma_{N,M}(p_1...p_N;q_1...q_M)$ 
are not themselves gauge-independent. 
Furthermore, although 1-$\phi$-I, they are 1-$A^\mu$-Reducible (1-$A^\mu$-R) by cutting a transverse $A_\mu$ gauge boson line.

The 1-$\phi$-I ${ \pi}$  and $h$ inverse propagators are:
\begin{eqnarray}
\label{InversePropagators}
\Gamma_{0,2}(;q,-q) &\equiv& \left[ \Delta_{\pi}(q^2) \right]^{-1} \nonumber \\
\Gamma_{2,0}(q,-q;) &\equiv& \left[ \Delta_{BEH}(q^2) \right]^{-1} 
\end{eqnarray}

We can now form the $\phi$-sector effective momentum-space Lagrangian in Lorenz gauge.
All perturbative quantum loop corrections, to all-loop-orders 
and including all UVQD, log-divergent and finite  contributions, 
are included in the $\phi$-sector effective Lagrangian:
1-$\phi$-I Green's functions $\Gamma_{N,M}(p_1...p_N;q_1...q_M)$; 
wavefunction renormalizations;  
renormalized $\phi$-scalar propagators (\ref{pNGBPropagator},\ref{BEHPropagator}); 
the Brout-Englert-Higgs (BEH) VEV $\HVEV$ (\ref{HVEV}); 
all gauge boson and ghost propagators. 
This includes the full all-loop-orders renormalization of the AHM $\phi$-sector,
originating in quantum loops containing transverse virtual gauge bosons, $\phi$-scalars and ghosts:
$A^\mu;h,\pi;{\bar \eta},\omega$ respectively. 
Because they arise entirely from global $U(1)_Y$ WTI, 
our results are independent of regularization-scheme \cite{KrausSiboldAHM}.

We want to classify operators arising in AHM loops,
and separate the finite operators %
from the divergent ones.
We focus on finite relevant operators, 
as well as quadratic and logarithmically divergent operators.

There are 3 classes of finite operators: 
\begin{itemize}
\item Finite  ${\cal O}_{AHM}^{1/\Lambda^2;Irrelevant}$ vanish as $m_{Weak}^2/ \Lambda^2 \to 0$;

\item ${\cal O}_{AHM}^{\Dim>4;Light}$ are finite dimension $\Dim>4$ operators, 
where only the light degrees of freedom $A^\mu;h,\pi;{\bar \eta},\omega$ 
contribute to   all-loop-orders renormalization;  

\item ${\cal O}_{AHM}^{\Dim\leq4;NonAnalytic}$ are finite dimension $\Dim\leq4$ operators 
that are non-analytic in momenta or in a renormalization scale $\mu^2$  
(e.g. finite renormalization-group logarithms). 
\end{itemize}

All such operators will be ignored.
\begin{eqnarray}
{\cal O}_{AHM}^{Ignore}&=&{\cal O}_{AHM}^{1/\Lambda^2;Irrelevant} + {\cal O}_{AHM}^{\Dim>4;Light} \nonumber \\
&+&{\cal O}_{AHM}^{\Dim\leq4;NonAnalytic}
\end{eqnarray}
Such finite operators appear throughout the $U(1)_Y$ Ward-Takahashi IDs (\ref{GreensWTIPrime}): 
\begin{itemize}
\item $N+M \geq 5$ is  ${\cal O}_{AHM}^{1/ \Lambda^2; Irrelevant}$ and ${\cal O}_{AHM}^{\Dim>4;Light}$;
\item the left hand side of  (\ref{GreensWTIPrime}) for $N+M=4$ is also  
${\cal O}_{AHM}^{1/ \Lambda^2; Irrelevant}$ and ${\cal O}_{AHM}^{\Dim>4;Light}$;
\item  $N+M\leq 4$ operators ${\cal O}_{AHM}^{\Dim\leq 4;NonAnalytic}$  appear in (\ref{GreensWTIPrime}). 
\end{itemize}

Finally, there are  $N+M\leq 4$ operators that are analytic in momenta. 
We expand these in powers of momenta, 
count the resulting dimension of each term in the operator Taylor-series, 
and ignore ${\cal O}_{AHM}^{\Dim>4;Light}$ and ${\cal O}_{AHM}^{1/ \Lambda^2; Irrelevant}$ terms in that series. 

Suppressing gauge fields, the
all-loop-orders renormalized scalar-sector effective Lagrangian 
with operator dimension less than or equal to $4$ is then formed
for ($h, \pi$) with CP=($1,-1$)
\begin{eqnarray}
\label{FormSchwingerPotential}
&& L^{Eff;Wigner,SI,Goldstone}_{AHM;\phi;Lorenz} \nonumber \\
&&\quad \quad =  \Gamma_{1,0}(0;)h +\frac{1}{2!} \Gamma_{2,0}(p,-p;)h^2 \nonumber \\
&&  \quad \quad + \frac{1}{2!} \Gamma_{0,2}(;q,-q)\pi^2 +\frac{1}{3!} \Gamma_{3,0}(000;)h^3  \nonumber \\ 
&& \quad \quad + \frac{1}{2!} \Gamma_{1,2}(0;00) h \pi^2  +\frac{1}{4!} \Gamma_{4,0}(0000;)h^4  \nonumber \\
&&  \quad \quad + \frac{1}{2!2!} \Gamma_{2,2}(00;00) h^2 \pi^2  \nonumber \\
&& \quad \quad +  \frac{1}{4!}\Gamma_{0,4}(;0000)\pi^4 + {\cal O}^{AHM}_{Ignore}   \,.
\end{eqnarray} 

The Ward-Takahashi IDs (\ref{GreensWTIPrime}) for Greens functions 
severely constrain the effective Lagrangian (\ref{FormSchwingerPotential}):
\begin{itemize}

\item $N=0, M=1$ WTI:
\begin{eqnarray}
\label{NM01}
\Gamma_{1,0}(0;) &=& \HVEV \Gamma_{0,2}(;00)\,,
\end{eqnarray}
since no momentum can run into the tadpoles.
\item $N=1, M=1$ WTI:
\begin{eqnarray}
\label{NM11}
\Gamma_{2,0}(-q,q;) &-& \Gamma_{0,2}(;q,-q)  \nonumber \\
&=&\HVEV \Gamma_{1,2}(-q;q0)  \nonumber \\
&=&\HVEV \Gamma_{1,2}(0;00)  + {\cal O}^{AHM}_{Ignore} \nonumber \\ 
\Gamma_{2,0}(00;) &=& \Gamma_{0,2}(;00) + \HVEV \Gamma_{1,2}(0;00) \,.
\end{eqnarray}
\item $N=2, M=1$ WTI:
\begin{eqnarray}
\label{NM21}
\HVEV\Gamma_{2,2}(00;00) &=& \Gamma_{3,0}(000;) -2\Gamma_{1,2}(0;00) 
\end{eqnarray}
\item $N=0, M=3$ WTI:
\begin{eqnarray}
\label{NM03}
\HVEV\Gamma_{0,4}(;0000) &=& 3 \Gamma_{1,2}(0;00)\,.
\end{eqnarray}
\item $N=1, M=3$ WTI:
\begin{eqnarray}
\label{NM13}
0&=&3\Gamma_{2,2}(00;00)-\Gamma_{0,4}(;0000)\,.
\end{eqnarray}

\item $N=3, M=1$ WTI:
\begin{eqnarray}
\label{NM31}
0&=&\Gamma_{4,0}(0000;)-3\Gamma_{2,2}(00;00)  \,.
\end{eqnarray}

\item The quadratic and quartic coupling constants are defined 
in terms of 2-point and 4-point 1-$\phi$-I Green's function:
\begin{eqnarray}
\label{SchwingerWignerDataA}
\Gamma_{0,2}(;00)  &\equiv& - \mpisq \nonumber \\
 \Gamma_{0,4}(;0000)  &\equiv& -6 \lambda_{\phi}^2 \,.
\end{eqnarray} 

\end{itemize}

The  all-loop-orders renormalized $\phi$-sector momentum-space effective Lagrangian (\ref{FormSchwingerPotential}) - 
constrained only by those $U(1)_Y$ WTI governing Greens functions (\ref{GreensWTIPrime}) - 
may be written
\begin{eqnarray}
\label{LEffectiveSM}
&&L^{Eff;Wigner,SI,Goldstone}_{AHM;\phi;Lorenz} = L^{Kinetic;Eff;Wigner,SI,Goldstone}_{AHM;\phi;Lorenz} \nonumber \\
&& \quad \quad \qquad \qquad -V^{Eff;Wigner,SI,Goldstone}_{AHM;\phi;Lorenz} + {\cal O}_{Ignore}^{AHM} \,, 
\end{eqnarray}
with
\begin{eqnarray}
&&L^{Kinetic;Eff;Wigner,SI,Goldstone}_{AHM;\phi;Lorenz}  \\
&&\quad \quad \qquad \qquad =\half \Big( \Gamma_{0,2}(;p,-p) - \Gamma_{0,2}(;00) \Big) h^2 \nonumber \\
&&\quad \quad \qquad \qquad +\half \Big( \Gamma_{0,2}(;q,-q)- \Gamma_{0,2}(;00)  \Big) { \pi}^2\,, \nonumber 
\end{eqnarray}
incorporating
finite non-trivial wavefunction renormalization 
\begin{eqnarray}
\label{WavefunctionA}
\Gamma_{0,2}(;q,-q)-\Gamma_{0,2}(;00) \sim q^2 \,,
\end{eqnarray}
and
\begin{eqnarray}
&&V^{Eff;Wigner,SI,Goldstone}_{AHM;\phi;Lorenz}= \mpisq \Big[ \frac{h^2 + { \pi}^2}{2} +\HVEV h \Big] \nonumber \\
&&\quad \quad \qquad \qquad + \lambda_{\phi}^2 \Big[ \frac{h^2 + { \pi}^2}{2} +\HVEV h \Big]^2 \,.
\end{eqnarray}

The $\phi$-sector effective Lagrangian (\ref{LEffectiveSM}) has insufficient boundary conditions 
to distinguish among the three modes \cite{Lee1970,Symanzik1970a,Symanzik1970b,Vassiliev1970}  
of the BRST-invariant Lagrangian $L_{AHM}$ in (\ref{LagrangianAHM}). 
For example, the effective potential $V^{Eff;Wigner;SI;Goldstone}_{AHM;\phi;Lorenz}$ 
becomes in various limits:%
\footnote{
	The inclusive Gell-Mann L${\acute e}$vy \cite{GellMannLevy1960} effective potential derived \cite{Lynnetal2012} from B.W. Lee's WTI \cite{Lee1970}, reduces to the three different effective potentials of the global $SU(2)_L \times SU(2)_R$ Schwinger model \cite{Schwinger1957}: 
	Schwinger Wigner mode $(\HVEV =0,\mpisq =m_{BEH}^2 \neq 0)$; 
	Schwinger Scale-Invariant point $(\HVEV =0,\mpisq = m_{BEH}^2 =0)$; 
	or Schwinger Goldstone mode $(\HVEV \neq 0,\mpisq = 0;m_{BEH}^2\neq 0)$.
	} 
AHM Wigner mode $(m_A^2=0;\HVEV =0;\mpisq =m_{BEH}^2\neq 0)$; 
AHM  ``Scale-Invariant" (SI) point $(m_A^2=0;\HVEV =0;\mpisq = m_{BEH}^2= 0)$; 
or AHM Goldstone mode $(m_A^2\neq0;\HVEV \neq 0;\mpisq = 0; m_{BEH}^2\neq 0)$;
with
\begin{eqnarray}
\label{WignerSIGoldstonePotentials}
V^{Eff;Wigner}_{AHM;\phi;Lorenz}&=& \mpisq \Big[ \frac{h^2 + { \pi}^2}{2} \Big] + \lambda_{\phi}^2 \Big[ \frac{h^2 + { \pi}^2}{2}  \Big]^2\, \nonumber \\
V^{Eff;ScaleInvariant}_{AHM;\phi;Lorenz}&=& \lambda_{\phi}^2 \Big[ \frac{h^2 + { \pi}^2}{2}  \Big]^2 \,, \\
V^{Eff;Goldstone}_{AHM;\phi;Lorenz}&=& \lambda_{\phi}^2 \Big[ \frac{h^2 + { \pi}^2}{2} +\HVEV h \Big]^2 \,.\nonumber
\end{eqnarray}

Eqn. (\ref{LEffectiveSM}) has exhausted the constraints 
(on the allowed terms in the $\phi$-sector effective Lagrangian) 
due to those $U(1)_Y$ WTIs 
that govern 1-$\phi$-I $\phi$-sector Green's functions $\Gamma_{N,M}$ 
(\ref{GreensWTIPrime}, \ref{GreensFWTI}). 
In order to provide boundary conditions 
that distinguish among the effective potentials in (\ref{WignerSIGoldstonePotentials}), 
we must turn to the $U(1)_Y$ WTIs 
that govern $\phi$-sector 1-$\phi$-R T-Matrix elements $T_{N,M}$.

\subsection{The Lee-Stora-Symanzik (LSS) Theorem:
IR finiteness
and automatic tadpole renormalization}
\label{LSSTheorem}

\begin{quote}
{\it ``Whether you like it or not, 
you have to include in the Lagrangian all possible terms consistent with locality and power counting, 
unless otherwise constrained by Ward identities."}
Kurt Symanzik, in a private letter to Raymond Stora \cite{SymanzikPC} 
\end{quote}

In strict obedience to K. Symanzik's edict, 
we now further constrain the allowed terms in the $\phi$-sector effective Lagrangian,
using  those $U(1)_Y$ Ward-Takahashi identities 
that govern 1-$\phi$-R T-Matrix elements $T_{N,M}$.

In Appendix A, we extend Adler's self-consistency condition  
(originally written for the global 
$SU(2)_L\times SU(2)_R$ \GMLfull Linear Sigma Model with PCAC \cite{Adler1965,AdlerDashen1968}), 
but now derived for the AHM gauge theory in Lorenz gauge (\ref{AdlerSelfConsistency})
\begin{eqnarray}
\label{AdlerSelfConsistencyPrime} 
&&\HVEV T_{N,M+1}(p_1...p_N;0q_1...q_M)\nonumber \\
&& \quad \quad \times (2\pi)^4\delta^4 \Big(\sum_{n=1}^N p_n +\sum_{m=1}^M q_m \Big) \Big\vert^{p_1^2 =p_2^2...=p_N^2=m_{BEH}^2}_{q_1^2 =q_2^2...=q_M^2=0}  \nonumber \\
&& \quad \quad =0 
\end{eqnarray}
The T-matrix elements vanish as one of the pion momenta goes to zero 
provided all other physical scalar particles are on mass-shell. 
In other words, these { are new} 1-soft-pion theorems.
Eqn. (\ref{AdlerSelfConsistencyPrime}) also 
\begin{quote}
{\it ``asserts the absence of infrared (IR) divergences 
in the scalar-sector (of AHM) Goldstone mode  (in Lorenz gauge). 
Although individual Feynman diagrams are IR divergent, 
those IR divergent parts cancel exactly in each order of perturbation theory. 
Furthermore, the Goldstone mode amplitude must vanish in the soft-pion limit.
B.W. Lee \cite{Lee1970}".}
\end{quote}
{\bf It is crucial to note that the external states in ${\mathbf{T_{N,M}}}$ 
are N h's and M ${\mathbf{\pi}}$'s,
not $\mathbf{\tilde\pi}$'s.
We are working in the soft-$\mathbf{\pi}$, not the soft-$\mathbf{\tilde\pi}$ limit.}

The $N=0,M=1$ case of (\ref{AdlerSelfConsistencyPrime}) 
is the LSS theorem  (\ref{TMatrixGoldstoneTheorem}): 
\begin{eqnarray}
\label{TMatrixGoldstoneTheoremPrime}
\HVEV T_{0,2}(;00)=0
\end{eqnarray}
This looks like the Goldstone Theorem
\footnote{
	B.W. Lee \cite{Lee1970} proves two towers of WTI 
	for the global $SU(2)_L\times SU(2)_R$ \GMLfull model (GML) \cite{GellMannLevy1960}
	 in the presence of the Partially Conserved Axial-vector Current (PCAC) hypothesis. 
	 PCAC conserves the vector current $\partial_\mu {\vec J}_{L+R}^{\mu;GML} =0$, 
	 but explicitly breaks the axial-vector current, 
	$\partial_\mu {\vec J}_{L-R}^{\mu;GML} = \gamma_{PCAC}^{GML} {\vec \pi} $.
	Lee identifies the all-loop-orders GML WTI 
	\begin{eqnarray}
	\label{LeePCACGoldstoneTheorem}
	\gamma_{PCAC}^{GML} = -\HVEV \Gamma_{0,2}^{GML}(;00)
	\end{eqnarray}
	as the ``Goldstone theorem in the presence of PCAC." 
	Exact conservation of ${\vec J}_{L-R}^{\mu;GML}$, 
	i.e. $\gamma_{PCAC}^{GML}=0$, 
	is restored for both GML's ``Wigner mode" 
	$\big(\HVEV \equiv 0,\Gamma_{0,2}^{GML}(;00) \neq 0\big)$ 
	and its ``Goldstone mode" $\big(\HVEV \neq 0,\Gamma_{0,2}^{GML}(;00) \equiv 0\big)$. 
	\hfil\break
	The PCAC analogy for the Lorenz-gauge $AHM$ would have been
	\begin{eqnarray}
	\label{StandardModelPCAC}
	\partial_\mu { J}_{L}^{\mu;AHM} &=& \gamma_{PCAC}^{AHM} { \pi} 
		+\HVEV\times ({\rm a~gauge-fixing~term})  \nonumber \\
	\gamma^{AHM}_{PCAC} &=& -\HVEV \Gamma_{0,2}^{AHM}(;00)  \,,
	\end{eqnarray}
	but the $AHM$ is a  local/gauge theory.
	This requires that $\gamma^{AHM}_{PCAC}\equiv 0$ exactly.
	SSB current conservation can be broken only softly 
	by gauge-fixing terms as in (\ref{DivergenceAHMCurrentPrime}), 
	in order to preserve renormalizability and unitarity \cite{JCTaylor1976}. 
	{\bf The Lorenz-gauge AHM LSS theorem
	therefore reads}
	\begin{eqnarray}
	\label{StandardModelNoPCAC}
	\mathbf{
	\gamma^{AHM}_{PCAC} =-\HVEV \Gamma_{0,2}^{AHM}(;00)\equiv 0
	\,,}
	\end{eqnarray}
	as in (\ref{GFGoldstoneTheoremPrime}). 
	The crucial fact here is that, 
	in the SSB Goldstone mode of the $AHM$ 
	(and SSB $E-AHM$, $SM_{Ghosts}^{Bosons}$, $\nu_D SM$ and $E-\nu_D SM$ 
		\cite{LSS-4Proof,SU(2)Proof}) 
	with $\HVEV \neq 0$,
	\begin{equation}
	\label{SSBStandardModelNoPCAC}
	 0\equiv \Gamma_{0,2}^{AHM}(;00)
	=\big[\Delta_\pi^{AHM}(0)\big]^{-1}
	=-\mpisq\,.
	\end{equation}
	This condition that the mass-squared of the pseudoscalar $\pi$ is exactly zero 
	is distinct from, and more powerful than, the more familiar condition $m_{\tilde \pi}^2=0$, 
	i.e. the massless-ness of the NGB $ {\tilde \pi}$. 
	\hfil\break\indent
	We see that (\ref{AdlerSelfConsistencyPrime}) 
	adds information to that contained 
	in Green's function WTI (\ref{GreensWTIPrime},\ref{GreensFWTI}). 
	Beyond IR finiteness \cite{Lee1970}, 
	on-shell T-Matrix WTI (\ref{AdlerSelfConsistencyPrime},\ref{TMatrixGoldstoneTheorem},
		\ref{AdlerSelfConsistency}) 
	provide absolutely crucial constraints on the gauge theory by insisting: 
	that $\gamma^{AHM}_{PCAC}\equiv0$ as in (\ref{StandardModelNoPCAC},\ref{SSBStandardModelNoPCAC}); 
	that the $U(1)_Y$ current is softly broken or conserved as in
	(\ref{DivergenceAHMCurrentPrime},\ref{GaugeConditionsPrimePrime},\ref{PhysicalAHMCurrentConservation}); 
	and that unitarity and renormalizability of the $AHM$ gauge theory 
	is preserved \cite{JCTaylor1976}.
} 
but, since it involves $\pi$ not $\tilde\pi$ it is quite distinct.

{
We will write the  LSS theorem 
(\ref{TMatrixGoldstoneTheoremPrime})
as a further constraint on the 1-$\phi$-I Greens function}
\footnote{
	{
	An SSB 1-$\phi$-R T-Matrix element $T_{N,M}$ consists
	of a sum of many possible diagrams, $T^{i}_{N,M}$, 
	where $i$ indexes all the possibilities.
	We can represent each such diagram 
	as a set of  1-$\phi$-I  vertices $\Gamma_{n,m}$ (which we term beads) 
	attached by $\phi$ propagators,
	in such a way as to leave
	$N$ external $h$ lines and $M$ external $\pi$ lines.
 
	Consider in particular $T_{0,2}(;q,-q)$.
	For any diagram $T^{i}_{0,2}(;q,-q)$ contributing to $T_{0,2}(;q,-q)$,
	there is a unique  ``string'' of $\phi$ propagators
	that threads from end to end through the diagram.
	Each bead on this string has  2 $\phi$-legs, with equal and opposite 4-momenta $q$ and $-q$.
	Since $\Gamma_{0,0}=\Gamma_{0,1}=\Gamma_{1,0}=0$, one cannot have additional
	$\phi$ legs connecting off this main $\phi$ line to another ``side bead''  
	unless they connect in groups of two or more.  But in this case, the main bead
	and the secondary bead cannot be separated by cutting one $\phi$ line,
	and so are part of  the same bead.
	Since $CP=(+1,-1)$ for $(h,\pi)$, and is conserved in this paper, 
	the 1-$h$-Reducible contribution vanishes,
	and so the beads must be connected only by $\pi$s, and each bead is just a
	$\Gamma_{0,2}(;q,-q)$.

	Thus the diagram corresponding to $T^i_{0,2}(;q,-q)$ 
	would appear to consist of $i+1$ copies of  $\Gamma_{0,2}(;q,-q)$
	irreducible vertices connected by $\pi$ propagators $\Delta_\pi(q^2)$,
	and so 
	$T^i_{0,2}(;q,-q) = \Gamma_{0,2}(;q,-q) \left[\Gamma_{0,2}(;q,-q)\Delta_\pi(q^2)\right]^i$.
	$T_{0,2}(;q,-q)$ would then consist of the sum over all such strings.

	However, $\Gamma_{0,2}(;q,-q)\Delta_\pi(q^2)=1$, and so, in fact,
	one should not separately count each $T^i_{0,2}(;q,-q)$, 
	but rather 
	\be
	T_{0,2}(;q,-q)=\Gamma_{0,2}(;q,-q) = \left[\Delta_\pi(q^2)\right]^{-1}\,.
	\ee
	}
}
\begin{eqnarray}
\label{GFGoldstoneTheoremPrime}
\HVEV \Gamma_{0,2}\left(;00\right) &=&\HVEV \big[ \Delta_\pi (0) \big]^{-1} =0
\end{eqnarray}
or, in terms of the $\pi$ mass
\begin{eqnarray}
\label{GFLSSTheorem}
\HVEV\mpisq &=&0
\end{eqnarray}

Evaluating the effective potential
\footnote
{
{In the AHM-forbidden case of $\HVEV \mpisq \neq 0$ imagined in (\ref{PreLSSGoldstoneModePotential}), $\lim_{k_\lambda \to 0} k^2\Delta_\pi (k^2,\mpisq \neq0) =0$ in (\ref{TMatrixIdentity}), so (\ref{SoftPionTMatrixID},\ref{GreensWTIPrime},\ref{AdlerSelfConsistencyPrime}) are still true for all 3 modes: these include Wigner mode and the Scale-Invariant point where $\HVEV=0$, and where the LSS theorem $\HVEV T_{0,2}(;00)=0$, and all the Adler self-consistency conditions, are satisfied trivially.}
}
in (\ref{LEffectiveSM}) with $\HVEV \neq 0$, 
and then in the Kibble representation
\begin{eqnarray}
\label{PreLSSGoldstoneModePotential}
&&V^{Eff;PreLSSGoldstoneMode}_{AHM;\phi;Lorenz}\\
&&= \mpisq \Big[ \frac{h^2 + { \pi}^2}{2} +\HVEV h\Big] 
+\lambda_{\phi}^2 \Big[ \frac{h^2 + { \pi}^2}{2}  +\HVEV h\Big]^2 \nonumber\\
&&=\mpisq \Big[ \phi^{\dagger} \phi -\half \HVEV ^2\Big]+\lambda_{\phi}^2 \Big[ \phi^{\dagger} \phi -\half \HVEV ^2\Big]^2 \nonumber \\
&&=\frac{\mpisq}{2} \Big[ {\tilde H}^2 -\HVEV ^2\Big]
+\frac{\lambda_{\phi}^2}{4} \Big[ {\tilde H}^2 -\HVEV ^2\Big]^2 \nonumber 
\end{eqnarray}
As expected, the NGB ${\tilde \pi}$ has disappeared from the effective potential, 
has purely derivative couplings through its kinetic term, 
and obeys the shift symmetry ${\tilde \pi}\to {\tilde \pi}+\HVEV \theta$ for constant $\theta$. 
In other words, the Goldstone theorem is, on the face of it, already properly enforced. 

Eqn. (\ref{PreLSSGoldstoneModePotential}) appears at first sight to embrace a disaster: 
the term linear in $\phi^{\dagger} \phi -\half \HVEV ^2$ (a remnant of Wigner mode in (\ref{WignerSIGoldstonePotentials})) persists, 
destroying the symmetry of the famous ``Mexican hat",  
and the AHM is not actually in Goldstone mode! 
To the rescue, the LSS theorem (\ref{GFLSSTheorem},\ref{TMatrixGoldstoneTheoremPrime}) 
(and not the Goldstone theorem)
forces the AHM gauge theory fully into its true Goldstone $\HVEV \neq 0$ mode
\footnote{
Ref. \cite{PseudoCW} shows that, { including $d>4$ operators,} the SSB AHM 
scalar potential may be written, from symmetry and WTI alone, in the form
\bea
\label{PseudoCW}
V_{\phi;AHM}^{eff} &=& 
-\sum_{n=2}^{\infty}
\frac{1}{(2n)!}\Gamma_{0,2n}(;0...0)\Big( {\tilde H}^2-\HVEV^2\Big)^{n}
\eea
So can the E-AHM. 
}
\begin{eqnarray}
\label{GoldstoneModePotential}
&&V^{Eff;LSSGoldstoneMode}_{AHM;\phi;Lorenz}= \frac{\lambda_{\phi}^2}{4} \Big[ {\tilde H}^2-\HVEV^2\Big]^2 \nonumber \\
&& \qquad \qquad =\lambda_{\phi}^2 \Big[ \phi^{\dagger} \phi -\half \HVEV ^2\Big]^2
\end{eqnarray}

A central result of this paper is to recognize that,
in order to force equation (\ref{PreLSSGoldstoneModePotential}) 
to equation (\ref{GoldstoneModePotential}),
{\bf the LSS theorem incorporates a ``new" on-shell T-Matrix symmetry,
which is not a full symmetry of the BRST-invariant AHM Lagrangian.}
AHM  physics, but not its Lagrangian, 
has  the $U(1)_Y\otimes$BRST symmetry of Section \ref{AbelianHiggsModelSymmetry},
a conserved current (\ref{AHMCurrentPrime},\ref{PhysicalAHMCurrentConservation}), 
un-deformed WTIs governing connected amputated Green's functions (\ref{GreensWTIPrime}), 
and un-deformed  WTIs 
governing connected amputated on-shell T-Matrix elements (\ref{AdlerSelfConsistencyPrime}).

A crucial effect of the LSS theorem (\ref{GFLSSTheorem}),
together with the  $N=0, M=1$ $U(1)_Y$ Ward-Takahashi Greens function identity (\ref{GreensWTIPrime}), 
is to automatically eliminate tadpoles in (\ref{FormSchwingerPotential})
\begin{eqnarray}
\label{ZeroTadpoles}
\Gamma_{1,0}(0;) &=& \HVEV \Gamma_{0,2}(;00) =0\,,
\end{eqnarray}
so that separate tadpole renormalization is un-necessary.

The proof of the Lee-Stora-Symanzik theorem for the AHM (in Appendix \ref{DerivationWTIAHM}) 
is extended to the E-AHM 
(which includes certain ``Beyond-the-AHM" scalars $\Phi$ 
and { CP-conserving} fermions $\psi$) 
in Appendix \ref{DerivationWTIE-AHM}.
The AHM LSS considerations in this Section \ref{AbelianHiggsModel} 
therefore have their direct corresponding analogs, for the E-AHM, 
in Sections \ref{E-AHM} and \ref{ParticlePhysicsAHM}.  
We shall not needlessly repeat ourselves there.

\subsection{Further constraints on the $\phi$-sector effective Lagrangian:
$m_{BEH}^2 = 2 \lambda _\phi ^2 \HVEV ^2$}
\label{TMatrixAHM}

We rewrite the Goldstone-mode effective Lagrangian (\ref{LEffectiveSM}) 
and effective potential (\ref{PreLSSGoldstoneModePotential}),
but now including the constraint from the LSS theorem:
 (\ref{TMatrixGoldstoneTheoremPrime}, \ref{GFGoldstoneTheoremPrime}, \ref{GFLSSTheorem}):
\begin{eqnarray}
\label{LEffectiveGoldstoneTheorem}
L^{Eff;Goldstone}_{AHM;\phi;Lorenz} \nonumber 
&=& L^{Kinetic;Eff;Goldstone}_{AHM;\phi;Lorenz} \nonumber \\
 &-&V^{Eff;Goldstone}_{AHM;\phi;Lorenz} \nonumber \\
&+&{\cal O}_{Ignore}^{AHM} \nonumber \\
V^{Eff;Goldstone}_{AHM;\phi;Lorenz} &=& \lambda_{\phi}^2  
\left[ {\frac{h^2+{\pi}^2}{2}} +\HVEV h\right] ^2 \,,
\end{eqnarray} 
with wavefunction renormalization
\begin{eqnarray}
\label{GoldstoneWavefunction}
\Gamma_{0,2}(;q,-q)-\Gamma_{0,2}(;00) = q^2 +{\cal O}_{Ignore}^{AHM}\,.
\end{eqnarray}
so the $\phi$-sector Goldstone-mode effective  coordinate-space Lagrangian becomes
\begin{eqnarray}
\label{GoldstoneLagrangian}
L^{Eff;Goldstone}_{AHM;\phi;Lorenz} &=& \vert D_{\mu}\phi \vert ^2 -\lambda_{\phi}^2  \Big[ {\frac{h^2+{\pi}^2}{2}} +\HVEV h\Big] ^2 \nonumber \\
&+&{\cal O}_{Ignore}^{AHM}\,.
\end{eqnarray} 

Eqn. (\ref{GoldstoneLagrangian}) is the $\phi$-sector effective Lagrangian 
of the  {\em spontaneously broken} Abelian Higgs model, in
Lorenz gauge, constrained by the LSS { Theorem}:%
\footnote{
	Imagine we suspected that $\pi$ is not all-loop-orders massless in 
	Lorenz gauge SSB AHM, and simply/naively wrote a mass-squared $m_{\pi:Pole}^2$ 
	into the  $\pi$ inverse-propagator
	\begin{eqnarray}
	\label{GoldstoneTheoremViolation}
	 \left[ \Delta_{\pi}(0)  \right]^{-1}  &\equiv& -\mpisq \\
	&=&-m_{\pi ;Pole}^2\Bigg[ 1+ m_{\pi ;Pole}^2\int dm^2 \frac{\rho_{\pi}(m^2)}{m^2 } \Bigg]^{-1} \,.{\quad}\nonumber
	\end{eqnarray}
	However, the LSS theorem (\ref{GFGoldstoneTheoremPrime}) insists instead that
	\begin{eqnarray}
	\label{GoldstoneTheoremNGBMass}
	&& \HVEV \left[ \Delta_{\pi}(0) \right]^{-1} \equiv -\HVEV \mpisq = \HVEV \Gamma_{0,2}(;00)=0 \quad \quad
	\end{eqnarray} 
	The $\pi$-pole-mass vanishes {\em exactly}.
	\begin{eqnarray}
	\label{PionPoleMass}
	m_{\pi;Pole}^2 &=& \mpisq\Bigg[ 1- \mpisq\int dm^2 \frac{\rho_{\pi}(m^2)}{m^2 } \Bigg]^{-1} =0 \quad \quad 
	\end{eqnarray}
}
\begin{itemize}
\item It is derived from the local BRST-invariant Lagrangian $L_{AHM}$ (\ref{LagrangianAHM}).
\item It includes all divergent
${\cal O}(\Lambda^2),{\cal O}(\ln \Lambda^2)$ and finite terms that arise 
to all perturbative loop-orders 
in the full $U(1)_Y$ gauge theory, due to virtual transverse gauge bosons, $\phi$ scalars and ghosts
($A^\mu;h,\pi;{\bar \eta},\omega$ respectively).
\item It obeys the LSS theorem  
	(\ref{TMatrixGoldstoneTheoremPrime},\ref{GFGoldstoneTheoremPrime}) 
and all other $U(1)_Y$ Ward-Takahashi Green's function and T-Matrix identities. 
\item It obeys the Goldstone theorem in the Lorenz gauge, 
having a massless derivatively coupled NGB, $\tilde\pi$.
\item It is minimized at $(H=\HVEV, {\pi}=0)$,
and obeys stationarity  \cite{ItzyksonZuber} of that true minimum.
\item It preserves the theory's renormalizability and unitarity, 
which require that wavefunction  renormalization, 
$\HVEV_{Bare}=\big[ Z^{\phi}_{AHM}\big]^{1/2}\HVEV$ \cite{LSS-2,Bjorken1965,ItzyksonZuber}, 
forbid UVQD, relevant, or any other dimension-2 operator corrections to $\HVEV$.
\item The LSS theorem (\ref{TMatrixGoldstoneTheoremPrime}) has caused all relevant operators in the spontaneously broken Abelian Higgs model to vanish!

\end{itemize}

In order to make manifest  that $\tilde  \pi$ 
is a true NGB \cite{JCTaylor1976,Georgi2009} in 
Lorenz gauge,
re-write (\ref{GoldstoneLagrangian})  in the  Kibble representation
\cite{Ramond2004,Georgi2009}, with $Y_\phi =-1$ the $\phi$ hypercharge. 
In coordinate space,
\begin{eqnarray}
\label{GoldstoneKibbleAHMLagrangian}
L_{AHM;\phi;Lorenz}^{Eff;Goldstone} &=& 
\frac{1}{2} \left( \partial_{\mu} {\tilde h} \right)^2 \\
&+& \frac{1}{2} e^2 \left(\HVEV+{\tilde h}\right)^2 
		\Big( A_\mu + \frac{1}{e\HVEV}\partial_\mu {\tilde \pi}\Big)^2\nonumber  \\
&-&\frac{\lambda_{\phi}^2}{4}  
	\Big( {\tilde h}^2 +2\HVEV {\tilde h} \Big)^2+{\cal O}_{Ignore}^{AHM} \nonumber
\end{eqnarray}
shows that $\tilde \pi$ has only derivative couplings and, 
for constant $\theta$, a shift symmetry
\begin{eqnarray}
\label{ShiftSymmetryAHM}
{\tilde \pi} &\to& {\tilde \pi} + \HVEV \theta\,.
\end{eqnarray}

The  Green's function WTI (\ref{GreensWTIPrime}) for $N=1,M=1$, 
constrained by the LSS theorem (\ref{GFGoldstoneTheoremPrime}), 
relates the BEH mass  to the coefficient of the $h{\pi}^2$ vertex 
\begin{eqnarray}
\label{WTIHiggsMass}
\Gamma_{2,0}(00;)=\HVEV\Gamma_{1,2}(0;00) \,.
\end{eqnarray} 
Therefore, the BEH mass-squared in (\ref{GoldstoneKibbleAHMLagrangian}),
\begin{eqnarray}
\label{BEHMassAHMS}
m_{BEH}^2 = 2\lambda_{\phi}^2\HVEV^2\,,
\end{eqnarray} 
arises entirely from SSB, 
as does (together with its AHM decays) 
the resonance pole-mass-squared,
\begin{eqnarray}
\label{BEHPoleMassPrime}
m^2_{BEH;Pole}  &=& 2\lambda_\phi^2 \HVEV^2\Big[ 1- 2\lambda_\phi^2 \HVEV^2 
	\int dm^2 \frac{{\rho}^{BEH}_{AHM}(m^2)}{m^2 - i\epsilon} \Big]^{-1} \nonumber \\
	&+& {\cal O}^{Ignore}_{AHM;\phi}\,.
\end{eqnarray}

\section{Extended Abelian Higgs Model: WTI-enforced decoupling of certain heavy matter representations}
\label{E-AHM}

If the Euclidean cutoff $\Lambda^2$ were a true proxy for very heavy $M_{Heavy}^2 \gg m_{Weak}^2$   spin $S=0$ scalars $\Phi$, and $S=\half$ fermions $\psi$, 
we would already be in a position to comment on their de-coupling.
Unfortunately, although the literature seems to cite such proxy, it is simply not true. 
``In order to prove theorems that reveal symmetry-driven results in gauge theories,
one must keep {\em all} of the terms arising from {\em all} Feynman graphs,
not just a selection of interesting terms from a representative subset of Feynman graphs" 
(Ergin Sezgin's dictum). 

\subsection{$\phi$-sector effective Lagrangian for the E-AHM}
\label{LagrangianE-AHM}

\subsubsection{1-$\phi$-I connected amputated $\phi$-sector Green's functions $\Gamma^{\Extended}_{N,M}$}

In Appendix \ref{DerivationWTIE-AHM} we derive 
a tower of recursive $U(1)_Y$ WTI (\ref{ExtendedGreensFWTI}) 
that govern connected amputated 1-$\phi$-I Green's functions  for the E-AHM:
\begin{eqnarray}
	\label{GreensWTIPrimeExtended}
	&&\HVEV\Gamma_{N,M+1}^{\Extended}(p_1 ...p_N;0q_1...q_M) \nonumber  \nonumber \\
	&&\quad \quad =\sum ^M_{m=1} \Gamma^{\Extended}_{N+1,M-1}(q_mp_1...p_N;q_1...{\widehat {q_m}}...q_M) \nonumber \\
	&&\quad \quad -\sum ^N_{n=1}\Gamma^{\Extended}_{N-1,M+1}(p_1 ...{\widehat {p_n}}...p_N;p_nq_1...q_M)
\end{eqnarray} 
valid for $N,M \ge0$.

$\Gamma_{N,M}^{\Extended}$ includes the all-loop-orders renormalization of the $\phi$-sector SSB E-AHM, 
including virtual transverse gauge bosons, $\phi$-scalars, ghosts, 
and new { CP-conserving} scalars and
fermions: 
$A^\mu$; $h$, $\pi$; ${\bar \eta}$, $\omega$; $\Phi$ and $\psi$, respectively.

In the full SSB E-AHM gauge theory, there are 4 classes of {\it finite} operators  
that cannot spoil the  
decoupling of heavy particles:
\begin{itemize}
\item Finite ${\cal O}_{E-AHM;\phi}^{1/\Lambda^2;Irrelevant} $  vanish as $m_{Weak}^2/ \Lambda^2 \to 0$
or $M_{Heavy}^2/ \Lambda^2 \to 0$.
\item Finite ${\cal O}_{E-AHM;\phi}^{\Dim>4;Light} $ are dimension $\Dim>4$ operators,  
where only the light degrees of freedom, 
($A^\mu;h,\pi;{\bar \eta},\omega$ and also $\Phi_{Light}$ and $\psi_{Light}$)
contribute to  all-loop-orders renormalization. 
\item
${\cal O}_{E-AHM;\phi}^{\Dim \leq 4;NonAnalytic;Light} $ are finite-dimension $\Dim \leq4$ operators 
that are non-analytic in momenta or in a renormalization scale $\mu^2$,
where only the light degrees of freedom 
contribute to   all-loop-orders renormalization.
\item ${\cal O}_{E-AHM;\phi}^{1/M_{Heavy}^2;Irrelevant} $  vanish as $m_{Weak}^2/ M_{Heavy}^2 \to 0$.
\end{itemize}

In addition
${\cal O}_{E-AHM;\phi}^{\Dim \leq 4;NonAnalytic;Heavy}$ 
are finite dimension $\Dim \leq4$ operators 
that are non-analytic in momenta or in a renormalization scale $\mu^2$, 
where the heavy degrees of freedom $\Phi_{Heavy};\psi_{Heavy}$
contribute to   all-loop-orders renormalization. 
Analysis of these operators lies outside the scope of this paper.

All such operators will be ignored
\begin{eqnarray}
\label{IgnoreE-AHMOperators}
&&{\cal O}_{E-AHM;\phi}^{Ignore} \nonumber \\
&&\qquad ={\cal O}_{E-AHM;\phi}^{1/\Lambda^2;Irrelevant}+{\cal O}_{E-AHM;\phi}^{\Dim>4;Light} \nonumber \\
&&\qquad +{\cal O}_{E-AHM\phi}^{\Dim \leq 4;NonAnalytic;Light} \nonumber \\
&&\qquad +{\cal O}_{E-AHM\phi}^{\Dim \leq 4;NonAnalytic;Heavy} \nonumber \\
&&\qquad +{\cal O}_{E-AHM;\phi}^{1/M_{Heavy}^2;Irrelevant} 
\end{eqnarray}

Such finite operators appear throughout the extended $U(1)_Y$ WTIs (\ref{GreensWTIPrimeExtended}):
\begin{itemize}
\item $N+M \geq 5$ is  ${\cal O}_{E-AHM;\phi}^{1/ \Lambda^2; Irrelevant}$, ${\cal O}_{E-AHM;\phi}^{\Dim>4;Light}$.
and ${\cal O}_{E-AHM;\phi}^{1/M_{Heavy}^2;Irrelevant}$;
\item The left hand side of  (\ref{GreensWTIPrimeExtended}) for $N+M=4$ is also  
${\cal O}_{E-AHM;\phi}^{1/ \Lambda^2; Irrelevant}$, ${\cal O}_{E-AHM;\phi}^{\Dim>4;Light}$ 
and ${\cal O}_{E-AHM;\phi}^{1/M_{Heavy}^2;Irrelevant}$.
\item  $N+M\leq 4$ operators ${\cal O}_{E-AHM;\phi}^{\Dim\leq 4;NonAnalytic;Light}$  also appear in (\ref{GreensWTIPrimeExtended}). 
\end{itemize}

Finally, there are  $N+M\leq 4$ operators that are analytic in momenta. 
We expand these in powers of momenta, 
count the resulting dimension of each term in the operator Taylor-series, 
and then ignore ${\cal O}_{E-AHM;\phi}^{\Dim>4;Light}$,
${\cal O}_{E-AHM;\phi}^{1/ \Lambda^2; Irrelevant}$ 
and ${\cal O}_{E-AHM;\phi}^{1/M_{Heavy}^2;Irrelevant}$ in that series.

 Suppressing gauge fields, the 
all-loop-orders renormalized $\phi$-sector effective momentum-space Lagrangian, with operator dimensions $\leq4$, for E-AHM is then formed
for ($h,\pi$) external particles with CP=($1,-1$)
\begin{eqnarray}
\label{FormSchwingerPotentialExtended}
&& L^{Eff;Wigner,SI,Goldstone}_{E-AHM;\phi} =  \Gamma_{1,0}^{\Extended}(0;)h \nonumber \\
&& \quad \quad +\frac{1}{2!} \Gamma^{\Extended}_{2,0}(p,-p;)h^2 \nonumber \\
&&  \quad \quad + \frac{1}{2!} \Gamma^{\Extended}_{0,2}(;q,-q)\pi^2 +\frac{1}{3!} \Gamma^{\Extended}_{3,0}(000;)h^3  \nonumber \\ 
&& \quad \quad + \frac{1}{2!} \Gamma^{\Extended}_{1,2}(0;00) h \pi^2  +\frac{1}{4!} \Gamma^{\Extended}_{4,0}(0000;)h^4  \nonumber \\
&&  \quad \quad + \frac{1}{2!2!} \Gamma^{\Extended}_{2,2}(00;00) h^2 \pi^2  \\
&& \quad \quad +  \frac{1}{4!}\Gamma^{\Extended}_{0,4}(;0000)\pi^4 + {\cal O}^{E-AHM}_{Ignore}\,.  \nonumber 
\end{eqnarray}

The $U(1)_Y$ Ward-Takahashi IDs (\ref{GreensWTIPrimeExtended})  
severely constrain the effective Lagrangian of the E-AHM:
\begin{itemize}
\item $N=0, M=1$ WTI:
\begin{eqnarray}
\label{NM01E-AHM}
\Gamma^{\Extended}_{1,0}(0;) &=& \HVEV \Gamma^{\Extended}_{0,2}(;00) 
\end{eqnarray}
since no momentum can run into the tadpoles.
\item $N=1, M=1$ WTI: 
\footnote{
	In previous papers on  $SU(2)_L\times SU(2)_R$ \GMLfull $L\Sigma M$ \cite{GellMannLevy1960}, 
	we have written the $N=1,M=1$  WTI 
	as a mass-relation between the BEH $h$ scalar 
	and the {\em pseudo}-Nambu-Goldstone boson ${\pi}$ pseudo-scalar. 
	In the K$\ddot a$ll$\acute e$n-Lehmann representation
	\begin{eqnarray}
	\label{MassRelation}
	&&m_{BEH}^2 = \mpisq + 2\lambda_{\phi}^2 \HVEV ^2 \\
	&&\mpisq  =  \Bigg[ \frac{1}{m_{\pi ;Pole}^2}+ \int dm^2 \frac{\rho_{\pi}(m^2)}{m^2 } \Bigg]^{-1} \nonumber \\
	&&m_{BEH}^2 =  \Bigg[ \frac{1}{m_{BEH;Pole}^2}+ \int dm^2 \frac{\rho_{BEH}(m^2)}{m^2 } \Bigg]^{-1} \nonumber 
	\end{eqnarray}
	so that
	\begin{eqnarray}
	\label{MpisqToZer0}
	m^{2}_{BEH}  \quad
	 {\buildrel  {\mpisq, m_{\pi ;Pole}^2 \to 0} \over {=\joinrel=\joinrel=\joinrel=\joinrel=\joinrel=\joinrel=\joinrel=\joinrel=\joinrel=\joinrel\Longrightarrow}} 
	\quad 2\lambda_{\phi}^2 \HVEV ^2\quad \quad
	\end{eqnarray}
	arises entirely from spontaneous symmetry breaking, in obedience to the $U(1)_Y$ on-shell T-Matrix WTI, i.e. the LSS theorem.
}
\begin{eqnarray}
\label{NM11E-AHM}
\Gamma^{\Extended}_{2,0}(-q,q;) - &\Gamma^{\Extended}_{0,2}(;q,-q)&  \nonumber \\
=\HVEV &\Gamma^{\Extended}_{1,2}(-q;q0)&   \\
=\HVEV &\Gamma^{\Extended}_{1,2}(0;00)&  + {\cal O}^{E-AHM}_{Ignore} \nonumber \\ 
\Gamma^{\Extended}_{2,0}(00;) = &\Gamma^{\Extended}_{0,2}(;00)&  \nonumber \\
+ \HVEV &\Gamma^{\Extended}_{1,2}(0;00)\,.\nonumber
\end{eqnarray}
\item $N=2, M=1$ WTI: 
\begin{eqnarray}
\label{NM21E-AHM}
\HVEV\Gamma^{\Extended}_{2,2}(00;00) &=& \Gamma^{\Extended}_{3,0}(000;)  \\
&-&2\Gamma^{\Extended}_{1,2}(0;00)\,. \nonumber
\end{eqnarray}
\item $N=0, M=3$ WTI: 
\begin{eqnarray}
\label{NM03E-AHM}
\HVEV\Gamma^{\Extended}_{0,4}(;0000) &=&3 \Gamma^{\Extended}_{1,2}(0;00)\,.
\end{eqnarray}
\item $N=1, M=3$ WTI:
\begin{eqnarray}
\label{NM13E-AHM}
0&=&3\Gamma^{\Extended}_{2,2}(00;00)-\Gamma^{\Extended}_{0,4}(;0000)\,.
\end{eqnarray}
\item $N=3, M=1$ WTI: 
\begin{eqnarray}
\label{NM31E-AHM}
0&=&\Gamma^{\Extended}_{4,0}(0000;)-3\Gamma^{\Extended}_{2,2}(00;00) \,.
\end{eqnarray}
\item The quadratic and quartic coupling constants are defined in terms of 2-point and  4-point 1-SPI connected amputated GF
\begin{eqnarray}
\label{SchwingerWignerDataB}
 \Gamma^{\Extended}_{0,2}(;00)  &\equiv& -\mpisq \,. \nonumber \\
 \Gamma^{\Extended}_{0,4}(;0000)  &\equiv& -6 \lambda_{\phi}^2 \,.
\end{eqnarray} 
\end{itemize}

 Still suppressing gauge fields, the  all-loop-orders renormalized $\phi$-sector effective Lagrangian (\ref{FormSchwingerPotentialExtended}), severely constrained only by the $U(1)_Y$ WTI
 governing connected amputated Greens functions (\ref{GreensWTIPrimeExtended}),  may be written
\begin{eqnarray}
\label{LEffectiveE-AHM}
&&L^{Eff;Wigner,SI,Goldstone}_{E-AHM;\phi} = \nonumber\\
&&\quad\quad L^{Kinetic}_{E-AHM;\phi}   
    - V^{Wigner,SI,Goldstone}_{E-AHM;\phi} +{\cal O}^{Ignore}_{E-AHM;\phi}  \nonumber \\
&&L^{Kinetic}_{E-AHM;\phi}  \\
&&\quad \quad =\half \Big( \Gamma^{\Extended}_{0,2}(;p,-p) - \Gamma^{\Extended}_{0,2}(;00) \Big) h^2 \nonumber \\
&&\quad \quad +\half \Big( \Gamma^{\Extended}_{0,2}(;q,-q)- \Gamma^{\Extended}_{0,2}(;00)  \Big) { \pi}^2 \nonumber \\
&&V^{Wigner,SI,Goldstone}_{E-AHM;\phi} = \mpisq \Big[ \frac{h^2 + { \pi}^2}{2} +\HVEV h \Big] \nonumber \\
&& \quad \quad +\lambda_{\phi}^2 \Big[ \frac{h^2 + { \pi}^2}{2} +\HVEV h \Big]^2 \nonumber
\end{eqnarray} 
with finite non-trivial wavefunction renormalization 
\be
\label{WavefunctionB}
\Gamma^{\Extended}_{0,2}(;q,-q)-\Gamma^{\Extended}_{0,2}(;00) \sim q^2 \,.
\ee

The $\phi$-sector effective Lagrangian (\ref{LEffectiveE-AHM}) for the E-AHM 
has insufficient boundary conditions to distinguish among the three modes 
of the  BRST-invariant Lagrangian $L_{E-AHM}$.%
\footnote{
	\label{InstructiveFootnote}
	It is instructive, and we argue dangerous, 
	to ignore vacuum energy and rewrite the potential in (\ref{LEffectiveE-AHM}) as:
	\begin{eqnarray}
		\label{VViolatesGoldstoneTheoremExtendedAHM}
		V^{Wigner;SI;Goldstone}_{E-AHM} &=& {\lambda_\phi^2}\left[ \phi^\dagger \phi - \half \left(\HVEV^2 - { \frac{\mpisq}{\lambda_\phi^2} } \right) \right]^2 \quad \quad
	\end{eqnarray} 
	using $\frac{h^2 + { \pi}^2}{2} +\HVEV h = \phi^\dagger \phi - \half \HVEV^2$.
	If one then minimizes 
	$V^{Wigner;SI;Goldstone}_{E-AHM} $ 
	while ignoring the crucial constraint imposed by the LSS Theorem,
	the resultant (incorrect and un-physical) minimum
	$ \big< H \big>_{unphysical}^2 \equiv \Big( \HVEV^2 - \frac{\mpisq} {\lambda_\phi^2}  \Big)$ 
	does not distinguish properly 
	among the three modes
	of (\ref{VViolatesGoldstoneTheoremExtendedAHM}).
	\newline \indent
	At issue is 
	renormalized
	\begin{eqnarray}
	\label{FTExtendeAHM}
	\mpisq &=& \mu_{\phi;Bare}^2 +C_\Lambda \Lambda^2+C_{BEH} m_{BEH}^2 +\delta 
m_{\pi ;Miscellaneous}^2 \nonumber \\
	&+& M_{Heavy}^2\Big[ C_{Heavy} +C_{Heavy;ln} \ln{(M_{Heavy}^2)}  \nonumber \\
	&+&C_{Heavy;\ln{\Lambda}} \ln{(\Lambda^2)} +++\Big] +\lambda_\phi^2 \HVEV^2
	\end{eqnarray}
	where the $C$'s are constants, 
$\delta m_{\pi ;Miscellaneous}^2$ sweeps up the remaining loop-corrections,
and $m_{BEH}^2=\mpisq +2\lambda^2\HVEV^2$. For pedagogical clarity, we display the linearized approximation to contributions $\sim M_{Heavy}^2$ explicitly.
	It is fashionable to simply drop the UVQD term $C_\Lambda \Lambda^2$ in (\ref{FTExtendeAHM}), 
	and argue that it is somehow an artifact of dimensional regularization (DR), 
	even though M.J.G. Veltman \cite{Veltman1981} showed 
	that UVQD appear at 1-loop in the SM
	and are properly handled by DR's poles at dimension $\Dim=2$.  
	We keep UVQD. 
	For pedagogical efficiency, 
	we have included in  (\ref{FTExtendeAHM}) 
	terms with $M_{Heavy}^2 \gg m_{Weak}^2$, 
	such as might arise in  Majorana neutrino or Beyond-AHM physics
	(cf. Sub-section  \ref{HeavyNeutrino} or \ref{HeavyScalar}).
	\newline \indent
	In the spontaneously broken (Goldstone) mode,
	where $\HVEV \neq0$, 
	as in AHM, so too in the E-AHM,
	in obedience to  the LSS theorem (\ref{TMatrixGoldstoneTheoremE-AHM})
	the bare counter-term $\mu_{\phi;Bare}^{2}$ in (\ref{FTExtendeAHM}) 
	is defined by 
	\begin{eqnarray}
	\label{GoldstoneFTExtendedAHM}
	\mpisq  \equiv 0 \,.
	\end{eqnarray} 
	We show  below that, 
	for constant $ \theta$, 
	the zero-value in  (\ref{GoldstoneFTExtendedAHM}) 
	is protected by the LSS theorem and a NGB shift symmetry 
	\begin{eqnarray}
		\label{BEHFTShiftSymmetry}
		{\tilde { \pi}}\to {\tilde { \pi}} + \HVEV {\ \theta} 
	\end{eqnarray}
	\newline\indent
	Minimization of (\ref{VViolatesGoldstoneTheoremExtendedAHM})
	violates stationarity of the true minimum at $\HVEV$ \cite{ItzyksonZuber} 
	and destroys the theory's renormalizability and unitarity, 
	which require that dimensionless wavefunction  renormalization 
	$\HVEV_{Bare}=\Big[ Z^{\phi}\Big] ^{1/2}\HVEV$ 
	contain no relevant operators \cite{Lynn2011,Bjorken1965,ItzyksonZuber}. 
	The crucial observation is that, in obedience to the  LSS theorem, 
	$Renormalized (\HVEV_{Bare}^2) \neq  \HVEV_{unphysical}^2$.
} 
The effective potential $V^{Wigner,SI,Goldstone}_{E-AHM;\phi}$ 
becomes in various limits: 
E-AHM Wigner mode $(m_A^2=0;\HVEV =0;\mpisq =m_{BEH}^2\neq 0)$; 
E-AHM scale-invariant point $(m_A^2=0;\HVEV =0;\mpisq = m_{BEH}^2= 0)$; 
or E-AHM Goldstone mode $(m_A^2\neq0;\HVEV \neq 0;\mpisq = 0; m_{BEH}^2\neq 0)$.
\begin{eqnarray}
\label{WignerSIGoldstonePotentialsE-AHM}
V^{Wigner}_{E-AHM;\phi} &=& \mpisq \Big[ \frac{h^2 + { \pi}^2}{2} \Big] + \lambda_{\phi}^2 \Big[ \frac{h^2 + { \pi}^2}{2}  \Big]^2 \nonumber \\
V^{ScaleInvariant}_{E-AHM;\phi} &=& \lambda_{\phi}^2 \Big[ \frac{h^2 + { \pi}^2}{2}  \Big]^2 \\
V^{Goldstone}_{E-AHM;\phi} &=& \lambda_{\phi}^2 \Big[ \frac{h^2 + { \pi}^2}{2} +\HVEV h \Big]^2 \nonumber\,.
\end{eqnarray}

Eqn. (\ref{LEffectiveE-AHM})  has exhausted the constraints 
on the allowed terms in the $\phi$-sector effective E-AHM Lagrangian 
due to those $U(1)_Y$ WTIs 
that govern 1-$\phi$-I connected amputated Green's functions $\Gamma^{\Extended}_{N,M}$. 

\subsubsection{1-$\phi$-R connected amputated $\phi$-sector T-Matrix elements $T^{\Extended}_{N,M}$:}

In order to provide such boundary conditions 
(which distinguish among the effective potentials in
(\ref{WignerSIGoldstonePotentialsE-AHM})), 
we turn to the off-shell  T-Matrix
and strict obedience to the wisdom of K. Symanzik's edict  at the top of Subsection \ref{TMatrixAHM}:
{\bf ``... unless otherwise constrained by Ward identies"}. 
We can further constrain the allowed terms in the $\phi$-sector effective E-AHM Lagrangian
with those $U(1)_Y$ Ward-Takahashi identities that govern 1-$\phi$-R T-Matrix elements.

In Appendix \ref{DerivationWTIE-AHM}, 
we derive three such identities governing 1-$\phi$-R connected amputated T-Matrix elements 
$T_{N,M}^{\Extended}$ in the $\phi$-sector of the E-AHM:
\begin{itemize}
\item  Adler self-consistency conditions  
(originally written for the {\em global} $SU(2)_L\times SU(2)_R$ \GMLfull model 
with PCAC \cite{Adler1965,AdlerDashen1968}) 
constrain  the E-AHM {\em gauge theory's} effective $\phi$-sector Lagrangian in 
Lorenz gauge (\ref{ExtendedAdlerSelfConsistency})
\begin{eqnarray}
\label{AdlerSelfConsistencyE-AHM} 
&&\quad\HVEV T^{\Extended}_{N,M+1}(p_1...p_N;0q_1...q_M) \\
&& \quad\quad \times (2\pi)^4\delta^4 \Big(\sum_{n=1}^N p_n +\sum_{m=1}^M q_m \Big) \Big\vert^{p_1^2 =p_2^2...=p_N^2=m_{BEH}^2}_{q_1^2 =q_2^2...=q_M^2=0}  \nonumber \\
&& \quad \quad =0\,. \nonumber
\end{eqnarray}
The E-AHM T-matrix vanishes as one of the pion momenta goes to zero (i.e. 1-soft-pion theorems), provided all other physical scalar particles are on mass-shell. 
Eqn. (\ref{AdlerSelfConsistencyE-AHM}) also shows that there are no infrared (IR) divergences in the ($\phi$-sector E-AHM) Goldstone mode 
(in Lorenz gauge) 
\cite{Lee1970}.
\item The  $N=0,M=1$ case of (\ref{AdlerSelfConsistencyE-AHM})  comprises the LSS theorem (\ref{TMatrixGoldstoneTheoremExtended}) \cite{Lee1970}: 
\begin{eqnarray}
\label{TMatrixGoldstoneTheoremE-AHM}
\HVEV T^{\Extended}_{0,2}(;00) &=&0  \\
\HVEV \Gamma^{\Extended}_{0,2}(;00) &\equiv&-\HVEV\mpisq =0\,.\nonumber\nonumber
\end{eqnarray}
\item Define $T_{N,M+1}^{\Extended;External}$ 
as the 1-$ \phi$-R $\phi$-sector T-Matrix  
with one soft $\pi (q_\mu =0)$ attached to an external leg,
as in  Figure \ref{fig:LeeFig10}.
\begin{figure}
\centering
\includegraphics[width=1\hsize,trim={0cm 5cm 0cm 5.5cm},clip]{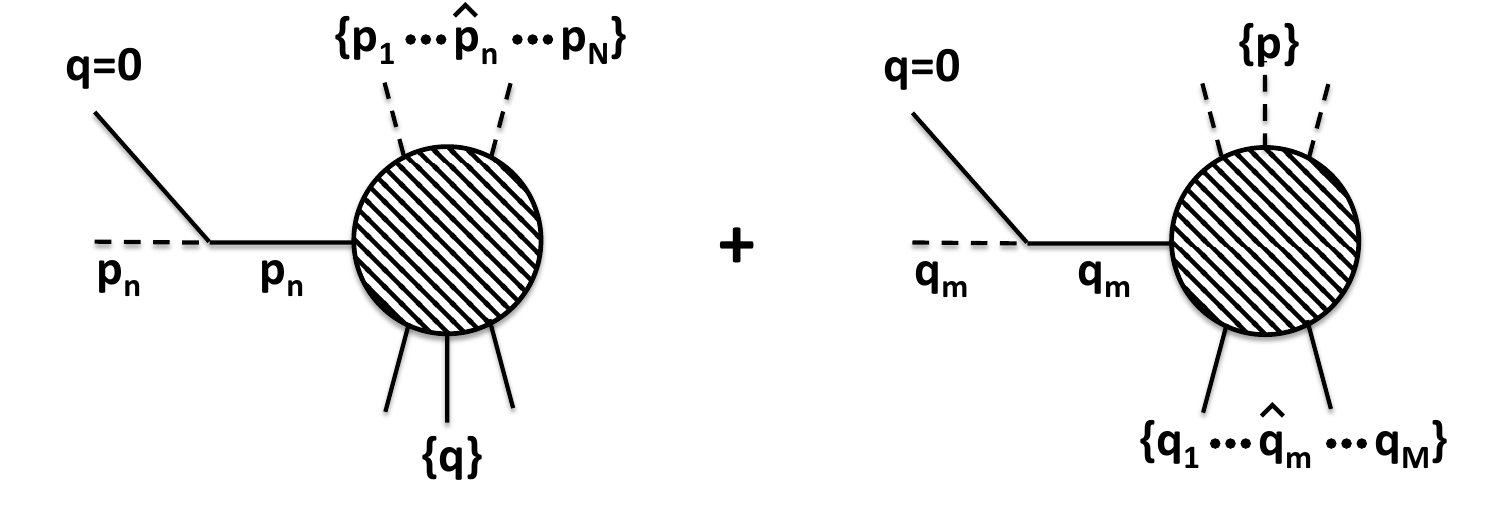}
\caption{
\label{fig:LeeFig10} 
$T_{N,M+1}^{\Extended;External}$: Hashed circles are 1-$\phi$-R $T^{\Extended}_{N,M}$, solid lines $\pi$, dashed lines $h$. One (zero-momentum) soft pion is attached to an external leg in all possible ways. $T^{\Extended}_{N,M}$ is 1-$A^\mu$-R by cutting an $A^\mu$ line, and also 1-$\Phi$-R by cutting a $\Phi$ line. Fig. \ref{fig:LeeFig10}  is the E-AHM analogy of B.W. Lee's Figure 10 \cite{Lee1970}. The same graph topologies, but without internal Beyond-AHM $\Phi \psi$ heavy matter, are used in the proof of (\ref{InternalTMatrix}) for the (unextended) AHM.
}
\end{figure} 
Now separate
\begin{eqnarray}
\label{DefineInternalTMatrixE-AHM}
&&T^{\Extended}_{N,M+1}(p_1...p_N;0q_1...q_M)  \\
&&\quad \quad =T_{N,M+1}^{\Extended;External}(p_1...p_N;0q_1...q_M) \nonumber \\
&&\quad \quad +T_{N,M+1}^{\Extended;Internal}(p_1...p_N;0q_1...q_M)  \nonumber
\end{eqnarray}
Appendix \ref{DerivationWTIE-AHM} (\ref{InternalTMatrixExtended})  proves that 
\begin{eqnarray}
\label{InternalTMatrixE-AHM}
&&\quad\HVEV T_{N,M+1}^{\Extended;Internal}(p_1...p_N;0q_1...q_M) \\
&&\quad \quad =\sum_{m=1}^M T^{\Extended}_{N+1,M-1}(q_mp_1...p_N;q_1....{\widehat{q_m}}...q_M)  \nonumber \\
&&\quad \quad -\sum_{n=1}^N T^{\Extended}_{N-1,M+1}(p_1...{\widehat{p_n}}...p_N;p_nq_1...q_M) \nonumber 
\end{eqnarray}
\end{itemize}

The $U(1)_Y$ WTIs  (\ref{GreensWTIPrimeExtended},\ref{ExtendedGreensFWTI}) governing  1-$\phi$-I connected amputated  Greens functions 
$\Gamma^{\Extended}_{N,M}$ are solutions to (\ref{InternalTMatrixE-AHM},\ref{InternalTMatrixExtended}).

We re-write the E-AHM effective $\phi$-sector Lagrangian (\ref{LEffectiveE-AHM}) but now include the constraint from the  LSS theorem (\ref{TMatrixGoldstoneTheoremE-AHM},\ref{TMatrixGoldstoneTheoremExtended}), in the SSB $\HVEV\neq 0$ case, $\mpisq=0$:
\begin{eqnarray}
\label{LEffectiveGoldstoneTheoremE-AHM}
L^{Eff;Goldstone}_{E-AHM;\phi} &=& L^{Kinetic}_{E-AHM;\phi}+{\cal O}_{Ignore}^{E-AHM} \nonumber \\
&-&V^{Eff;Goldstone}_{E-AHM;\phi}  \nonumber \\
V^{Eff;Goldstone}_{E-AHM;\phi} &=&\lambda_{\phi}^2  \left[ {\frac{h^2+{\pi}^2}{2}} +\HVEV h\right] ^2 \quad \quad
\end{eqnarray} 
and wavefunction renormalization
\begin{eqnarray}
\label{GoldstoneWavefunctionE-AHM}
&&\Gamma^{\Extended}_{0,2}(;q,-q)-\Gamma^{\Extended}_{0,2}(;00) \nonumber \\
&&\quad \quad = q^2 +{\cal O}_{Ignore}^{E-AHM}
\end{eqnarray}

A crucial effect of the LSS theorem,
together with the  $N=0, M=1$ Ward-Takahashi Greens function identity (\ref{GreensWTIPrimeExtended}), is to automatically eliminate tadpoles in (\ref{LEffectiveGoldstoneTheoremE-AHM})
\begin{eqnarray}
\label{ZeroTadpolesE-AHM}
\Gamma^{\Extended}_{1,0}(0;) &=& \HVEV \Gamma^{\Extended}_{0,2}(;00) =0\,,
\end{eqnarray}
so that separate tadpole renormalization is un-necessary.

We form the effective Goldstone-mode  Lagrangian governing low-energy $\phi$-sector physics 
in coordinate space%
\footnote{
	It is not lost on the authors that, 
	since we derived it from connected amputated Greens functions 
	(where all vacuum energy and disconnected vacuum bubbles are absorbed into an overall phase, 
	which cancels exactly in the S-matrix \cite{Bjorken1965,ItzyksonZuber}), 
	the vacuum energy in  $V^{Eff;Goldstone}_{E-AHM;\phi}$ in (\ref{GoldstoneLagrangianE-AHM}) 
	is exactly zero.
} 
\begin{eqnarray}
\label{GoldstoneLagrangianE-AHM}
L_{E-AHM;\phi}^{Eff;Goldstone} &=& \vert D_{\mu}\phi \vert ^2 -V_{E-AHM;\phi}^{Eff;Goldstone}  \nonumber \\
&+&{\cal O}^{Ignore}_{E-AHM;\phi} \nonumber \\
V_{E-AHM;\phi}^{Eff;Goldstone} &=& \lambda_{\phi}^2  \Big[ {\frac{h^2+{\pi}^2}{2}} +\HVEV h\Big] ^2
\,.
\end{eqnarray}

Eqn.  (\ref{GoldstoneLagrangianE-AHM}) is the $\phi$-sector effective  Lagrangian 
of the  {\em spontaneously broken} E-AHM in 
Lorenz gauge:
\begin{itemize}
\item It obeys the LSS theorem (\ref{TMatrixGoldstoneTheoremE-AHM},\ref{TMatrixGoldstoneTheoremExtended}) and all other $U(1)_Y$ WTI (\ref{GreensWTIPrimeExtended},\ref{AdlerSelfConsistencyE-AHM},\ref{TMatrixGoldstoneTheoremE-AHM},\ref{InternalTMatrixE-AHM},\ref{ExtendedAdlerSelfConsistency},\ref{TMatrixGoldstoneTheoremExtended},\ref{InternalTMatrixExtended},\ref{ExtendedGreensFWTI}).
\item It obeys the Goldstone theorem in the Lorenz gauge, 
having a massless derivatively coupled NGB, $\tilde\pi$.
\item It is minimized at $(H=\HVEV, {\pi}=0)$, and obeys stationarity  \cite{ItzyksonZuber} of that true minimum.
\item It preserves the theory's renormalizability and unitarity, 
which require that wavefunction  renormalization, 
$\HVEV_{Bare} =\Big[ Z^{\phi}_{E-AHM}\Big]^{1/2}\HVEV$ \cite{LSS-2,Bjorken1965,ItzyksonZuber}, 
forbid {\em any} relevant operator corrections to $\HVEV$.
\item It includes all divergent
${\cal O}(\Lambda^2),{\cal O}(\ln \Lambda^2)$ and finite terms that arise 
to all perturbative loop-orders in the full $U(1)_Y$ theory, 
due to virtual transverse gauge bosons, AHM scalars, ghosts, 
and new { CP-conserving} scalars and  fermions
($A^\mu$; $h,\pi$;${\bar \eta},\omega$; and $\Phi$, $\psi$ respectively). 

\item {\bf The LSS theorem
(\ref{TMatrixGoldstoneTheoremE-AHM},\ref{TMatrixGoldstoneTheoremExtended})
has caused all relevant operators in (\ref{GoldstoneLagrangianE-AHM}) to vanish!}
\end{itemize}

\subsubsection{The LSS theorem comes from exact $U(1)_Y$ symmetry. Minimization of the effective potential does not.}
\label{MinimizationVsLSS}

It's important to compare the results of our LSS theorem to those of the mainstream literature.
 For pedagogical simplicity, 
 in this sub-subsection we suppress mention of vacuum energy 
 and ${\cal O}^{Ignore}_{E-AHM;\phi}$. 
 After renormalization, but before application of the LSS theorem, 
 the effective potential (\ref{LEffectiveE-AHM}), 
 which is  derived entirely from Green's function WTIs, can be written 
\bea
\label{MainstreamPotential}
&&V_{E-AHM;\phi}^{Eff;Wigner,SI,Goldstone} =\mu_{\phi}^2\Big(\phi^\dagger \phi\Big)+\lambda_{\phi}^2\Big(\phi^\dagger \phi\Big)^2  \nonumber \\
&&\qquad =\Big( \mu_{\phi}^2+\lambda^2_\phi \HVEV^2 \Big)\Big(\phi^\dagger \phi-\half\HVEV^2\Big)\nonumber \\
&&\qquad +\lambda_{\phi}^2\Big(\phi^\dagger \phi-\half\HVEV^2\Big)^2 \nonumber \\
&&\qquad =\Big( \mu_{\phi}^2+\lambda^2_\phi \HVEV^2 \Big)\Big(\frac{h^2+\pi^2}{2}+\HVEV h\Big)\nonumber \\
&&\qquad +\lambda_{\phi}^2\Big(\frac{h^2+\pi^2}{2}+\HVEV h\Big)^2 \,,
\eea
where $V_{E-AHM;\phi}^{Eff;Wigner,SI,Goldstone}$, $\phi$, $\mu_{\phi}^2$,
$\lambda_{\phi}^2$ and $\HVEV^2$ in (\ref{MainstreamPotential}) 
are all renormalized quantities. 

The vanishing of relevant operators due to heavy $\Phi,\psi$ in the effective E-AHM theory 
is therefore not itself controversial.
The mainstream literature minimizes (\ref{MainstreamPotential}) to find the vacuum:
\bea
\label{MinimizeVacuum}
&&\frac{\partial}{\partial h}V_{E-AHM;\phi}^{Eff;Wigner,SI,Goldstone}\Big\vert_{h=\pi=0} \nonumber \\
&&\qquad \qquad =\HVEV\Big( \mu_{\phi}^2+\lambda^2_\phi \HVEV^2 \Big) =0
\eea
which, for the SSB case, gives
\bea
\label{MinimizeGoldstoneVacuum}
\frac{\partial}{\partial h}V_{E-AHM;\phi}^{Eff;Goldstone}\Big\vert_{h=\pi=0} &=&0\nonumber \\
 \mu_{\phi}^2+\lambda^2_\phi \HVEV^2  &=&0\,.
\eea
This is conventionally interpreted 
as a calculation of $\HVEV^2$
\bea
\label{MinimizeGoldstoneVacuumHiggsVEV}
 \HVEV^2 =-\frac{\mu_{\phi}^2}{\lambda^2_\phi}\,,
\eea
where, in renormalized $\mu^2_{\phi}$, 
UVQD and all other relevant contributions, 
such as those due to $\Phi,\psi$ in loops, 
are regarded as having cancelled against a bare counter-term $\delta\mu^2_{\phi;Bare}$.

In contrast, 
we have derived a tower of Adler self-consistency conditions 
(\ref{AdlerSelfConsistencyE-AHM}) in Lorenz gauge in Appendix \ref{DerivationWTIE-AHM}: i.e. derived directly from the exact $U(1)_Y$ symmetry obeyed by gauge-independent {\it on-shell} T-Matrix elements. 
One of these, the $N=0,M=1$ case, is the LSS theorem:
\bea
\label{LSSInsteadOfMinimize}
\HVEV\mpisq=\HVEV \Big( \mu_{\phi}^2+\lambda^2_\phi \HVEV^2 \Big) =0
\eea
which, for the SSB case, gives
\bea
\label{LSSInsteadOfMinimizeSSB}
\mpisq =\mu_{\phi}^2+\lambda^2_\phi \HVEV^2 =0
\eea
whose practical effect is the same as  minimization  of the effective potential, 
as captured in (\ref{MinimizeGoldstoneVacuum}).

So, we agree with the mainstream literature  
that all relevant operators vanish in the effective low-energy E-AHM theory.


\subsubsection{Decoupling of heavy matter representations:}
\label{DecouplingHeavyMatter} 

Adding  a $U(1)_Y$ local/gauge invariant Lagrangian
$L_{BeyondAHM}^{GaugeInvariant}(A_\mu,\phi;\Phi;\psi_L ,\psi_R)$ to (\ref{LagrangianAHM}) forms the  E-AHM.  

In order to force renormalized connected amplitudes with {\bf an odd number of $\pi$s to vanish, the new particles $\Phi,\psi_L,\psi_R$ are taken in this paper to conserve $CP$}.

In sub-subsections \ref{DecouplingHeavyMatter} 
through \ref{3rdDecouplingTheorem}, we take 
all of the new scalars $\Phi$, 
left-handed fermions $\psi_L$ and right-handed fermions $\psi_R$ to be very heavy.
\begin{eqnarray}
\label{HeavyScaleScalar}
&&M_{\psi_L}^2 , M_{\psi_R}^2 ,M_\Phi^2 \sim M_{Heavy}^2 \\
&&\qquad \gg \Big( \big\vert q^2 \big\vert,m_A^2, m_{BEH}^2 \Big) \sim m_{Weak}^2 \sim (100GeV)^2 \nonumber
\end{eqnarray}
with $q_\mu$ typical for a studied low-energy process. 
Fermion $U(1)_Y$ hypercharges are chosen so that the axial anomaly is zero.
To remain perturbative, we keep the Yukawa couplings  
$y_{\phi \psi},y_{\Phi \psi} \lapp 1$,
but take the Majorana masses-squared
\begin{eqnarray}
\label{GlobalInvariantBeyondAHMpsi}
&&L^{Majorana}_{BeyondAHM;\psi} = -\half M_{\psi_L} \Big( {\psi_L^{Weyl} }{\psi_L^{Weyl}}+{\bar \psi}_L^{Weyl} {\bar \psi}_L^{Weyl}\Big) \nonumber \\ 
&& \quad \quad  -\half M_{\psi_R} \Big( {\psi_R^{Weyl} }{\psi_R^{Weyl}}+{\bar \psi}_R^{Weyl} {\bar \psi}_R^{Weyl}\Big) \nonumber
\end{eqnarray}
heavy.
We keep all Yukawas and masses real for pedagogical simplicity.

Some comments are in order:
\begin{itemize}
\item We have ignored finite  ${\cal O}_{E-AHM;\phi}^{1/M_{Heavy}^2;Irrelevant}$ 
that decouple and vanish as $m_{Weak}^2/ M_{Heavy}^2 \to 0$.

\item Among the terms included in (\ref{GoldstoneLagrangianE-AHM}) are finite relevant operators 
dependent on the heavy matter representations:
{
\begin{eqnarray}
\label{HeavyRelevantOperators}
&&\quad {\cal O}\left(M_{Heavy}^2\right), 
{\cal O}\left(M_{Heavy}^2 \ln{ \big(M_{Heavy}^2\big)}\right),  \\
&& \quad {\cal O}\left(M_{Heavy}^2 \ln{ \big(m_{Weak}^2\big)}\right), 
{\cal O}\left(m_{Weak}^2 \ln{ \big(M_{Heavy}^2\big)}\right) \,, \quad \nonumber
\end{eqnarray}}
but they have {become invisible to us} because of the LSS  theorem (\ref{TMatrixGoldstoneTheoremE-AHM},\ref{TMatrixGoldstoneTheoremExtended})! That fact is one of the central results of this paper.

\item Marginal operators $\sim \ln{ \big(M_{Heavy}^2\big)}$ have been absorbed in  (\ref{GoldstoneLagrangianE-AHM}): i.e. in the renormalization of gauge-independent observables (i.e. the quartic-coupling constant 
$\lambda_\phi^2$ calculated in the Kibble representation, and the BEH VEV $\HVEV$), and in un-observable wavefunction renormalization (\ref{GoldstoneWavefunctionE-AHM}).
\end{itemize}

No trace of $M_{Heavy}$-scale $\Phi,\psi$ survives in (\ref{GoldstoneLagrangianE-AHM})! All the heavy Beyond-AHM matter representations have completely decoupled.

\subsubsection{1st decoupling theorem: 
\label{1stDecouplingTheorem}
SSB $\phi$-sector connected amputated 1-$\phi$-I Green's functions.}
We take ${\cal O}^{1/\Lambda^2;Irrelevant}_{E-AHM;\phi} \to0$ (i.e. to un-encumber our notation) and work  
in the ${{m_{Weak}^2}/{M_{Heavy}^2} \to 0}$ limit.  

In the SSB E-AHM, $\Gamma_{N,M}^{\Extended}$ with:
\begin{itemize}
\item $N+M \geq 5$ obey the Appelquist-Carazzone decoupling theorem \cite{AppelquistCarazzone};
\item $N+M =3, 4$ are absorbed by coupling constant renormalization;
\item $N+M =2$ are absorbed by wavefunction renormalization, vanish due to the LSS theorem
$\mpisq=0$, or contribute to SSB origination of $m_{BEH}^2=2\lambda_\phi^2\HVEV^2$ (see below).
\end{itemize}
Therefore, {\it including the contributions to relevant operators 
from  heavy { CP-conserving} $\Phi,\psi$ matter in virtual loops} 
\begin{eqnarray}
\label{SSBGreensFunctionDecouplingTheorem}
\Gamma_{N,M}^{\Extended} \quad
 {\buildrel  {{m_{Weak}^2}/{M_{Heavy}^2} \to 0} \over {=\joinrel=\joinrel=\joinrel=\joinrel=\joinrel=\joinrel=\joinrel=\joinrel=\joinrel=\joinrel\Longrightarrow}} 
\quad \Gamma_{N,M}^{AMH}\,.
\end{eqnarray}

\subsubsection{2nd decoupling theorem: 
\label{2ndDecouplingTheorem}
SSB $\phi$-sector connected amputated 1-$\phi$-R T-Matrices.}
In the limit ${{m_{Weak}^2}/{M_{Heavy}^2} \to 0}$
\begin{eqnarray}
\label{SSBTMatrixDecouplingTheorem}
T_{N,M}^{Extended} \quad
 {\buildrel  {{m_{Weak}^2}/{M_{Heavy}^2} \to 0} \over {=\joinrel=\joinrel=\joinrel=\joinrel=\joinrel=\joinrel=\joinrel=\joinrel=\joinrel=\joinrel\Longrightarrow}} 
\quad T_{N,M} \,.
\end{eqnarray}
including  heavy { CP-conserving} 
$\Phi,\psi$ matter contributions to relevant operators.

\subsubsection{3rd decoupling theorem: SSB $\phi$-sector BEH pole-mass-squared}
\label{3rdDecouplingTheorem}

The  $N=1,M=1$ connected amputated Green's function $U(1)_Y$ WTI (\ref{GreensWTIPrimeExtended}), 
augmented by the  LSS theorem (\ref{TMatrixGoldstoneTheoremE-AHM}) reads
\begin{eqnarray}
\label{BEHMassE-AHM}
\Gamma^{\Extended}_{2,0}(00;)&=&\HVEV\Gamma^{\Extended}_{1,2}(0;00) \nonumber \\
&=&-2\lambda^2_\phi \HVEV^2 \nonumber \\
\lim_{\HVEV \to 0}\Gamma^{\Extended}_{2,0}(00;)&=&0 
\end{eqnarray} 
shows that {\bf the BEH pole-mass-squared arises entirely from SSB.} Define
\begin{eqnarray}
\label{E-AHMBEHPropagator}
\Delta^{BEH}_{E-AHM}(q^2) &=& 
\frac{1}{q^2-m^2_{BEH;Pole} + i\epsilon} \nonumber \\
&+& \int dm^2 \frac{\rho^{BEH}_{E-AHM}(m^2)}{q^2-m^2 + i\epsilon} \qquad \qquad
\end{eqnarray}

$m^2_{BEH;Pole}$ is the 
BEH resonance pole-mass-squared. 
In analogy with (\ref{SpectralDensityPropagators}), 
the spectral density $\rho^{BEH}_{E-AHM}(M_{Heavy}^2) \sim 1/M_{Heavy}^2$.  
Thus
\begin{eqnarray}
\label{E-AHMBEHPropagatorAtZero}
&& \rho^{BEH}_{E-AHM}(m^2) = \rho^{BEH}_{AHM}(m^2)  + {\cal O}^{1/M_{Heavy}^2;Irrelevant}_{E-AHM;\phi} \nonumber \\
&&\Gamma^{\Extended}_{2,0}(00;)\equiv\Big[ \Delta^{BEH}_{E-AHM}(0) \Big]^{-1}\nonumber \\ 
&&\quad \quad =-2\lambda_\phi^2 \HVEV^2 \nonumber \\
&&\quad \quad =-m^2_{BEH;Pole} \Big[ 1+  m^2_{BEH;Pole} \int dm^2 \frac{\rho^{BEH}_{AHM}(m^2)}{m^2 - i\epsilon} \Big]^{-1} \nonumber \\
&&\quad \quad + {\cal O}^{1/M_{Heavy}^2;Irrelevant}_{E-AHM;\phi} \qquad \qquad
\end{eqnarray}
and we have
\begin{eqnarray}
\label{GaugeIndependentBEHPoleMass}
&&m^2_{BEH;Pole} \nonumber = 2\lambda_\phi^2 \HVEV^2\Big[ 1- 2\lambda_\phi^2 \HVEV^2 \int dm^2 \frac{\rho^{BEH}_{AHM}(m^2)}{m^2 - i\epsilon} \Big]^{-1} \nonumber \\
&&\qquad + {\cal O}^{1/M_{Heavy}^2;Irrelevant}_{E-AHM;\phi} 
\end{eqnarray}

Because $\lambda^2_\phi , Z^{\phi}_{ExrendedAHM}$ are dimensionless, $\lambda^2_\phi$ and
\begin{eqnarray}
\label{BEHVEVE-AHM}
\HVEV=\Big[Z^{\phi}_{ExrendedAHM}\Big]^{-\half}\HVEV_{Bare}
\end{eqnarray}
absorb no relevant operators, 
Eqn.  (\ref{GaugeIndependentBEHPoleMass}) shows 
that  the  BEH pole-mass-squared $m_{BEH;Pole}^2$ also absorbes no relevant 
operators.

No trace of $M_{Heavy}$-scale $\Phi,\psi$, including their contributions to relevant operators, survives in (\ref{GaugeIndependentBEHPoleMass})! All the heavy Beyond-AHM matter representations have completely decoupled, and the BEH-pole masses-squared
\begin{eqnarray}
\label{SSBBEHMassDecouplingTheorem}
m^{2;E-AHM}_{BEH;Pole}  \quad
 {\buildrel  {{m_{Weak}^2}/{M_{Heavy}^2} \to 0} \over {=\joinrel=\joinrel=\joinrel=\joinrel=\joinrel=\joinrel=\joinrel=\joinrel=\joinrel=\joinrel\Longrightarrow}} 
\quad m^{2;AHM}_{BEH;Pole} \quad \quad
\end{eqnarray}
become equal in the limit ${{m_{Weak}^2}/{M_{Heavy}^2} \to 0}$. 
We call (\ref{SSBBEHMassDecouplingTheorem}) the {\bf ``SSB BEH-Mass Decoupling Theorem."}

By dimensional analysis, heavy $\Phi,\psi$ also decouple from the $\pi$ spectral functions
\begin{eqnarray}
\label{E-AHMBEHPropagatorAtZeroB}
&& \Delta^{\pi;Spectral}_{E-AHM}(q^2) = \Delta^{\pi;Spectral}_{AHM}(q^2) + {\cal O}\Big(1/M_{Heavy}^2\Big) \quad \quad
\end{eqnarray}

\subsection{Example: Decoupling of gauge singlet $M^2_S \gg m_{Weak}^2$ real scalar field $S$ with discrete $Z_2$ symmetry and $\SVEV=0$}
\label{HeavyScalar}

We consider  a $U(1)_Y$ gauge singlet  real scalar $S$, with  ($S\to-S$) $Z_2$ symmetry, 
$M_S^2\gg m_{Weak}^2$, and $\SVEV =0$. 
We add to the renormalized theory
\begin{eqnarray}
\label{SingletScalarLagrangian}
&&L_S=\half(\partial_{\mu}S)^2 -V_{\phi S} \nonumber \\
&&V_{\phi S} = \half M_S^2 S^2 + \frac {\lambda_S^2}{4} S^4 + \half \lambda_{\phi S}^2 S^2 \left[ \phi^\dagger\phi -\half \HVEV^2 \right] \nonumber \\
&&\phi^\dagger\phi -\half \HVEV^2=\frac{h^2 +\pi^2}{2}+\HVEV h
\end{eqnarray}
Since $S$ is a gauge singlet, it is also a rigid/global singlet. Its $U(1)_Y$ hypercharge,  transformation and current 
\begin{eqnarray}
\label{MajoranaCurrent}
Y_S=0;\quad \delta_{U(1)_Y} S(t,{\vec y})&=&0 \nonumber \\
{J}^{\mu;S}_{BeyondAHM}&=&0
\end{eqnarray}
therefore satisfy all of the de-coupling criteria in Appendix \ref{DerivationWTIE-AHM}:

\begin{itemize}
\item Since it is massive, $S$ cannot carry information to the surface $z^{3-surface}\to \infty$ of the (all-space-time) 4-volume $\int d^4 z$, and so satisfies (\ref{ExtendedSurfaceIntegral}).
\item The equal-time commutators satisfy (\ref{EqTimeCommE-AHM})
\begin{eqnarray}
\label{ExtendedEqTimeCommAHMA}
 \delta(z_0-y_0)\left[ {J}^{0;S}_{BeyondAHM}(z),H(y)\right] &=&0 \nonumber \\
 \delta(z_0-y_0)\left[ {J}^{0;S}_{BeyondAHM}(z),\pi(y)\right] &=&0\,.
\end{eqnarray} 
\item The classical equation of motion  
\begin{eqnarray}
\label{DivergenceCurrentAHMSA}
&&\partial_{\mu} \Big( { J}^{\mu;S}_{BeyondAHM} + { J}^{\mu}_{AHM} \Big)  \\
&&\qquad \qquad =\partial_{\mu}   { J}^{\mu}_{AHM} = m_A H \partial_{\beta}{A}^{\beta}   \nonumber
\end{eqnarray}
restores conservation of the rigid/global $U(1)_Y$ extended current 
for $\phi$-sector physical states, and satisfies (\ref{ExtendedQuantumCurrentConservation})
\begin{eqnarray}
\label{ExtendedQuantumCurrentConservationSA}
&&\Big< 0\vert T\Big[  \partial_{\mu} \Big( { J}^{\mu;S}_{BeyondAHM} + { J}^{\mu}_{AHM} \Big)(z)  \\
&&\quad \quad \times h(x_1)...h(x_N) \pi(y_1)...\pi(y_M)\Big]\vert 0\Big>_{\rm connected} \nonumber \\
&&\quad \quad=0\,. \nonumber 
\end{eqnarray}
\item The zero VEV $\big< S \big>=0$ satisfies (\ref{E-AHMVEV}).

\end{itemize}

The $U(1)_Y$ WTI  governing  the extended $\phi$-sector transition matrix $T_{N,M}^{\Extended;S}$ are therefore true, namely: the extended Adler self-consistency conditions (\ref{AdlerSelfConsistencyE-AHM},\ref{ExtendedAdlerSelfConsistency}), together with their proof of infra-red finiteness in the presence of massless NGB;
the extended 1-soft-$\pi$ theorems (\ref{InternalTMatrixE-AHM},\ref{InternalTMatrixExtended});
the extended $U(1)_Y$ WTI  (\ref{GreensWTIPrimeExtended},\ref{ExtendedGreensFWTI}) governing  connected amputated $\phi$-sector Green's functions $\Gamma_{N,M}^{\Extended;S}$ are also true.
The $U(1)_Y\otimes$BRST symmetry of Section \ref{AbelianHiggsModelSymmetry}
is faithfully represented by these, 
and the tower of on-shell T-Matrix extended WTI 
(\ref{AdlerSelfConsistencyE-AHM},\ref{ExtendedAdlerSelfConsistency}) 
$T_{N,M}^{\Extended;S}\vert_{on-shell}=0$, 
and its extended LSS theorem 
(\ref{TMatrixGoldstoneTheoremE-AHM},\ref{TMatrixGoldstoneTheoremExtended}). 

The three decoupling theorems 
(\ref{SSBTMatrixDecouplingTheorem},\ref{SSBGreensFunctionDecouplingTheorem},\ref{SSBBEHMassDecouplingTheorem})
therefore follow, so that no trace of the $M_S^2 \sim M^2_{Heavy}$ scalar $S$ survives the 
${m_{Weak}^2}/{M_{Heavy}^2} \to 0$ limit: 
i.e. it has completely decoupled! 
The $\phi$-sector connected amputated T-Matrices and Green's functions, 
and the BEH pole masses-squared 
\begin{eqnarray}
\label{PhiSectorDecouplingTheorem}
T_{N,M}^{\Extended;S} \quad
& {\buildrel  {{m_{Weak}^2}/{M_S^2} \to 0} \over {=\joinrel=\joinrel=\joinrel=\joinrel=\joinrel=\joinrel=\joinrel=\joinrel=\joinrel=\joinrel
\Longrightarrow}} &
\quad T_{N,M} \\
\Gamma_{N,M}^{\Extended;S} \quad
& {\buildrel  {{m_{Weak}^2}/{M_S^2} \to 0} \over {=\joinrel=\joinrel=\joinrel=\joinrel=\joinrel=\joinrel=\joinrel=\joinrel=\joinrel=\joinrel \Longrightarrow}} &
\quad \Gamma_{N,M} \nonumber  \\
m^{2;E-AHM;S}_{BEH;Pole;\phi}  \quad 
&{\buildrel  {{m_{Weak}^2}/{M_{Heavy}^2} \to 0} \over {=\joinrel=\joinrel=\joinrel=\joinrel=\joinrel=\joinrel=\joinrel=\joinrel=\joinrel=\joinrel
\Longrightarrow}} &
\quad m^{2;AHM}_{BEH;Pole;\phi} \nonumber
\end{eqnarray}
become equal in the limit ${{m_{Weak}^2}/{M_{Heavy}^2} \to 0}$,
including all contributions to relevant operators from heavy $S$ in virtual loops.

\subsection{One generation of Standard Model quarks and leptons, 
augmented by a right-handed neutrino $\nu_R$ with 
Dirac mass, gauged hypercharge and global colors}
\label{SMQuarksLeptons}

{ 
We consider the addition of one Standard Model generation 
of spin $S=\half$ fermions -- 
$t_L$, $b_L$, $t_R$, $b_R$, $\tau{e}_L$, $\nu_{\tau_L}$, $\tau_R$ --
augmented by one right-handed neutrino$\nu_{\tau_R}$,
with global $SU(3)$ colors $c=$red, white, blue, and gauged $U(1)_Y$ hypercharge.
These are regarded here as E-AHM matter representations.

Baryon-number and lepton-number-conserving Dirac masses-squared 
arise entirely from SSB and are light, in the sense that 
$m_{Quark}^2,m_{Lepton}^2  \lapp m_{Weak}^2$.  
The so-extended $U(1)_Y$ AHM gauge theory has zero axial-anomaly because quark/lepton  AHM quantum numbers are chosen to be their SM hypercharges
(including $Y_{\nu_R}=0$).
This addition  also retains the CP-conservation of the AHM.
We choose the third generation mostly for definiteness, 
but also slightly to emphasize that we are not relying in any way on the smallness of quark Yukawas.

Adding Beyond-AHM Dirac quarks, augments $L_{AHM}^{Lorenz}$ of (\ref{LagrangianAHM}) with
\begin{eqnarray}
\label{BeyondAHMQuark}
&&L_{BeyondAHM;q}^{GlobalInvariant} =  L_{BeyondAHM;q}^{Kinetic} +L_{BeyondAHM;q}^{Yukawa}\quad \quad \\
&&L_{BeyondAHM;q}^{Kinetic}= i \sum_{color}^{r,w,b} \sum^{t,b}_{flavor}  \Big( {\bar q}_L^c \gamma^\mu D_\mu q_L^c 
+  {\bar q}_R^c \gamma^\mu D_\mu q_R^c \Big) \nonumber \\
&&L_{BeyondAHM;q}^{Yukawa}= - \sum_{color}^{r,w,b} \sum^{t,b}_{flavor} y_q  \Big( {\bar q}_L^c \phi q_R^c 
+  {\bar q}_R^c \phi^\dagger q_L^c \Big) \nonumber 
\end{eqnarray}

The $U(1)_Y$ quark current  and transformation properties are
\begin{eqnarray}
\label{BeyondAHMQuarkCurrent}
J_{BeyondAHM;q}^{\mu;Dirac}&=& -\sum_{color}^{r,w,b} \sum^{t,b}_{flavor} \nonumber \\
&\times&  \Big( Y_{q_L}{\bar q}_L^c \gamma^\mu q_L^c 
+  Y_{q_R}{\bar q}_R^c \gamma^\mu  q_R^c \Big) \quad \nonumber \\
\delta_{U(1)_Y} q^c_L(t,{\vec x}) &=& -iY_{q_L} q^c_L(t,{\vec x})\theta \nonumber \\
\delta_{U(1)_Y} q^c_R(t,{\vec x}) &=& -iY_{q_R} q^c_R(t,{\vec x})\theta \nonumber \\
Y_{t_L}=\frac{1}{3};Y_{b_L}&=&\frac{1}{3};Y_{t_R}=\frac{4}{3};Y_{b_R}=-\frac{2}{3};
\end{eqnarray}

Adding Beyond-AHM Dirac leptons, futher adds to $L_{AHM}^{Lorenz}$:
\begin{eqnarray}
\label{BeyondAHMpsi}
&&L_{BeyondAHM;l}^{GlobalInvariant} =  L_{BeyondAHM;l}^{Kinetic} +L_{BeyondAHM;l}^{Yukawa}\quad \quad \\
&&L_{BeyondAHM;l}^{Kinetic}= i \sum^{\nu_\tau,\tau}_{flavor}  \Big( {\bar l}_L \gamma^\mu D_\mu l_L 
+  {\bar l}_R \gamma^\mu D_\mu l_R \Big) \nonumber \\
&&L_{BeyondAHM;l}^{Yukawa}= - \sum^{\nu_\tau,\tau}_{flavor} y_l  \Big( {\bar l}_L \phi l_R 
+  {\bar l}_R \phi^\dagger l_L \Big) \nonumber 
\end{eqnarray}

The lepton $U(1)_Y$  current and transformation properties are
\begin{eqnarray}
\label{BeyondAHMLeptonCurrent}
J_{BeyondAHM;l}^{\mu;Dirac}&=&  - \sum^{\nu,e}_{flavor}  \Big( Y_{l_L}{\bar l}_L \gamma^\mu l_L
+  Y_{l_R}{\bar l}_R \gamma^\mu  l_R \Big) \quad \nonumber \\
\delta_{U(1)_Y} l_L(t,{\vec x}) &=& -iY_{l_L} l_L(t,{\vec x})\theta \nonumber \\
\delta_{U(1)_Y} l_R(t,{\vec x}) &=& -iY_{l_R} l_R(t,{\vec x})\theta \nonumber \\
Y_{\nu_{\tau,L}}=-1;Y_{\tau_L}&=&-1;Y_{\nu_{\tau,R}}=0;Y_{e_R}=-2; 
\end{eqnarray}
}

With these Standard Model quark and lepton hypercharges $Y_i$, our $U(1)_Y$ WTI have zero axial anomaly. 

We now prove applicability of  our $U(1)_Y$ WTI 
for connected amputated $\phi$-sector Greens functions 
$\Gamma_{N,M}^{\Extended}$ and for on-shell T-Matrix elements $T_{N,M}^{\Extended}$.

\begin{itemize}
\item {\bf The equal-time quantum commutators} satisfy (\ref{EqTimeCommE-AHM})
\begin{eqnarray}
\label{ExtendedEqTimeCommAHMB}
 \delta(z_0-y_0)\left[ J_{BeyondAHM;q}^{0;Dirac}(z),H(y)\right] &=&0 \nonumber \\
 \delta(z_0-y_0)\left[ J_{BeyondAHM;q}^{0;Dirac},\pi(y)\right] &=&0 \nonumber \\
\delta(z_0-y_0)\left[ J_{BeyondAHM;l}^{0;Dirac}(z),H(y)\right] &=&0 \nonumber \\
 \delta(z_0-y_0)\left[ J_{BeyondAHM;l}^{0;Dirac},\pi(y)\right] &=&0
\end{eqnarray} 
\item {\bf The classical equation of motion} 
\begin{eqnarray}
\label{DivergenceCurrentAHMSB}
&&\partial_{\mu} \Big( J_{BeyondAHM;l}^{\mu;Dirac}+ J_{BeyondAHM;q}^{\mu;Dirac} + { J}^{\mu}_{AHM} \Big) \nonumber \\
&&\quad \quad = m_A H \partial_{\beta}{A}^{\beta}   
\end{eqnarray}
restores conservation of the rigid/global $U(1)_Y$ extended current for $\phi$-sector physical states, and satisfies (\ref{ExtendedQuantumCurrentConservation})
\begin{eqnarray}
\label{ExtendedQuantumCurrentConservationSB}
&&\Big< 0\vert T\Big[  \partial_{\mu} \Big( J_{BeyondAHM;l}^{\mu;Dirac}+ J_{BeyondAHM;q}^{\mu;Dirac} + { J}^{\mu}_{AHM}\Big)(z)  \nonumber \\
&&\quad \quad \times h(x_1)...h(x_N) \pi_{t_1}(y_1)...\pi_{t_M}(y_M)\Big]\vert 0\Big>_{\rm connected} \nonumber \\
&& \quad \quad  =0 
\end{eqnarray}
\item  {\bf Dirac-mass-quark surface terms vanish.}
Since the quarks $t$ and $b$ are taken to have Dirac masses,
$m_t=\frac{1}{\sqrt 2}y_u\HVEV$ and $m_b=\frac{1}{\sqrt 2}y_d\HVEV$, 
and since we need only {\em connected} graphs, 
the  quarks  cannot carry information to the 3-surface at timelike infinity of the 4-volume
of space-time, and so do not spoil equation  (\ref{ExtendedSurfaceIntegral}). 
In contrast, 
massless quarks could carry  $U(1)_Y$ information on the light-cone to this surface, 
would therefore violate  (\ref{ExtendedSurfaceIntegral}),
and so destroy  the spirit, results and essence of our $U(1)_Y$-WTI-based heavy particle decoupling results here in Section \ref{E-AHM}.

\item  {\bf Charged-lepton surface terms also vanish.} 
Since $\tau$ is massive, $m_\tau=\frac{1}{\sqrt 2}y_e\HVEV$, 
and we need only connected graphs, 
the charged lepton $\tau$ also cannot carry information to the 3-surface at time-like infinity
of  the 4-volume of spacetime, and so satisfies (\ref{ExtendedSurfaceIntegral});

\item  {\bf Dirac-neutrino surface terms:} 
Since  $\nu_\tau$ is taken to be massive in deference to observed SM neutrino mixing,
$m^{Dirac}_{\nu}=\frac{1}{\sqrt 2}y_{\nu}\HVEV$, 
$\nu_\tau$ also satisfies (\ref{ExtendedSurfaceIntegral}). 
In contrast, 
a massless neutrino {\em could} carry  $U(1)_Y$ information 
on the light-cone to the 3-surface at infinity and would violate (\ref{ExtendedSurfaceIntegral})
\footnote{
	Our proof of axial-vector WTI in Appendix \ref{DerivationWTIE-AHM} 
	requires that neutrinos be incapable of carrying information to the 3-surface at timeline infinity of the 4-volume of spacetime. 
	We have worked here  within SSB E-AHM, 
	with its explicit Dirac neutrino mass, 
	for this purely mathematical reason. 
	\newline
	Imagine, however, that we are able to extend this work 
	to the $CP$-conserving standard electroweak model 
	with {\it two generations} of quarks, charged leptons, and $\nu_L,\nu_R$, 
	with neutrino Dirac masses, but zero Majorana masses. 
	(Ref. \cite{LSS-4Proof} analyses local $SU(2)\otimes U(1)_Y$ 
	with one such generation and non-zero Majorana $\nu_R$ mass.) 
	With its gauge group $SU(2)_L \times U(1)_Y$, 
	we would build two sets of {\em rigid/global} WTIs: 
	unbroken electromagnetic  $U(1)_{QED}$; and spontaneously broken $SU(2)_L$.  
	It is then amusing to elevate such rigid/global WTIs to a principle of nature, 
	so as to give them predictive power for actual experiments and observations. 
	The $U(1)_{QED}$ WTIs would be unbroken vector-current identities. 
	Focus instead on the spontaneously broken $SU(2)_L$. 
	Start with Yukawa couplings which generate, after SSB, 
	masses and mixings among weak-eigenstate neutrinos. 
	The observable $2\times 2$ PMNS matrix would then rotate those 
	to mass eigenstates $m^{Dirac}_{\nu_1}, m^{Dirac}_{\nu_2}$. 
	\newline
	The axial-vector current WTIs 
	from the spontaneously broken $SU(2)_L$ 
	will require and demand a neutrino Dirac mass 
	for each and every one of the mass eigenstates 
	$m^{Dirac}_{\nu_1}, m^{Dirac}_{\nu_2} \neq 0$. 
	Would we then claim that SSB $SU(2)_L$ WTIs {\em predict} neutrino oscillations? 
	To make a possible connection with Nature, 
	although current experimental neutrino-mixing data 
	cannot rule out an exactly zero mass for the lightest neutrino \cite{KarolLang}, 
	the mathematical self-consistency of $SU(2)_L$ WTIs would!
}, 
and so destroy  the spirit, results and essence of our $U(1)_Y$-WTI-based  heavy particle decoupling results here in Section \ref{E-AHM}.

\end{itemize}

Having satisfied all of the criteria in Appendix \ref{DerivationWTIE-AHM},
the $U(1)_Y$ WTI  governing  the extended $\phi$-sector transition matrix $T_{N,M}^{\Extended;q,l}$ are therefore true, namely: the extended Adler self-consistency conditions (\ref{AdlerSelfConsistencyE-AHM},\ref{ExtendedAdlerSelfConsistency}), together with their proof of infra-red finiteness in the presence of massless NGB;
the extended 1-soft-$\pi$ theorems (\ref{InternalTMatrixE-AHM},\ref{InternalTMatrixExtended});
the extended $U(1)_Y$ WTI  (\ref{GreensWTIPrimeExtended},\ref{ExtendedGreensFWTI}) governing  connected amputated $\phi$-sector Green's functions $\Gamma_{N,M}^{\Extended;q,l}$.
The $U(1)_Y\otimes$BRST symmetry of Section \ref{AbelianHiggsModelSymmetry}
is faithfully represented by these, and the tower of on-shell T-Matrix extended WTI (\ref{AdlerSelfConsistencyE-AHM},\ref{ExtendedAdlerSelfConsistency}) $T_{N,M}^{\Extended;q,l}\vert_{on-shell}=0$, and its extended LSS theorem (\ref{TMatrixGoldstoneTheoremE-AHM},\ref{TMatrixGoldstoneTheoremExtended}).

\subsection{(Practical) decoupling 
of a gauge-singlet right-handed Type I See-saw Majorana neutrino 
with $M_{\nu_R}^2\gg m_{BEH}^2 \sim m_{Weak}^2$
(as in the $\nu$AHM)}
\label{HeavyNeutrino}
We consider here the addition to the AHM of 
a heavy $U(1)_Y$ gauge-singlet right-handed Majorana neutrino $\nu_R$, 
with $M_{\nu_R}^2\gg m_{Weak}^2$, 
involved in a Type 1 see-saw with a left-handed neutrino $\nu_L$, 
through a Yukawa coupling $y_{\nu}$, 
with resulting Dirac mass $m_\nu^{Dirac}=y_{\nu}\HVEV /\sqrt{2}$. 

We add to the renormalized theory in Subsection \ref{SMQuarksLeptons} a Majorana mass
\begin{eqnarray}
\label{NeutrinoMajoranaMass}
L^{Majorana}_{\nu_R}&=& -\half M_{\nu_R} \Big( {\nu_R^{Weyl} }{\nu_R^{Weyl}}
+{\bar \nu}_R^{Weyl} {\bar \nu}_R^{Weyl}\Big) \qquad 
\end{eqnarray}
Since $\nu_R$ is a gauge singlet, 
it is also a rigid/global singlet. 
Its hypercharge $U(1)_Y$ transformation and current 
\begin{eqnarray}
\label{MajoranaCurrentB}
Y_{\nu_R}&=&0; \quad \delta_{U(1)_Y} \nu_R(t,{\vec y})=0 \nonumber \\
{J}^{\mu;Majorana}_{BeyondAHM;\nu_R}&=&0
\end{eqnarray}
therefore satisfy all of the de-coupling criteria in Appendix \ref{DerivationWTIE-AHM}:
\begin{itemize}
\item Since it has a Dirac mass, the neutrino
$\nu$ cannot carry information to the surface $z^{3-surface}\to \infty$ 
of the (all-space-time) 4-volume $\int d^4 z$, 
and so satisfies (\ref{ExtendedSurfaceIntegral}).
\item The equal-time quantum commutators satisfy (\ref{EqTimeCommE-AHM})
\begin{eqnarray}
\label{ExtendedEqTimeCommAHMC}
 \delta(z_0-y_0)\left[ {J}^{0;Majorana}_{BeyondAHM;\nu_R}(z),H(y)\right] &=&0  \\
 \delta(z_0-y_0)\left[ {J}^{0;Majorana}_{BeyondAHM;\nu_R}(z),\pi(y)\right] &=&0 \nonumber
\end{eqnarray}

\item The classical equation of motion  
\begin{eqnarray}
\label{DivergenceCurrentAHMSC}
&&\partial_{\mu} \Big( {J}^{\mu;Majorana}_{BeyondAHM;\nu_R} +J_{BeyondAHM;l}^{\mu;Dirac}  \\
&&\qquad \qquad + J_{BeyondAHM;q}^{\mu;Dirac}  + { J}^{\mu}_{AHM} \Big) \nonumber \\
&&\qquad \qquad=\partial_{\mu} \Big( J_{BeyondAHM;l}^{\mu;Dirac}+ J_{BeyondAHM;q}^{\mu;Dirac} + { J}^{\mu}_{AHM} \Big) \nonumber \\
&&\qquad \qquad= {m_A H \partial_{\beta}{A}^{\beta} }  \nonumber
\end{eqnarray}

restores conservation of the extended rigid/global $U(1)_Y$  current 
for $\phi$-sector physical states, 
and satisfies (\ref{ExtendedQuantumCurrentConservation})
\begin{eqnarray}
\label{ExtendedQuantumCurrentConservationSC}
&&\Big< 0\vert T\Big[  \partial_{\mu} \Big( {J}^{\mu;Majorana}_{BeyondAHM;\nu_R} +J_{BeyondAHM;l}^{\mu;Dirac}\\
&&\quad \quad +J_{BeyondAHM;q}^{\mu;Dirac}  + { J}^{\mu}_{AHM}\Big)(z)  \nonumber \\
&&\quad \quad \times h(x_1)...h(x_N) \pi_{t_1}(y_1)...\pi_{t_M}(y_M)\Big]\vert 0\Big>_{\rm connected} \nonumber \\
&&\quad \quad =0 \nonumber
\end{eqnarray}

\end{itemize}

Having satisfied all of the criteria in Appendix \ref{DerivationWTIE-AHM},
the $U(1)_Y$ WTI  governing  the extended $\phi$-sector transition matrix $T_{N,M}^{\Extended;q,l,M_{\nu_R}}$ are therefore true, namely: the extended Adler self-consistency conditions (\ref{AdlerSelfConsistencyE-AHM},\ref{ExtendedAdlerSelfConsistency}), together with their proof of infra-red finiteness in the presence of massless NGB;
the extended 1-soft-$\pi$ theorems (\ref{InternalTMatrixE-AHM},\ref{InternalTMatrixExtended});
the extended $U(1)_Y$ WTI  (\ref{GreensWTIPrimeExtended},\ref{ExtendedGreensFWTI}) governing  connected amputated $\phi$-sector Green's functions $\Gamma_{N,M}^{\Extended;q,l,M_{\nu_R}}$ are also true.
The $U(1)_Y\otimes$BRST symmetry of Section \ref{AbelianHiggsModelSymmetry}
is faithfully represented by these, and the tower of on-shell T-Matrix extended WTI (\ref{AdlerSelfConsistencyE-AHM},\ref{ExtendedAdlerSelfConsistency}) $T_{N,M}^{\Extended;q,l,M_{\nu_R}}\vert_{on-shell}=0$, and its extended LSS theorem (\ref{TMatrixGoldstoneTheoremE-AHM},\ref{TMatrixGoldstoneTheoremExtended}). 

The three decoupling theorems   
(\ref{SSBGreensFunctionDecouplingTheorem},
\ref{SSBTMatrixDecouplingTheorem},
\ref{SSBBEHMassDecouplingTheorem}) follow, 
but there is a ``non-decoupling subtlety." 
The vanishing of the $\nu_L$ surface terms 
requires a non-zero neutrino Dirac mass
\begin{eqnarray}
\label{DiracNeutrinoMass}
m_{\nu ;Dirac}=\frac{1}{\sqrt 2}y_\nu\HVEV \neq 0
\end{eqnarray}
The light and heavy Type I See-saw $\nu$ masses are 
\begin{eqnarray}
\label{TotalNeutrinoMass}
m_{\nu ;Light} &\sim& m_{\nu ;Dirac}^2 / M_{\nu_R}  \\
m_{\nu ;Heavy} &\sim& M_{\nu_R},\nonumber
\end{eqnarray}
but, in obedience to our proof of $U(1)_Y$ WTI,
$m_{Light}$ must not vanish. 
Therefore Type I See-saw $\nu$'s 
do not allow the  $M_{\nu_R}\to \infty$ limit!
For the decoupling theorems, we instead imagine huge,  
but finite, $M_{\nu_R}$ with
\begin{eqnarray}
\label{NeutrinoMassLimit}
1 \gg m_{\nu ;Dirac}^2/M_{\nu_R}^2 \neq 0\,.
\end{eqnarray}

No practical trace of the $M_{\nu_R}^2 \sim M^2_{Heavy}$ 
right-handed neutrino $\nu_R$ survives. 

{
Still, our $U(1)_Y$ WTIs insist that in principle 
a very heavy Majorana mass $M_{\nu_R}$ cannot completely decouple. 
It may still have some measureable or observational effect that we have not identified..

\section{SSB E-AHM's physical particle spectrum  excludes the NGB $\tilde \pi$:}
\label{ParticlePhysicsAHM}

G.S. Guralnik, C.R. Hagan and T.W.B. Kibble \cite{Guralnik1964} first showed  in the spontaneously broken Abelian Higgs model that, although there are no massless particles in the $(A^0=0,{\vec \nabla}\cdot {\vec  A} =0)$ ``radiation gauge",  there is a Goldstone theorem, and a true massless NGB,   in the covariant $\partial_\mu A^\mu =0$ Lorenz gauge. T.W.B. Kibble then showed \cite{Kibble1967} that the results of experimental measurements are nevertheless the same in radiation and Lorenz gauges, 
and that the spectrum and dynamics of the {\em observable particle states} are gauge-independent.

\subsection{SSB E-AHM's physical particle spectrum  excludes the NGB $\tilde \pi$, whose S-Matrix elements all vanish}
\label{NGBDisappearsExtendeAHM}

The BRST-invariant Lagrangian for the E-AHM in Lorenz gauge is
\begin{eqnarray}
\label{LagrangianE-AHMLorenzA}
&&L_{E-AHM}^{Lorenz}=L_{AHM}^{Lorenz} \nonumber  \\
&& \quad \quad +L_{BeyondAHM}^{GaugeInvariant}(A_\mu,\phi;\Phi,\psi)
\end{eqnarray}
with $L_{AHM}^{Lorenz}$ in (\ref{LagrangianAHM}).

\subsubsection{ Lagrangian governing dynamics of observable particles} 
We now identify the {\em observable particle spectrum} of Lorenz gauge E-AHM by re-writing (\ref{LagrangianE-AHMLorenzA})
in terms of a new gauge field
\begin{eqnarray}
\label{ProcaField}
B_\mu&\equiv& A_\mu +\frac{1}{e\HVEV}\partial_\mu {\tilde \pi}
\end{eqnarray}
and transforming to the Kibble representation \cite{Ramond2004}
\begin{itemize}

\item Gauge field
\begin{eqnarray}
\label{BFieldStrength}
A_{\mu\nu}&\equiv&\partial_\mu A_\nu - \partial_\nu A_\mu \nonumber \\
&=&\partial_\mu B_\nu - \partial_\nu B_\mu \equiv B_{\mu\nu}
\end{eqnarray}

\item AHM scalar
\begin{eqnarray}
\label{KibbleRepresentationAHM}
\tilde \pi &=& \HVEV \vartheta \nonumber \\
\phi &=& \frac{1}{\sqrt 2}{\tilde H}e^{-iY_\phi \vartheta}; \quad {\tilde H}={\tilde h}+\HVEV \nonumber \\
D_\mu \phi &=& \frac{1}{\sqrt 2}\Big[ \partial_\mu -ieY_\phi A_\mu \Big] {\tilde H}e^{-iY_\phi \vartheta} \nonumber \\ 
&=&\frac{1}{\sqrt 2}\Big[ \partial_\mu {\tilde H}-ieY_\phi {\tilde H}\Big(A_\mu +\frac{1}{e}\partial_\mu\vartheta \Big)\Big] e^{-iY_\phi \vartheta} \nonumber \\
&=&\frac{1}{\sqrt 2}\Big[ \partial_\mu {\tilde H}-ieY_\phi {\tilde H} B_\mu \Big] e^{-iY_\phi \vartheta}  
\end{eqnarray}

\item Beyond-AHM scalar
\begin{eqnarray}
\label{KibbleScalarsBeyondAHM}
\Phi &=& {\tilde \Phi}e^{-iY_\Phi \vartheta} \nonumber \\
\big< {\tilde \Phi} \big> &=& 0 \nonumber \\
D_\mu \Phi &=& \Big[ \partial_\mu -ieY_\Phi A_\mu \Big] {\tilde \Phi}e^{-iY_\Phi \vartheta} \nonumber \\ 
&=&\Big[ \partial_\mu {\tilde \Phi}-ieY_\Phi {\tilde \Phi}\Big(A_\mu +\frac{1}{e}\partial_\mu\vartheta \Big)\Big] e^{-iY_\Phi \vartheta} \nonumber \\
&=&\Big[ \partial_\mu {\tilde \Phi}-ieY_\Phi {\tilde \Phi} B_\mu \Big] e^{-iY_\Phi \vartheta} 
\end{eqnarray}

\item Beyond-AHM fermion(s)
\begin{eqnarray}
\label{KibbleFermionsBeyondAHM}
\psi &=& {\tilde \psi}e^{-iY_\psi \vartheta} \nonumber \\
D_\mu \psi &=& \Big[ \partial_\mu -ieY_\psi A_\mu \Big] {\tilde \psi}e^{-iY_\psi \vartheta} \nonumber \\ 
&=&\Big[ \partial_\mu {\tilde \psi}-ieY_\psi {\tilde \psi} \Big( A_\mu +\frac{1}{e}\partial_\mu \vartheta \Big) \Big] e^{-iY_\Phi \vartheta} \nonumber \\
&=&\Big[ \partial_\mu {\tilde \psi}-ieY_\psi {\tilde \psi} B_\mu \Big] e^{-iY_\Phi \vartheta}
\end{eqnarray}

\end{itemize}

The E-AHM Lagrangian, which governs the spectrum and dynamics of {\em particle physics}, is
\begin{eqnarray}
\label{LagrangianE-AHMParticlePhysics}
&&L_{E-AHM}^{ParticlePhysics} \Big(  B_\mu; {\tilde H} ; {\tilde \Phi}; {\tilde \psi}\Big) \nonumber \\
&& \quad =L_{AHM;{\tilde H},B_\mu}^{Lorenz} \Big(  B_\mu; {\tilde H;{\bar \eta},\omega} \Big)  \nonumber \\
&& \quad +L_{BeyondAHM; {\tilde \Phi} }^{GaugeInvariant}+L_{BeyondAHM;{\tilde \psi} }^{GaugeInvariant} 
\end{eqnarray}
where the spin $S=1$ field $B_\mu$ 
\begin{eqnarray}
\label{GaugeInvariantLagrangianAHMKibble}
&&L_{AHM}^{Lorenz} \Big(  B_\mu; {\tilde H};  {\bar \eta},\omega \Big) = L^{GaugeInvariant}_{AHM;{\tilde H},B_\mu}  \nonumber \\
&&\quad +L_{AHM;B_\mu}^{GaugeFix;Lorenz} +L_{AHM;B_\mu}^{Ghost;Lorenz}  \nonumber \\
&&L^{GaugeInvariant}_{AHM;{\tilde H},B_\mu} = -\frac{1}{4}B_{\mu \nu}B^{\mu \nu} \nonumber +\half e^2Y_\phi^2 \HVEV^2 B_\mu B^\mu \nonumber \\
&& \quad + \half\Big(\partial_\mu {\tilde H}\Big)^2+\half e^2Y_\phi^2 \Big( {\tilde H}^2 -\HVEV^2 \Big) B_\mu B^\mu  -V_{AHM} \nonumber \\
&&L_{AHM;B_\mu}^{GaugeFix;Lorenz}=-\lim_{\xi \to 0}\frac{1}{2\xi}\Big( \partial_\mu B^\mu \Big)^2\nonumber \\
&&L_{AHM;B_\mu}^{Ghost;Lorenz}=-{\bar \eta}\partial^2 \omega  \nonumber \\
&&V_{AHM}=\frac{1}{4}\lambda^2_\phi \Big( {\tilde H}^2-\HVEV^2 \Big) 
\end{eqnarray}

For the Beyond-AHM scalar(s)
\begin{eqnarray}
\label{BeyondAHMPhi}
&&L_{BeyondAHM;{\tilde \Phi}}^{GaugeInvariant} = \Big\vert  D_\mu {\tilde \Phi} \Big\vert ^2 -V_{{\tilde \Phi}}-V_{ {\tilde \phi} {\tilde \Phi}}   \\
&& \quad \quad D_\mu {\tilde \Phi} = \Big[ \partial_\mu-ieY_\Phi B_\mu \Big] {\tilde \Phi} \nonumber \\ 
&&V_{\tilde \Phi} = M_\Phi^2\Big( {\tilde \Phi}^\dagger {\tilde \Phi} \Big) + \lambda_\Phi^2\Big( {\tilde \Phi}^\dagger {\tilde \Phi} \Big) ^2 \nonumber \\
&&V_{ {\tilde \phi} {\tilde \Phi}}=\half \lambda_{\phi \Phi}^2 \Big({\tilde H}^2 \Big)  \Big( {\tilde \Phi}^\dagger {\tilde \Phi} \Big) \nonumber
\end{eqnarray}
while, for Beyond-AHM fermions, we take a Standard Model generation of fermions with anomaly-cancelling hypercharges
\begin{eqnarray}
\label{BeyondAHMpsiB}
&&L_{BeyondAHM;{\tilde \psi}}^{GaugeInvariant} =  i{\bar {\tilde \psi} }_L D_\mu {\tilde \psi}_L 
+  i{\bar {\tilde \psi}}_R D_\mu {\tilde \psi}_R  \nonumber  \\
&& \quad \quad +L_{BeyondAHM;{\tilde \psi}}^{Yukawa}+L_{BeyondAHM;{\tilde \nu}_R}^{Majorana} \\
&& \quad \quad D_\mu {\tilde \psi} _L = \Big[ \partial_\mu -ieY_{\psi_L} B_\mu \Big] {{\tilde \psi}} _L\nonumber \\ 
&& \quad \quad D_\mu {\tilde \psi} _R = \Big[ \partial_\mu -ieY_{\psi_R} B_\mu \Big] {\tilde \psi} _R\nonumber \\ 
&&L_{BeyondAHM;{\tilde \psi}}^{Yukawa}=-\frac{1}{\sqrt 2}y_{\phi \psi} \Big(  {\bar {\tilde \psi}}_L  {\tilde \psi}_R 
+ {\bar {\tilde \psi}}_R  {\tilde \psi}_L \Big) {\tilde H} \nonumber \\
&& \quad \quad -y_{\Phi \psi} \Big(  {\bar {\tilde \psi} }_L {\tilde \Phi} {\tilde \psi}_R 
+ {\bar {\tilde \psi}}_R {\tilde \Phi}^\dagger {\tilde \psi}_L \Big) \nonumber \\
&&L^{Majorana}_{BeyondAHM; {\tilde \nu}_R } =  -\half M_{\nu_R} \Big( {{\tilde \nu}_R^{Weyl} }{{\tilde \nu}_R^{Weyl}}
+{\bar {\tilde \nu}}_R^{Weyl} {\bar {\tilde \nu}}_R^{Weyl}\Big) \nonumber 
\end{eqnarray}
For $y_{\Phi \psi}\neq 0$, the heavy scalar hypercharge $Y_\Phi =-1$.

The $B_\mu$ mass-squared in (\ref{GaugeInvariantLagrangianAHMKibble}) arises entirely from SSB
\begin{eqnarray}
\label{ProcaMass}
m_B^2 = m_A^2 = e^2 \HVEV ^2
\end{eqnarray}
Dimensional analysis 
shows that the contribution of a state  of mass/energy $\sim M_{Heavy}$  
to the spectral function $\Delta_{E-AHM}^{B;Spectral}$ 
gives  terms $\sim 1/{M_{Heavy}^2}$, so that 
\begin{eqnarray}
\label{VectorPropagatorExtendedDecoupled}
&&\Delta_{E-AHM}^{B}(q^2) = \Delta^{B}_{AHM}(q^2) + {\cal O}\Big(1/M_{Heavy}^2 \Big) \nonumber \\
&&\Delta^{B}_{AHM}(q^2) =\frac{1}{q^2-m_{B;Pole}^2 + i\epsilon} 
+ \int dm^2 \frac{\rho^{B}_{AHM}(m^2)}{q^2-m^2 + i\epsilon} \nonumber \\
&&Z^{B}_{E-AHM} =  Z^{B}_{AHM} + {\cal O}\Big(1/M_{Heavy}^2 \Big) 
\end{eqnarray}
Therefore the $B_\mu$  pole-mass-squared is
\begin{equation}
\label{BPoleMassExtended}
\Big[ \Delta^{B}_{E-AHM}(0) \Big]^{-1} =-m_{B}^2=-e^2\HVEV^2 \nonumber
\end{equation}
with
\begin{eqnarray}
m^2_{B;Pole} &=& e^2 \HVEV^2\Big[ 1- e^2 \HVEV^2 
\int dm^2 \frac{{\rho}^{B}_{AHM}(m^2)}{m^2 - i\epsilon} \Big]^{-1} \nonumber \\
&&\qquad \qquad+ {\cal O}\Big(1/M_{Heavy}^2 \Big) \,.
\end{eqnarray}

\subsubsection{Decoupling of NGB $\tilde \pi$, particle spectrum and dynamics:} 
The Lagrangian (\ref{LagrangianE-AHMParticlePhysics}) 
is guaranteed to generate all of the results in Sections \ref{AbelianHiggsModel} 
and \ref{E-AHM}, and Appendices \ref{DerivationWTIAHM} and \ref{DerivationWTIE-AHM}. 
In practice, this is done via the manifestly renormalizeable E-AHM Lagrangian 
(\ref{LagrangianE-AHMLorenzA}).

G. Guralnik, T. Hagan and T.W.B. Kibble \cite{Guralnik1964}, and T.W.B. Kibble \cite{Kibble1967}, showed that, in the Kibble representation in Lorenz gauge, the $U(1)_Y$ AHM quantum states factorize. In the analogous $U(1)_Y$ E-AHM, and in the
$m_{Weak}^2/M_{Heavy}^2 \to 0$ limit 
the  analogous $U(1)_Y$ E-AHM also factorizes
\begin{eqnarray}
\label{E-AHMQuantumStates}
&&\Big\vert \Psi \Big( A^\mu;\phi; {\bar \eta},\omega;\Phi;\psi \Big) \Big> \to  
	\Big\vert \Psi^{Particles} \Big( B^\mu;{\tilde H} \Big) \Big> \\
&&\quad \times \Big\vert \Psi^{Ghost} \Big( {\bar \eta},\omega \Big) \Big>
	\Big\vert \Psi^{Goldstone} \Big( {\tilde \pi} \Big) \Big>
	\Big\vert \Psi^{B-AHM} \Big( {\tilde \Phi};  {\tilde \psi}\Big) \Big>\nonumber
\end{eqnarray}

With $\partial^2\omega =0; \partial^2{\bar \eta} =0$,
the ghost $\omega$  and anti-ghost ${\bar \eta}$ are free and massless 
and de-couple in Lorenz gauge. 

It is crucial for SSB gauge theories \cite{Guralnik1964,Kibble1967} to remember the additional gauge-fixing term inside (\ref{LagrangianE-AHMLorenzA}).
The E-AHM Lorenz gauge condition is re-written
\begin{eqnarray}
\label{LagrangianE-AHMLorenzB}
&&{L}^{GaugeFix;Lorenz}_{E-AHM}=-\lim_{\xi \to 0} \frac{1}{2\xi}\Big( \partial_\mu A^\mu \Big)^2 \\
&& \qquad \qquad = -\lim_{\xi \to 0} \frac{1}{2\xi}\Big( \partial_\mu B^\mu \Big)^2  \nonumber\\
&& \qquad \qquad -\lim_{\xi \to 0} \frac{1}{2\xi}\Big( \frac{1}{e\HVEV}\partial^2{\tilde \pi} \Big) \Big( \frac{1}{e\HVEV}\partial^2{\tilde \pi} - 2\partial_\mu B^\mu\Big) \nonumber
\end{eqnarray}
Besides enforcing the new Lorenz gauge-fixing constraint $\partial_\mu B^\mu =0$
in (\ref{GaugeInvariantLagrangianAHMKibble}), 
the auxiliary solution to the gauge-fixing condition (\ref{LagrangianE-AHMLorenzB}) is 
$\partial^2  {\tilde \pi} =0$, 
which  forces $\tilde \pi$ to be a free massless particle.
The NGB $\tilde \pi$
therefore completely decouples from, and disappears from,  
the observable particle spectrum and  its dynamics \cite{Guralnik1964,Kibble1967}, 
whose states factorize as in (\ref{E-AHMQuantumStates}).

In the $m_{Weak}^2/M_{Heavy}^2 \to 0$ limit, all physical measurements and observations 
are then entirely predicted by the AHM Lagrangian (\ref{GaugeInvariantLagrangianAHMKibble}) and its states in (\ref{E-AHMQuantumStates})
\begin{eqnarray}
\label{ParticleDynamics}
&&L_{AHM;B_\mu}^{Lorenz}\Big( {\tilde H};B_\mu; {\bar \eta},\omega \Big); \nonumber \\
&&\Big\vert \Psi^{ParticlePhysics} \Big( B^\mu;{\tilde H} ;{\bar \eta},\omega \Big) \Big>\\
&&\to  \Big\vert \Psi^{Particles} \Big( B^\mu;{\tilde H} \Big) \Big>\Big\vert \Psi^{Ghost} \Big( {\bar \eta},\omega \Big) \Big>\nonumber
\end{eqnarray}

What has become of our SSB $U(1)_Y$ Ward-Takahashi identities?
Although the NGB ${\tilde \pi}$ has de-coupled, 
it still governs the SSB dynamics and particle spectrum of (\ref{ParticleDynamics});
it is simply {\em hidden} from explicit view.
Still, that decoupling NGB  causes powerful  hidden constraints 
on (\ref{ParticleDynamics}) to arise from its hidden shift symmetry
\begin{eqnarray}
\label{HiddenShiftSymmetry}
{\tilde \pi}\to {\tilde \pi} +\HVEV \theta
\end{eqnarray}
for constant $\theta$.

Our SSB $U(1)_Y$ WTIs, and all of the results of Section \ref{AbelianHiggsModel}, Section \ref{E-AHM}, 
Appendix \ref{DerivationWTIAHM} and Appendix \ref{DerivationWTIE-AHM}
are also hidden but still in force: connected amputated Greens functions $\Gamma_{N,M}$ (\ref{GreensWTIPrimeExtended},\ref{ExtendedGreensFWTI}); connected amputated  T-Matrix elements $T_{N,M}$ (\ref{InternalTMatrixE-AHM},\ref{InternalTMatrixExtended}); 
Adler self-consistency conditions (\ref{AdlerSelfConsistencyE-AHM},\ref{ExtendedAdlerSelfConsistency}) together with their proof of IR finiteness;
 LSS theorem  (\ref{TMatrixGoldstoneTheoremE-AHM},\ref{TMatrixGoldstoneTheoremExtended});
1-soft-$\pi$ theorems (\ref{InternalTMatrixE-AHM},\ref{ExtendedAdlerSelfConsistency},\ref{InternalTMatrixExtended}); 
decoupling theorems for Green's functions and T-Matrix elements (\ref{SSBTMatrixDecouplingTheorem},\ref{SSBGreensFunctionDecouplingTheorem});
and the decoupling theorem for the BEH pole-mass-squared $m_{BEH;Pole}^2$ (\ref{GaugeIndependentBEHPoleMass}).
These still govern the SSB dynamics and 
particle spectrum of (\ref{ParticleDynamics}): they are simply hidden from explicit view.
We call this  {\bf ``the hidden ${\mathbf U(1)_Y\otimes}$BRST symmetry of the SSB AHM."}

\subsection{SSB causes decoupling of heavy $M_{Heavy}^2 \gg m_{Weak}^2$ particles. This  fact is hidden, from the  observable particle spectrum of the $U(1)_Y$ E-AHM and its dynamics, by the decoupling of the NGB $\tilde \pi$}
\label{DecouplingInGaugedE-AHM}

We now take
all of the new scalars $\tilde \Phi$ and fermions $\tilde \psi$ in the E-AHM  
to be very heavy, and are only interested in low-energy processes:
\begin{eqnarray}
\label{TildeHeavy}
&&M^2_{\tilde \Phi}, M^2_{\tilde \psi} \sim M_{Heavy}^2 \gg m_{Weak}^2 \nonumber \\
&&\big\vert q^2 \big\vert {\buildrel {<} \over {\sim}} m^2_{Weak}
\end{eqnarray} 
where $q_\mu$ is a typical momentum transfer.
In the limit
$m_{Weak}^2 / M_{Heavy}^2 \to 0$
the effective Lagrangian of the spontaneously broken E-AHM gauge theory
obeys the Appelquist-Carazzone decoupling theorem \cite{AppelquistCarazzone}
\begin{eqnarray}
\label{EffLagrangianLimit}
&&L^{Eff;SSB}_{E-AHM} \Big( k_\mu;B_\nu;{\tilde H};{\tilde \Phi};{\tilde \psi} \Big) \nonumber \\
&&\quad \rightarrow L^{Eff;SSB}_{AHM} \Big( k_\mu;B_\nu;{\tilde H} \Big) + {\cal O}\Big( m_{Weak}^2 / M_{Heavy}^2\Big) \quad \qquad
\end{eqnarray}

\subsubsection{4th decoupling theorem: SSB Abelian Higgs model} 
The $\phi$-sector of the extended theory is subject to all of the results of Sections \ref{AbelianHiggsModel} and \ref{E-AHM}, and Appendices \ref{DerivationWTIAHM} and \ref{DerivationWTIE-AHM}.
Therefore we know that
the BEH pole-mass-squared (\ref{GaugeIndependentBEHPoleMass}) arises entirely from SSB and (un-extended) AHM decays. We also know that
\begin{eqnarray}
\label{E-AHMEffectivePotential}
&&V_{E-AHM}^{Eff}=\lambda^2_\phi \Big(  \phi^\dagger \phi -\half \HVEV^2 \Big)^2 +{\cal O}^{Ignore}_{E-AHM} \nonumber \\ \nonumber \\
&&\qquad =\frac{{\lambda^2_\phi}}{4} \Big(  {\tilde H}^2 - \HVEV^2 \Big)^2 + {\cal O}^{Ignore}_{E-AHM} \nonumber \\
&&\qquad =\frac{{\lambda^2_\phi}}{4} \Big(  {\tilde h}^2 + 2 \HVEV {\tilde h} \Big)^2 + {\cal O}^{Ignore}_{E-AHM} 
\end{eqnarray}

\begin{itemize}
\item In (\ref{GaugeIndependentBEHPoleMass},\ref{E-AHMEffectivePotential}) finite  ${\cal O}_{E-AHM;\phi}^{1/M_{Heavy}^2;Irrelevant}$ 
decouple and vanish as $m_{Weak}^2/ M_{Heavy}^2 \to 0$.

\item Among the terms included in (\ref{E-AHMEffectivePotential}) are finite relevant operators dependent on the heavy matter representations:
\begin{eqnarray}
\label{HeavyRelevantOperatorsB}
&&M_{Heavy}^2, M_{Heavy}^2 \ln{ \big(M_{Heavy}^2\big)}, \nonumber \\
&& \quad \quad M_{Heavy}^2 \ln{ \big(m_{Weak}^2\big)}, m_{Weak}^2 \ln{ \big(M_{Heavy}^2\big)}  \quad 
\end{eqnarray}
but the  LSS theorem (\ref{TMatrixGoldstoneTheoremE-AHM}) has made them vanish! That fact is a central point of this paper.

\item Marginal operators $\sim \ln{ \big(M_{Heavy}^2\big)}$ have been absorbed in  (\ref{E-AHMEffectivePotential}): i.e. in the renormalization of gauge-independent observables (i.e. the quartic-coupling constant
$\lambda_\phi^2$ calculated in the Kibble representation, and the BEH VEV $\HVEV$), and in the un-observable wavefunction renormalization 
$Z_{E-AHM}^\phi$ 
(\ref{GoldstoneWavefunctionE-AHM}).
\end{itemize}

Therefore, no trace of $M_{Heavy}$-scale $\Phi,\psi$, {\it including their virtual loop-contributions to relevant operators,} survives in (\ref{GaugeIndependentBEHPoleMass},\ref{E-AHMEffectivePotential})! All the heavy Beyond-AHM matter representations have completely decoupled, and the two SSB gauge theories
\begin{eqnarray}
\label{AHMGaugeTheoryDecouplingTheorem}
E-AHM\quad  {\buildrel  {{m_{Weak}^2}/{M_{Heavy}^2} \to 0} \over {=\joinrel=\joinrel=\joinrel=\joinrel=\joinrel=\joinrel=\joinrel=\joinrel=\joinrel=\joinrel\Longrightarrow}} \quad AHM \quad \quad
\end{eqnarray}
become equivalent  in the limit ${{m_{Weak}^2}/{M_{Heavy}^2} \to 0}$, 
a central result of this paper.

\subsubsection{Gauge-independence of our  results}
\label{GaugeIndependence}

S.-H. Henry Tye and Y. Vtorov-Karevsky \cite{Tye1996} 
show that, calculated in the Kibble representation of Lorenz gauge 
(i.e. their ``polar gauge" \cite{Tye1996} ), 
the  effective potential is gauge-independent.
Nielsen \cite{Nielsen:2014spa} went on to prove that any gauge dependence
of the effective potential can be reabsorbed by a field re-definition. 
(For more details see \cite{Espinosa:2015qea}.)
With $\lambda_\phi^2$ calculated in the Kibble representation, e.g. taken from experiment, the dimension-4 AHM effective potential,
\begin{eqnarray}
\label{E-AHMEffectivePotentialSHHTye}
V_{AHM}^{Eff}&=&\frac{{\lambda^2_\phi}}{4} \Big(  {\tilde h}^2 + 2 \HVEV {\tilde h} \Big)^2 
\end{eqnarray}
is therefore all-loop-orders  gauge-independent.
The renormalized
experimentally measured gauge-coupling-constant-squared at zero momentum $e^2\equiv e^2(0)$ is also gauge-independent. 

With our 4 decoupling theorems 
(\ref{SSBTMatrixDecouplingTheorem},\ref{SSBGreensFunctionDecouplingTheorem},\ref{GaugeIndependentBEHPoleMass},\ref{AHMGaugeTheoryDecouplingTheorem}), 
so are 
$\lambda_\phi^2, \HVEV^2$ and $V_{E-AHM}^{Eff}$ 
in (\ref{E-AHMEffectivePotential}). 
and the $B_\mu$ pole-mass-squared  (\ref{BPoleMassExtended}),
when calculated in the polar gauge. 
These all appear in the 
decoupled particle physics
(\ref{ParticleDynamics}) of E-AHM.

After the $\tilde \pi$ NGB decouples, 
the all-loop-orders effective (dimension $\leq4$ operator) Lagrangian 
that governs low-energy scalar-sector E-AHM physics becomes,
in the ${m_{Weak}^2}/{M_{Heavy}^2} \to 0$ decoupling limit,
\bea
\label{leffAHMLorenzB}
{L}^{Eff}_{\phi;E-AHM} &\to& \frac{1}{2}\Big\vert (\partial_{\mu}+ieB_\mu){\tilde H}\Big\vert^{2} 
-V_{\phi;E-AHM}^{eff} \nonumber \\
V_{\phi;E-AHM}^{eff} &=& \frac{\lambda_{\phi}^{2}}{4}\Big( {\tilde H}^2-\HVEV^2\Big)^{2}  \nonumber\\
\phi^{\dagger}\phi - \frac{\HVEV^{2}}{2} &=& \half\Big( {\tilde H}^2-\HVEV^2\Big) \nonumber \\
{\tilde H} &=& {\tilde h} +\HVEV; \qquad \big<{\tilde h}\big>=0;
\eea
Eqn. (\ref{leffAHMLorenzB}) is proved gauge-independent by extension of the work of Tye and Vtorov-Karevsky \cite{Tye1996} and of Nielsen \cite{Nielsen:2014spa}
to the E-AHM.

\section{BWL \& GDS: 
This research, viewed through the prism of mathematical rigor demanded by Raymond Stora}
\label{RigueurMathematiqueExigeante}

Raymond Stora regarded Vintage-QFT as incomplete, 
fuzzy in its definitions, and primitive in technology. 
For example, he worried 
whether the off-shell T-Matrix could be mathematically rigourously  defined 
to exist in Lorenz gauge: e.g. without running into some infra-red (IR) sublety.
The Adler self-consistency conditions proved here guarantee the IR finiteness of the $\phi$-sector on-shell T-Matrix. 

Although he agreed on the correctness of the results presented here, 
Raymond might complain that we fall short of a strict mathematically rigourous proof 
(according to his  exacting mathematical standards).
He reminded us that much has been learned about Quantum Field Theory, via modern path integrals, 
in the recent $\sim 45$ years.
In the time up to his passing, he was intent on improving this work by focussing on 3 issues: 
\begin{itemize}
\item properly defining and proving the
Lorenz gauge results presented here with modern path integrals; 
\item tracking our central results directly to SSB,  via BRST methods,  
in an arbitrary manifestly IR finite 't Hooft $R_\xi$ gauge: 
i.e.  proving to his satisfaction that they are not an artifact of 
Lorenz gauge;
\item  tracking our central  results  
directly to those Slavnov-Taylor identities 
governing the SSB Goldstone mode of the BRST-invariant E-AHM Lagrangian.
\end{itemize}

Any errors, wrong-headedness, mis-understanding, 
or mis-representation appearing in  this paper are solely our fault.

\section{Conclusion}
\label{Conclusions}

AHM and E-AHM physics (e.g. on-shell T-Matrix elements) 
have more symmetry than their BRST-invariant Lagrangians. 
We introduced  global $U(1)_Y\otimes$BRST symmetry in Section \ref{AbelianHiggsModelSymmetry}, 
and showed in Sections \ref{E-AHM}, \ref{ParticlePhysicsAHM} and Appendix \ref{DerivationWTIE-AHM} 
that the low-energy weak-scale effective SSB E-AHM Lagrangian is protected 
(i.e. against loop-contributions from certain heavy $M_{Heavy}^2 \gg m_{Weak}^2$ Beyond-AHM particles $\Phi,\psi$) by hidden 1-soft-$\pi$ theorems for gauge theories: 
\begin{itemize}
\item A tower of  rigid SSB $U(1)_Y$ WTIs governing relations among Green's functions.
\item A new tower of rigid SSB $U(1)_Y$  WTI which force on-shell T-Matrix elements to vanish,
and represent the new on-shell behavior of the $U(1)_Y\otimes$BRST symmetry.  
\item A new Lee-Stora-Symanzik theorem. 
\item Four new decoupling theorems 
(\ref{SSBTMatrixDecouplingTheorem},
\ref{SSBGreensFunctionDecouplingTheorem},
\ref{GaugeIndependentBEHPoleMass} and
\ref{AHMGaugeTheoryDecouplingTheorem}).

\end{itemize}

What is remarkable is that heavy-particle decoupling 
is obscured/hidden from the physical particle spectrum (\ref{ParticleDynamics}) and its dynamics. 
The decoupling of the NGB $\tilde \pi$ 
has famously spared the AHM an observable massless particle \cite{Guralnik1964,Englert1964,Higgs1964}. 
It has also hidden from that physical particle spectrum and dynamics 
our $U(1)_Y$ WTI
 (\ref{GreensWTIPrimeExtended}, \ref{AdlerSelfConsistencyE-AHM}, \ref{TMatrixGoldstoneTheoremE-AHM},
\ref{InternalTMatrixE-AHM}, \ref{ExtendedAdlerSelfConsistency}, \ref{TMatrixGoldstoneTheoremExtended},
\ref{InternalTMatrixExtended} and \ref{ExtendedGreensFWTI}) 
and their severe constraints on the effective low-energy E-AHM Lagrangian.  
In particular, the weak-scale E-AHM SSB gauge theory 
has a hidden $U(1)_Y$ shift symmetry, 
for constant $\theta$
\begin{eqnarray}
\label{HiddenshiftSymmetryPrime}
{\tilde \pi} \to {\tilde \pi} +\HVEV \theta
\end{eqnarray}
which, together with the LSS theorem,  has caused the  complete %
\footnote{
	Modulo special cases: e.g. heavy Majorana $\nu_R$ in Subsection \ref{HeavyNeutrino}, 
	and possibly 
	${\cal O}_{E-AHM\phi}^{\Dim \leq 4;NonAnalytic;Heavy}$ in (\ref{IgnoreE-AHMOperators}).
	}
decoupling of certain heavy $M^2_{Heavy} \gg m_{Weak}^2$ $U(1)_Y$ matter-particles.


Such heavy-particle decoupling is historically 
(i.e. except  for high-precision electro-weak S,T and U 
\cite{Kennedy1988, PeskinTakeuchi, Ramond2004}) 
the usual physics experience, at each energy scale,
as experiments probed smaller and smaller distances. 
After all, Willis Lamb did not need to know the top quark or BEH mass 
in order to interpret theoretically the experimentally observed 
${\cal O}(m_e \alpha^5 \ln \alpha)$ 2S-2P splitting in the spectrum of hydrogen. 

Such heavy-particle decoupling may be the reason why the Standard Model \cite{SU(2)Proof,LSS-4Proof}, 
viewed as an effective low-energy weak-scale theory, 
is the most experimentally and observationally successful and accurate 
theory of Nature known to humans
(when augmented by classical General Relativity and neutrino mixing). 
That ``Core  Theory" \cite{WilcekCoreTheory} 
has no known experimental or observational counter-examples.

\section{Acknowledgments}
\label{Acknowledgements}
We thank Raymond Stora as an active co-researcher in this work during 2014-15, even increasing his effort on this specific paper in the last months of his life.
Although he assured us that he intended to join the paper as a full co-author, and his
contributions fully merit his inclusion, 
his untimely passing prevented us from 
obtaining his official approval of the manuscript.
In particular, we acknowledge: his specific contribution of much of the original thinking behind Section II's discussion of the role of BRST symmetry; his co-discovery that BRST commutes with spontaneously broken rigid $U(1)$ symmetry in $R_\xi$ gauges, and that there exists a conserved rigid $U(1)_Y$ current for T-Matrix elements, together with 2 towers of rigid $U(1)_Y$ WTI; his co-discovery that this behavior is ubiquitous to spontaneously broken rigid,  gauge and structure groups ${\cal G}$, and therefore to SSB gauge theories.

BWL thanks: 
Jon Butterworth and the Department of Physics and Astronomy at University College London for support as a UCL Honorary Senior Research Associate;
Chris Pope, the George P. and Cynthia W. Mitchell Center for Fundamental Physics and Astronomy, 
and Texas A\&M University,  
for support and hospitality during the  2010-2011 academic year, when this work began;
Saba Parsa for a clarifying discussion.
GDS thanks the CERN Theory group for their support and hospitality in 2012-13.
GDS is partially supported by CWRU grant DOE-SC0009946.  

\appendix
\section{$U(1)_Y$ Ward-Takahashi identities in the SSB Abelian Higgs Model}
\label{DerivationWTIAHM}

We present here the full self-contained and detailed derivation of our $U(1)_Y$ WTI for the SSB AHM. We
begin by focussing on the rigid/global  current ${J}^{\mu}_{AHM}$ of the Abelian Higgs model, the spontaneously broken gauge theory of a complex scalar $\phi = \frac{1}{{\sqrt 2}}\big( H+i\pi \big)= \frac{1}{{\sqrt 2}}{\tilde H}e^{i{\tilde \pi}/\HVEV}$, 
and a massive $U(1)_Y$ gauge field $A_\mu$. 

Construct the rigid/global $U(1)_Y$ current with (\ref{U(1)Transformations})
\begin{eqnarray}
\label{AHMCurrent}
 {J}^{\mu}_{AHM}&=& \pi \partial^\mu H-H\partial^\mu \pi-eA^\mu\Big(\pi^2 + H^2 \Big) \quad \quad
\end{eqnarray}
The classical equations of motion reveal the crucial fact: due to gauge-fixing terms in the BRST-invariant Lagrangian, the 
classical axial-vector current 
(\ref{AHMCurrent}) is 
not conserved.
Lorenz gauge  
\begin{eqnarray}
\label{DivergenceAHMCurrent}
\partial_{\mu} {J}^{\mu}_{AHM}&=& H m_A F_A   \nonumber \\
m_A &=& {eY_\phi \HVEV} \nonumber \\
F_A &=& \partial_{\beta}{A}^{\beta} 
\end{eqnarray}
with $F_A$ the gauge fixing function.  Still, the {\em physical states} $A_\mu;h,\pi$ of the theory (but not the BRST-invariant Lagrangian) obey $F_A=0$.
In 
Lorenz gauge, $A_\mu$ is transverse and $\tilde \pi$ is a massless Nambu-Goldstone Boson (NGB).

The purpose of this Appendix \ref{DerivationWTIAHM} 
is to derive a tower of quantum $U(1)_Y$ Ward-Takahashi identities 
that exhausts the information content of (\ref{DivergenceAHMCurrent}) 
and severely constrains the dynamics 
(i.e. the connected time-ordered products) of the physical states 
of the spontaneously broken Abelian Higgs model. 

{\bf 1) We study a total differential of a certain connected time-ordered product}
\begin{eqnarray}
\label{TotalDerivativeAHM}
&&\partial_{\mu} \Big< 0\vert T\Big[ {J}^{\mu}_{AHM}(z)  \\
&&\quad \quad \times h(x_1)...h(x_N) \pi_(y_1)...\pi(y_M)\Big]\vert 0\Big>_{\rm connected} \nonumber
\end{eqnarray}
written in terms of the {\em physical states} of the complex scalar $\phi$.
Here we have N external renormalized scalars $h=H-\HVEV$ (coordinates x, momenta p), 
and M external ($CP=-1$) renormalized pseudo-scalars ${ \pi}$ (coordinates y, momenta q). 

{\bf 2) Conservation of the global $U(1)_Y$ current for the {\em physical states}:} 
Strict quantum constraints are imposed 
that force the relativistically-covariant theory of gauge bosons 
to propagate {\it only} its true number of quantum spin $S=1$ degrees of freedom. 
These constraints are implemented 
by use of spin $S=0$ fermionic Fadeev-Popov ghosts $({\bar \eta},\omega)$ 
and, in Lorenz gauge,  $S=0$ massless $ \pi$.  
Physical states and their connected time-ordered products, 
but not the BRST-invariant Lagrangian,  
obey \cite{tHooft1971} the gauge-fixing condition 
${F}_A = \partial_{\beta}{A}^{\beta}=0$ 
in  Lorenz gauge:
\begin{eqnarray}
\label{GaugeConditions}
&&\big< 0\vert T\Big[ \Big( \partial_{\beta}{A}^{\beta}(z)\Big) \nonumber \\
&&\quad \quad \times h(x_1)...h(x_N)\pi(y_1)...\pi(y_M)\Big]\vert 0\big>_{\rm connected} \nonumber \\
&&\quad \quad =0 \,.
\end{eqnarray}
This restores conservation of the rigid/global $U(1)_Y$ current for physical states
\begin{eqnarray}
\label{QuantumCurrentConservation}
&&\Big< 0\vert T\Big[ \Big( \partial_{\mu}{J}^{\mu}_{AHM}(z) \Big) \\
&&\quad \quad \times h(x_1)...h(x_N) \pi(y_1)...\pi(y_M)\Big]\vert 0\Big>_{\rm connected}  \nonumber \\
&& \quad \quad =0\,.
\end{eqnarray}
It is in this ``time-ordered-product" sense 
that the ``physical" rigid global $U(1)_Y$ current ${J}^{\mu}_{AHM}$ is ``conserved", 
and it is this conserved current that generates 2 towers of quantum $U(1)_Y$ WTI. 
These WTI severely constrain the dynamics of the $\phi$-sector.

{\bf 3) Vintage QFT and canonical quantization:} Equal-time commutators are imposed on the exact renormalized fields, yielding equal-time quantum commutators at space-time points $y, z$.
\begin{eqnarray}
\label{EqTimeCommAHM}
 \delta(z_0-y_0)\left[ {J}^0_{AHM}(z),H(y)\right] &=& -i{\pi}(y)\delta^4(z-y) \nonumber \\
 \delta(z_0-y_0)\left[ {J}^0_{AHM}(z),\pi(y)\right] &=& iH(y)\delta^4(z-y) \nonumber \\
 \delta(z_0-y_0)\left[ {J}^0_{AHM}(z),A^\mu (y)\right] &=& 0 \nonumber \\
 \delta(z_0-y_0)\left[ {J}^0_{AHM}(z),\omega (y)\right] &=& 0 \nonumber \\
 \delta(z_0-y_0)\left[ {J}^0_{AHM}(z),{\bar \eta} (y)\right] &=& 0 
\end{eqnarray} 
Non-trivial commutators include
\begin{eqnarray}
\label{ZeroEqTimeCommAHM}
  \delta(z_0-y_0)\left[ \partial^0 H(z),H(y)\right] &=& -i\delta^4(z-y) \nonumber \\
 \delta(z_0-y_0)\left[ \partial^0 \pi (z),\pi(y)\right] &=& -i\delta^4(z-y) 
\end{eqnarray}

{\bf 4) Certain surface integrals vanish:} 
As appropriate to our study of the massless $\pi$, we use pion pole dominance to derive 1-soft-pion theorems, and form the surface integral
\begin{eqnarray}
\label{SurfacePionPoleDominance}
&&\lim_{k_\lambda \to 0} \int d^4z e^{ikz} \partial_{\mu} \Big< 0\vert T\Big[ \Big({J}^{\mu}_{AHM}+\HVEV \partial^\mu \pi\Big)(z)  \nonumber \\
&&\quad \quad \times h(x_1)...h(x_N) \pi_(y_1)...\pi(y_M)\Big]\vert 0\Big>_{\rm connected} \nonumber \\
&& =\int d^4z \partial_{\mu} \Big< 0\vert T\Big[ \Big({J}^{\mu}_{AHM}+\HVEV \partial^\mu \pi\Big)(z)  \nonumber \\
&&\quad \quad \times h(x_1)...h(x_N) \pi_(y_1)...\pi(y_M)\Big]\vert 0\Big>_{\rm connected} \nonumber \\
&& =\int_{3-surface\to\infty}  d^3z \quad {\widehat {z}_\mu}^{3-surface} \nonumber \\
&&\quad \quad \times \Big< 0\vert T\Big[ \Big({J}^{\mu}_{AHM}+\HVEV \partial^\mu \pi\Big)(z)  \nonumber \\
&&\quad \quad \times h(x_1)...h(x_N) \pi_(y_1)...\pi(y_M)\Big]\vert 0\Big>_{\rm connected} \nonumber \\
&& =0
\end{eqnarray}
where we have used Stokes' theorem, and $ {\widehat {z}_\mu}^{3-surface}$ is a unit vector normal to the $3-surface$. The time-ordered-product constrains the $3-surface$ to lie on, or inside, the light-cone. 

At a given point on the surface of a large enough 4-volume $\int d^4z$ (i.e. the volume of all space-time): all fields are asymptotic in-states and out-states, properly quantized as free fields, with each field species orthogonal to the others,
and  they are evaluated at equal times, making time-ordering un-necessary at $(z^{3-surface}\to \infty)$. 
Input the global AHM current (\ref{AHMCurrent}) to  (\ref{SurfacePionPoleDominance}), using $\partial_\mu \HVEV =0$
\begin{eqnarray}
\label{SurfaceIntegral}
&& \int_{3-surface\to\infty}  d^3z \quad {\widehat {z}_\mu}^{3-surface} \Big< 0\vert T\Big[  \nonumber \\
&&\quad \quad \times \Big(\pi \partial^\mu h-h\partial^\mu \pi-eA^\mu(\pi^2 + H^2 )\Big)(z)  \nonumber \\
&&\quad \quad \times h(x_1)...h(x_N) \pi_(y_1)...\pi(y_M)\Big]\vert 0\Big>_{\rm connected} \nonumber \\
&&\quad \quad =0
\end{eqnarray}

The surface integral (\ref{SurfaceIntegral}) vanishes because both $(h, A^\mu)$ are massive in the spontaneously broken $U(1)_Y$ AHM, with $(m_{BEH}^2\neq 0, m_A^2=e^2\HVEV^2)$ respectively. Propagators connecting $(h, A^\mu)$, from points on $z^{3-surface}\to \infty$ to the localized interaction points $(x_1...x_N;y_1...y_M)$, must stay inside the light-cone, die off exponentially with mass,
and are incapable of carrying information that far.

It is very important for ``pion-pole-dominance"  and this paper, that {\em this argument fails} for the remaining term in $J^\mu_{AHM}$ in (\ref{AHMCurrent}):
\begin{eqnarray}
\label{NGBSurfaceIntegral}
&& \int_{3-Surface\to\infty}  d^3z \quad {\widehat {z}_\mu}^{\rm 3-surface}  \nonumber \\
&&\quad \quad \times \Big< 0\vert T\Big[  \Big(-\HVEV \partial^\mu \pi(z)\Big) \nonumber \\
&&\quad \quad \times h(x_1)...h(x_N) \pi_(y_1)...\pi(y_M)\Big]\vert 0\Big>_{\rm connected} \nonumber \\
&&\quad \quad \neq 0
\end{eqnarray}
$\pi$ is massless in the SSB AHM, capable of carrying (along the light-cone) long-ranged pseudo-scalar forces out to the $3$-surface  $(z^{\rm 2-surface}\to \infty)$: i.e. the very ends of the light-cone (but not inside it).
That massless-ness is the basis of our pion-pole-dominance-based $U(1)_Y$ WTIs, 
which give 1-soft-pion theorems (\ref{SoftPionLimitPropagator}), 
infra-red finiteness for $\mpisq =0$ (\ref{AdlerSelfConsistency}),  
and the Lee-Stora-Symanzik (LSS)  theorem ({\ref{TMatrixGoldstoneTheorem}).

{\bf 5) Master equation:}
Using (\ref{QuantumCurrentConservation},\ref{ZeroEqTimeCommAHM}) in (\ref{TotalDerivativeAHM}) to form the right-hand-side, and (\ref{SurfaceIntegral}) in (\ref{TotalDerivativeAHM}) to form the left-hand-side, we write the master equation
\begin{eqnarray}
\label{MasterEquation}
&&\lim_{k_\lambda\to 0}\int d^4z e^{ikz} \nonumber \\
&&\quad \times \Big\{-\HVEV\partial_{\mu}^z \big< 0\vert T\Big[  \big(\partial^\mu\pi(z)\big)  \nonumber \\
&& \quad \quad \quad \quad \times h(x_1)...h(x_N) \pi_(y_1)...\pi(y_M)\Big]\vert 0\big>_{\rm connected} \nonumber \\
&&\quad - \sum_{m=1}^M \quad  i\delta^4(z-y_m) \big< 0\vert T\Big[ h(z) h(x_1)...h(x_N) \nonumber \\
&&\quad \quad \quad \quad \times  \pi(y_1)...{\widehat {\pi (y_m)}}...\pi(y_M)\Big]\vert 0\big>_{\rm connected} \nonumber \\
&&\quad + \sum_{n=1}^N \quad  i\delta^4(z-x_n) \big< 0\vert T\Big[ h(x_1)...{\widehat {h(x_n)}}...h(x_N) \nonumber \\
&&\quad \quad \quad \quad \times  \pi(z){\pi}(y_1)...\pi(y_M)\big]\vert 0\big>_{\rm connected}\Big\} \nonumber \\
&& \quad =0
\end{eqnarray}
where the ``hatted" fields ${\widehat {h(x_n)}}$ and ${\widehat {\pi (y_m)}}$ are to be removed. 
We have also thrown away a sum of $M$ terms, proportional to $\HVEV$,
that corresponds entirely to disconnected graphs.

{\bf 6) $\phi$-sector connected amplitudes:} Connected momentum-space amplitudes, with $N$ external BEHs and $M$ external $\pi$s, are defined in terms of $\phi$-sector connected time-ordered products
\begin{eqnarray}
\label{Amplitudes} 
&&iG_{N,M}(p_1...p_N;q_1...q_M)(2\pi)^4\delta^4 \Big(\sum_{n=1}^N p_n +\sum_{m=1}^M q_m \Big) \nonumber \\
&& \quad =\prod_{n=1}^N\int d^4x_n e^{ip_nx_n} \prod_{m=1}^M\int d^4y_m e^{iq_my_m}  \\
&&\quad \times \big< 0\vert T\Big[ h(x_1)...h(x_N) \pi_(y_1)...\pi(y_M)\Big]\vert 0\big>_{\rm connected} \nonumber
\end{eqnarray}

The master eqn. (\ref{MasterEquation}) can then be re-written
\begin{eqnarray}
\label{AmplitudeIdentity} 
&&\lim_{k_\lambda\to 0}\Big\{i\HVEV k^2 G_{N,M+1}(p_1...p_N;kq_1...q_M) \nonumber \\
&&\quad -\sum_{n=1}^N G_{N-1,M+1}(p_1...{\widehat {p_n}}...p_N;(k+p_n)q_1...q_M) \nonumber \\
&&\quad +\sum_{m=1}^M G_{N+1,M-1}((k+q_m)p_1...p_N;q_1...{\widehat {q_m}}...q_M)\Big\} \nonumber \\ 
&& \quad =0
\end{eqnarray}
with the ``hatted" momenta $({\widehat {p_n}},{\widehat {q_m}})$ removed  in (\ref{AmplitudeIdentity}), and an overall momentum conservation factor of $(2\pi)^4\delta^4 \Big(k+\sum_{n=1}^N p_n +\sum_{m=1}^M q_m \Big)$. 

{\bf 7) $\phi$-propagators:} Special cases of (\ref{Amplitudes}) are the BEH and $\pi$ propagators
\begin{eqnarray}
\label{ConnectedAmplitudePropagators} 
iG_{2,0}(p_1,-p_1;)&=&i\int \frac{d^4p_2}{(2\pi)^4}  G_{2,0}(p_1,p_2;)\nonumber \\
&=&\int d^4x_1 e^{ip_1x_1} \big< 0\vert T \Big[ h(x_1)h(0)\Big]\vert 0 \big> \nonumber \\
&\equiv& i\Delta_{BEH}(p_1^2) \nonumber \\
iG_{0,2}(;q_1,-q_1)&=&i\int \frac{d^4q_2}{(2\pi)^4}  G_{0,2}(;q_1,q_2) \nonumber \\
&=&\int d^4y_1 e^{iq_1y_1}\big< 0\vert T \Big[ \pi (y_1)\pi(0)\Big]\vert 0\big> \nonumber \\
&\equiv& i\Delta_{\pi}(q_1^2)
\end{eqnarray}

{\bf 8) $\phi$-sector connected amputated 1-$(h,\pi)$-Reducible (1-$\phi$-R) transition matrix (T-matrix):} With an overall momentum conservation factor 
$(2\pi)^4\delta^4 \Big(\sum_{n=1}^N p_n +\sum_{m=1}^M q_m \Big)$, the $\phi$-sector connected amplitudes are related to $\phi$-sector connected amputated T-matrix elements
\begin{eqnarray}
\label{TMatrix} 
&&G_{N,M}(p_1...p_N;q_1...q_M)  \\
&&\equiv\prod_{n=1}^N\Big[i\Delta_{BEH}(p_n^2)\Big] \prod_{m=1}^M\Big[i\Delta_{\pi}(q_m^2)\Big] T_{N,M}(p_1...p_N;q_1...q_M) \nonumber 
\end{eqnarray}
so that the master equation (\ref{MasterEquation}) can be written
\begin{eqnarray}
\label{TMatrixIdentity} 
&&\lim_{k_\lambda\to 0}\Big\{i\HVEV k^2 \Big[i\Delta_\pi(k^2)\Big]T_{N,M+1}(p_1...p_N;kq_1...q_M) \nonumber \\
&&\quad -\sum_{n=1}^N T_{N-1,M+1}(p_1...{\widehat {p_n}}...p_N;(k+p_n)q_1...q_M) \nonumber \\
&&\quad \quad \quad \quad \times \Big[i\Delta_\pi((k+p_n)^2)\Big] \Big[i\Delta_{BEH}(p_n^2)\Big]^{-1} \nonumber \\
&&\quad +\sum_{m=1}^M T_{N+1,M-1}((k+q_m)p_1...p_N;q_1...{\widehat {q_m}}...q_M) \nonumber \\
&&\quad\quad \quad \quad  \times \Big[i\Delta_{BEH}((k+q_m)^2)\Big] \Big[i\Delta_{\pi}(q_m^2)\Big]^{-1} \Big\} \nonumber \\
&&\quad =0
\end{eqnarray}
with the ``hatted" momenta $({\widehat {p_n}},{\widehat {q_m}})$ removed  in (\ref{TMatrixIdentity}), and an overall momentum conservation factor of $(2\pi)^4\delta^4 \Big(k+\sum_{n=1}^N p_n +\sum_{m=1}^M q_m \Big)$.

{\bf 9) ``Pion pole dominance" 
and ``1-soft-$\pi$ theorems" for the T-matrix:} Consider the 1-soft-pion limit 
\begin{eqnarray}
\label{SoftPionLimitPropagator}
\lim_{k_\lambda \to 0} k^2\Delta_\pi (k^2) =1 
\end{eqnarray}
where the $\pi$ is hypothesized to be all-loop-orders massless, and written in the K$\ddot a$ll$\acute e$n-Lehmann representation \cite{Bjorken1965} with spectral density $\rho^{\pi}_{AHM}$
\begin{eqnarray}
\label{NGBPropagator}
\Delta_{\pi}(k^2) &=& \frac{1}{k^2
+ i\epsilon} + \int dm^2 \frac{\rho^{\pi}_{AHM}(m^2)}{k^2-m^2 + i\epsilon}
\end{eqnarray} 
The master equation (\ref{MasterEquation}) then becomes
\begin{eqnarray}
\label{SoftPionTMatrixID} 
&&-\HVEV T_{N,M+1}(p_1...p_N;0q_1...q_M) \nonumber \\
&&\quad =\sum_{n=1}^N T_{N-1,M+1}(p_1...{\widehat {p_n}}...p_N;p_nq_1...q_M) \nonumber \\
&&\quad \quad \quad \quad \times \Big[i\Delta_\pi(p_n^2)\Big] \Big[i\Delta_{BEH}(p_n^2)\Big]^{-1} \nonumber \\
&&\quad -\sum_{m=1}^M T_{N+1,M-1}(q_m p_1...p_N;q_1...{\widehat {q_m}}...q_M) \nonumber \\
&&\quad \quad \quad \quad \times \Big[i\Delta_{BEH}(q_m^2)\Big] \Big[i\Delta_{\pi}(q_m^2)\Big]^{-1} 
\end{eqnarray}
in the 1-soft-pion limit.
As usual the ``hatted" momenta $({\widehat {p_n}},{\widehat {q_m}})$ and associated fields are removed  in (\ref{SoftPionTMatrixID}), and an overall momentum conservation factor  $(2\pi)^4\delta^4 \Big(\sum_{n=1}^N p_n +\sum_{m=1}^M q_m \Big)$ applied.

The set of 1-soft-pion theorems (\ref{SoftPionTMatrixID}) have the form 
\begin{eqnarray}
\HVEV T_{N,M+1}\sim T_{N-1,M+1} - T_{N+1,M-1}
\end{eqnarray}
relating, by the addition of a zero-momentum pion, an $N+M+1$-point function to $N+M$-point functions.

{\bf 10) The Adler self-consistency relations (but now for a gauge theory} 
rather than global $SU(2)_L \times SU(2)_R$ \cite{Adler1965,AdlerDashen1968} 
are obtained by putting the remainder of the (\ref{SoftPionTMatrixID}) particles on mass-shell
\begin{eqnarray}
\label{AdlerSelfConsistency} 
&&\HVEV T_{N,M+1}(p_1...p_N;0q_1...q_M)\nonumber \\
&& \quad \quad \times (2\pi)^4\delta^4 \Big(\sum_{n=1}^N p_n +\sum_{m=1}^M q_m \Big) \Big\vert^{p_1^2 =p_2^2...=p_N^2=m_{BEH}^2}_{q_1^2 =q_2^2...=q_M^2=0}  \nonumber \\
&& \quad \quad =0 \,,
\end{eqnarray}
which guarantees the infra-red (IR) finiteness 
of the $\phi$-sector on-shell T-matrix in the SSB AHM gauge theory in 
Lorenz gauge, with massless $\pi$ in the 1-soft-pion limit.
These ``1-soft-pion" theorems \cite{Adler1965,AdlerDashen1968} force the T-matrix to vanish as one of the pion momenta goes to zero, 
provided all other physical scalar particles are on mass-shell. 
Eqn.
(\ref{AdlerSelfConsistency})  asserts the absence of infrared divergences 
in the physical-scalar sector in Goldstone mode.
``Although individual Feynman diagrams may be IR divergent, 
those IR divergent parts cancel exactly in each order of perturbation theory. 
Furthermore, the Goldstone mode amplitude must vanish in the soft-pion limit \cite{Lee1970}".

{\bf 11) 1-$(h,\pi)$ Reducibility (1-$\phi$-R) and 1-$(h,\pi)$ Irreducibility (1-$\phi$-I):} With some exceptions, the $\phi$-sector connected amputated transition matrix $T_{N,M}$ can be cut apart by cutting an internal $h$ or $\pi$ line, and are designated 1-$\phi$-R. In contrast, the $\phi$-sector connected amputated Green's functions $\Gamma_{N,M}$ are defined to be 1-$\phi$-I: i.e. they cannot be cut apart by cutting an internal $h$ or $\pi$ line.
\begin{eqnarray}
\label{1SPReducibility}
T_{N,M} = \Gamma_{N,M} + (1-\phi-R)
\end{eqnarray}

Both $T_{N,M}$ and $\Gamma_{N,M}$ are 1-$(A_\mu)$-Reducible (1-$A^\mu$-R): i.e. they can be cut apart by cutting an internal transverse-vector $A_\mu$ gauge-particle line.

{\bf 12) $\phi$-sector 2-point functions, propagators and a 3-point vertex:} The special 2-point functions $T_{0,2}(;q,-q)$ and
$T_{2,0}(p,-p;)$, and the 3-point vertex  $T_{1,2}(q;0,-q)$, are 1-$\phi$-I (i.e. they are not 1-$\phi$-R), and are therefore equal to the corresponding 1-$\phi$-I connected amputated Green's functions. The 2-point functions
\begin{eqnarray}
\label{TMatrix2PointA}
T_{2,0}(p,-p;)&=&\Gamma_{2,0}(p,-p;)=\big[\Delta_{BEH}(p^2)\big]^{-1} \nonumber \\
T_{0,2}(;q,-q)&=&\Gamma_{0,2}(;q,-q)=\big[\Delta_{\pi}(q^2)\big]^{-1}  
\end{eqnarray}
are related to the $(1h,2\pi)$ 3-point $h\pi^2$ vertex 
\begin{eqnarray}
\label{3PointVertex}
T_{1,2}(p;q,-p-q) = \Gamma_{1,2}(p;q,-p-q) 
\end{eqnarray}
by a 1-soft-pion theorem (\ref{SoftPionTMatrixID})
\begin{eqnarray}
\label{TMatrix2and3Point}
&&\HVEV T_{1,2}(q;0,-q) -T_{2,0}(q,-q;)+T_{0,2}(;q,-q) \nonumber \\
&&\quad \quad =\HVEV T_{1,2}(q;0,-q) -\big[\Delta_{BEH}(q^2)\big]^{-1} +\big[\Delta_{\pi}(q^2)\big]^{-1} \nonumber \\
&&\quad \quad =\HVEV \Gamma_{1,2}(q;0,-q) -\Gamma_{2,0}(q,-q;)+\Gamma_{0,2}(;q,-q) \nonumber \\
&&\quad \quad =\HVEV \Gamma_{1,2}(q;0,-q) -\big[\Delta_{BEH}(q^2)\big]^{-1} +\big[\Delta_{\pi}(q^2)\big]^{-1} \nonumber \\
&&\quad \quad =0 
\end{eqnarray}

{\bf 13)  The LSS theorem, in the spontaneously broken AHM in 
Lorenz gauge,} is a special case of that SSB gauge theory's Adler self-consistency relations (\ref{AdlerSelfConsistency})
\begin{eqnarray}
\label{TMatrixGoldstoneTheorem} 
\HVEV T_{0,2}(;00)&=&0 \nonumber \\
\HVEV \Gamma_{0,2}(;00)&=&0 \nonumber \\
\HVEV \big[ \Delta_\pi (0) \big]^{-1} &=&0
\end{eqnarray}
proving  that $\pi$ is massless $\mpisq =0$ 
(i.e. not just the much weaker theorem 
that the Nambu-Goldstone boson $\tilde \pi$ is massless). 
That all-loop-orders renormalized masslessness 
is protected/guarranteed by the global $U(1)_Y$ symmetry 
of the {\em physical states} of the gauge theory after spontaneous symmetry breaking.

{\bf 14) $T_{N,M+1}^{External}$ $\phi$-sector T-Matrix with one soft $\pi (q_\mu =0)$ attached to an external-leg:} 
Figure \ref{fig:LeeFig10}  shows that
\begin{eqnarray}
\label{ExternalLegTMatrix}
&&\HVEV T_{N,M+1}^{External}(p_1...p_N;0q_1...q_M) \nonumber \\
&&\quad \quad =\sum_{n=1}^N \Big[ i\HVEV\Gamma_{1,2}(p_n;0,-p_n)\Big] \Big[ i\Delta_\pi (p_n^2)\Big]  \nonumber \\
&&\quad \quad \times T_{N-1,M+1}(p_1...{\widehat{p_n}}...p_N;p_nq_1...q_M) \nonumber \\
&&\quad \quad +\sum_{m=1}^M \Big[ i\HVEV\Gamma_{1,2}(q_m;0,-q_m)\Big] \Big[ i\Delta_{BEH} (q_m^2)\Big]  \nonumber \\
&&\quad \quad \times T_{N+1,M-1}(q_mp_1...p_N;q_1....{\widehat{q_m}}...q_M) \nonumber \\
&&\quad \quad =\sum_{n=1}^N \Big(1-\Big[ i\Delta_\pi (p_n^2)\Big]\Big[ i\Delta_{BEH} (p_n^2)\Big]^{-1} \Big) \nonumber \\
&&\quad \quad \times T_{N-1,M+1}(p_1...{\widehat{p_n}}...p_N;p_nq_1...q_M) \nonumber \\
&&\quad \quad -\sum_{m=1}^M \Big(1-\Big[ i\Delta_{BEH} (q_m^2)\Big]\Big[ i\Delta_\pi (q_m^2)\Big]^{-1} \Big)  \nonumber \\
&&\quad \quad \times T_{N+1,M-1}(q_mp_1...p_N;q_1....{\widehat{q_m}}...q_M)  
\end{eqnarray}
where we used (\ref{TMatrix2and3Point}). Now separate
\begin{eqnarray}
\label{DefineInternalTMatrixA}
&&T_{N,M+1}(p_1...p_N;0q_1...q_M) \nonumber \\
&&\quad \quad =T_{N,M+1}^{External}(p_1...p_N;0q_1...q_M) \nonumber \\
&&\quad \quad +T_{N,M+1}^{Internal}(p_1...p_N;0q_1...q_M)  
\end{eqnarray}
so that 
\begin{eqnarray}
\label{InternalTMatrix}
&&\HVEV T_{N,M+1}^{Internal}(p_1...p_N;0q_1...q_M) \nonumber \\
&&\quad \quad =\sum_{m=1}^M T_{N+1,M-1}(q_mp_1...p_N;q_1....{\widehat{q_m}}...q_M)  \nonumber \\
&&\quad \quad -\sum_{n=1}^N T_{N-1,M+1}(p_1...{\widehat{p_n}}...p_N;p_nq_1...q_M) \quad \quad 
\end{eqnarray}

{\bf 15) Recursive $U(1)_Y$ WTI  for 1-$\phi$-I connected amputated Green's functions $\Gamma_{N,M}$:} Removing the 1-$\phi$-R  graphs from both sides of (\ref{InternalTMatrix}) yields the recursive identity
\begin{eqnarray}
\label{GreensFWTI}
&&\HVEV \Gamma_{N,M+1}(p_1...p_N;0q_1...q_M) \nonumber \\
&&\quad \quad =\sum_{m=1}^M \Gamma_{N+1,M-1}(q_mp_1...p_N;q_1....{\widehat{q_m}}...q_M)  \nonumber \\
&&\quad \quad -\sum_{n=1}^N \Gamma_{N-1,M+1}(p_1...{\widehat{p_n}}...p_N;p_nq_1...q_M) \quad \quad
\end{eqnarray}

B.W. Lee \cite{Lee1970} gave an inductive proof for the corresponding recursive $SU(2)_L \times SU(2)_R$ WTI in the global  Gell-Mann L${\acute e}$vy model with PCAC \cite{GellMannLevy1960}. 
Specifically, he proved  that, given the global $SU(2)_L \times SU(2)_R$ analogy of (\ref{InternalTMatrix}), the global $SU(2)_L \times SU(2)_R$ analogy of (\ref{GreensFWTI}) follows. This he did by examination of the explicit reducibility/irreducibility of the various  Feynman graphs involved. 

That proof also works for the $U(1)_Y$ SSB AHM, thus establishing our tower of 1-$\phi$-I connected amputated Green's functions' recursive
$U(1)_Y$ WTI (\ref{GreensFWTI}) for a local/gauge theory. 

Rather than including the lengthy proof here, we paraphrase \cite{Lee1970} as follows: 
(\ref{TMatrix2and3Point}) shows that (\ref{GreensFWTI}) is true for $(N=1,M=1)$. Assume it is true for all $(n,m)$ such that $n+m < N+M$.
Consider (\ref{InternalTMatrix}) for $n=N,m=M$. The two classes of graphs contributing to $T_{N,M+1}^{Internal}(p_1...p_N;0q_1...q_M)$ are displayed in Figure \ref{fig:LeeFig11}. 

\begin{figure}
\centering
\includegraphics[width=1\hsize,trim={0cm 4cm 0cm 3cm},clip]{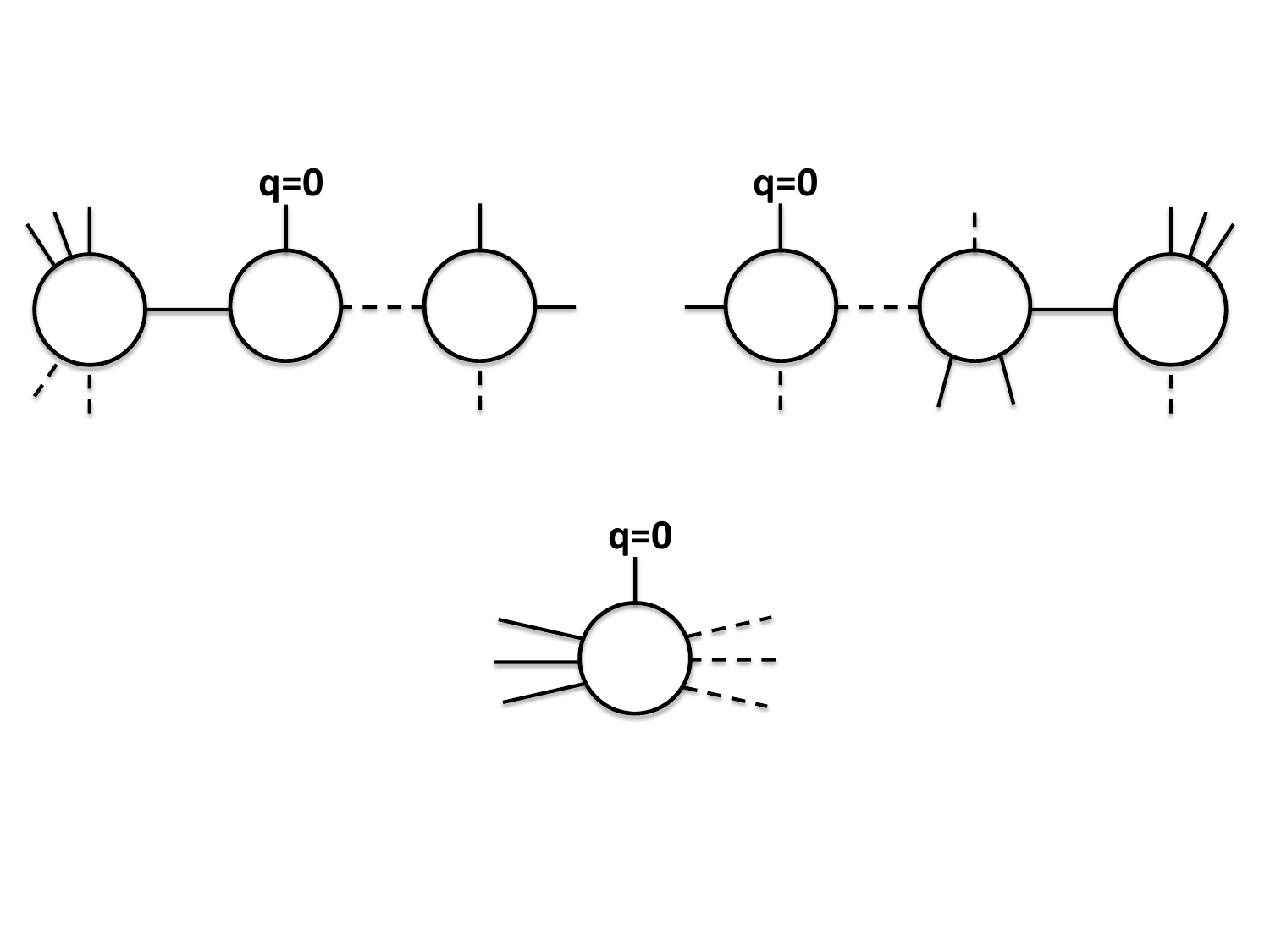}
\caption{
\label{fig:LeeFig11} 
Circles are 1-$\phi$-I $\Gamma^{E-AHM}_{n,m}$, solid lines $\pi$, dashed lines $h$, with $n+m<N+M$. 1 (zero-momentum) soft pion emerges in all possible ways from the  connected amputated Green's functions. $\Gamma^{E-AHM}_{n,m}$ is 1-$A^\mu$-R by cutting an $A^\mu$ line, and also 1-$\Phi$-R by cutting a $\Phi$ line. Fig. \ref{fig:LeeFig11}  is the E AHM analogy of B.W. Lee's Figure 11 \cite{Lee1970}. The same graph topologies, but without internal Beyond-AHM $\Phi,\psi$ heavy matter, are used in the proof of (\ref{GreensFWTI}) for the (unextended) AHM.
}
\end{figure}

The top graphs
 in Figure \ref{fig:LeeFig11} are 1-$\phi$-R. 
 For $(n,m;n+m<N+M)$ we may use (\ref{GreensFWTI}), 
 for those 1-$\phi$-I Green's functions $\Gamma_{n,m}$ that contribute to (\ref{InternalTMatrix}),
 to show that the contribution of 1-$\phi$-R graphs 
 to both sides of  (\ref{InternalTMatrix}) are identical.

The bottom graphs in  Figure \ref{fig:LeeFig11} are 1-$\phi$-I and so already obey (\ref{GreensFWTI}). 

 {\bf 16) LSS theorem makes tadpoles vanish:} 
\begin{eqnarray}
\label{Tadpoles}
&&\big<0\vert h(x=0)\vert0\big>_{\rm connected} \nonumber \\
&& \qquad \qquad = i \Big[i\Delta_{BEH}(0)\Big]\Gamma_{1,0}(0;) \qquad \quad
\end{eqnarray}
but the $N=0,M=1$ case of (\ref{GreensFWTI}) reads
\begin{eqnarray}
\label{GoldstoneTadpoles}
\Gamma_{1,0}(0;)&=& \HVEV \Gamma_{0,2}(;00) \nonumber \\
&=&0 
\end{eqnarray}
where we used (\ref{TMatrixGoldstoneTheorem}), so that tadpoles all vanish automatically, and separate tadpole renormalization is un-necessary.
Since we can choose the origin of coordinates anywhere we like
\begin{eqnarray}
\label{GoldstoneTadpolesVanishExtended}
\big<0\vert h(x)\vert0\big>_{\rm connected} &=& 0
\end{eqnarray}

{\bf 17) Renormalized  gauge-independent observable $\HVEV$}.
\begin{eqnarray}
\label{HVEV}
\big<0\vert H(x)\vert0\big>_{\rm connected} &=&\big<0\vert h(x)\vert0\big>_{\rm connected} +\HVEV \nonumber \\
&=& \HVEV \nonumber \\
\partial_\mu \HVEV &=&0
\end{eqnarray}

{\bf 18) Benjamin W. Lee's 1970 Cargese summer school lectures' \cite{Lee1970}} 
proof of $\phi$-sector WTI 
focuses on the global $SU(2)_L \times SU(2)_R$ Gell-Mann L${\acute e}$vy theory 
and Partially Conserved Axial-vector Currents (PCAC), 
but  gives a detailed pedagogical account of 
the appearance of the Goldstone theorem and its true massless Nambu-Goldstone bosons,
and especially of the {\bf emergence of the Lee-Stora-Symanzik (LSS) theorem}, 
in {\em global} theories, and is recommended reading. 
We include a translation guide in Table 1.

\newpage
\bigskip
\vbox{
\baselineskip=15pt
\halign{
\hfil\sl#&\hfil\quad\it#\hfil&\quad#\hfil\cr
\multispan3\hfil\bf Table 1: Derivation of Ward-Takahashi identities \hfil\cr
{\bf Property}&{\bf This paper}&{\bf B.W.Lee \cite{Lee1970}}\cr
LagrangianInvariant&BRST&global group\cr
structure group&$U(1)_Y$&$SU(2)_L \times SU(2)_R$\cr
local/gauge group&$U(1)_Y$& \cr
rigid/global group&$U(1)_Y$&$SU(2)_L \times SU(2)_R$\cr
global currents&${J}_{AHM}^\mu$&${\vec V}^\mu ;{\vec A}^\mu$ \cr
PCAC&no&yes\cr
current divergence&$Hm_A\partial_\beta A^\beta$&$0; f_\pi\mpisq{\vec \pi}$\cr
$L_{GaugeFixing}$&$Lorenz$&\cr
gauge&Lorenz&\cr
ghosts ${\bar \eta},\omega$&decouple&\cr
conserved current&physical states&Lagrangian \cr
physical states&$A_\mu,h, \pi , \Phi,\psi$&$s,{\vec \pi}$\cr
interaction&weak&strong \cr
fields&$A_\mu,H, \pi , {\bar \eta},\omega,\Phi,\psi$&$\sigma,{\vec \pi}$\cr
BEH scalar&$h=H-\HVEV$&$s=\sigma-<\sigma>$\cr
VEV&$\HVEV$&$<\sigma>=v=f_\pi$\cr
particles in loops&$Physical\&Ghosts$&$s,{\vec \pi}$\cr
renormalization&all-loop-orders&all-loop-orders\cr
amplitudes&&G\cr
ConnectedAmplitudes&$G_{N,M}$&H\cr
NoPionPoleSingularity&&$\bar H$\cr
1-$\phi$-I or R&$h,{ \pi}$&$s,{\vec \pi}$\cr
connected $\Gamma_{N,M}$&amputated&amputated\cr
connected $T_{N,M}$&amputated&amputated\cr
NGB after SSB&$\tilde \pi$&$\tilde{\vec \pi}$\cr 
LSS Theorem&$\HVEV\Gamma^{t_1t_2}_{0,2}(;00)=0$&$f_\pi\Gamma^{t_1t_2}_{0,2}(;00)$\cr
&&$=\epsilon\delta^{t_1t_2}=0$\cr
explicit breaking&&$\epsilon=f_\pi\mpisq$\cr
$\phi$-sector $T_{N,M}$&1-$\phi$-R&1-$\phi$-R\cr
&1-$A_\mu$-R, 1-$\Phi$-R&\cr
$\phi$-sector $\Gamma_{N,M}$&1-$\phi$-I&1-$\phi$-I\cr
&1-$A_\mu$-R, 1-$\Phi$-R&\cr
T-Matrix&$T_{N,M}$&$T$\cr
$\phi$SectorGreensF's&$\Gamma_{N,M}$&$\Gamma_{N,M}$\cr 
External $\pi (q_\mu =0)$&$T^{External}_{N,M+1}$&$T_1$\cr
Internal $\pi (q_\mu =0)$&$T^{Internal}_{N,M+1}$&$T_2$\cr
BEH propagator&$\Delta_{BEH}$&$\Delta_\sigma$\cr
TransversePropagator&$\Delta_{A}^{\mu \nu}$&\cr
Pion propagator&$\Delta_{\pi}$&$\delta^{t_i t_j}\Delta_{\pi}$\cr
SSB&GoldstoneMode&GoldstoneMode\cr
Goldstone theorem&physical states&GoldstoneMode\cr
LSS theorem&1-D line&1-D boundary of\cr
&&2-D quarter-plane\cr
}
}
\bigskip

\newpage

\section{$U(1)_Y$ $\phi$-sector WTIs 
which  include the all-loop-orders contributions 
of certain additional virtual $U(1)_Y$ $CP$-conserving matter representations
 $\Phi,\psi$
in the Extended Abelian Higgs Model (E-AHM)}
\label{DerivationWTIE-AHM}

We focus on the rigid/global  extended-AHM current 
\begin{eqnarray}
\label{extendedAHMCurrent}
{ J}^{\mu}_{E-AHM}&=&{J}^{\mu}_{AHM}(A^\mu,\phi) \nonumber \\
&+&{J}^{\mu}_{BeyondAHM}(\Phi,\Psi) \quad \quad
\end{eqnarray}
 of the  ``extended Abelian Higgs model", the spontaneously broken gauge theory of a complex spin $S=0$ scalar $\phi = \frac{1}{{\sqrt 2}}\big( H+i\pi \big)$, 
a massive $U(1)_Y$ $S=1$ transverse gauge field $A_\mu$, and certain $S=0$ scalars $\Phi$ and anomaly-cancelling $S=\half$  fermions $\psi$
originating in Beyond-AHM  models.

In order to force renormalized connected amplitudes with an odd number of $\pi$s to vanish, the new particles $\Phi,\psi$ are taken in this paper to conserve $CP$.

The classical equations of motion reveal that, due to gauge-fixing terms in the BRST-invariant Lagrangian, the 
classical current 
(\ref{extendedAHMCurrent}) is 
not conserved. In 
Lorenz gauge  
\begin{eqnarray}
\label{ExtendedDivergenceAHMCurrent}
\partial_{\mu}  { J}^{\mu}_{E-AHM} &=& H m_A F_A   \nonumber \\
m_A &=& {eY_\phi\HVEV} \nonumber \\
F_A &=& \partial_{\beta}{A}^{\beta} 
\end{eqnarray}
with $F_A$ the gauge fixing function.  

The purpose of this Appendix is to derive a tower of $U(1)_Y$ extended WTIs 
that exhausts the information content of (\ref{ExtendedDivergenceAHMCurrent}), 
and severely constrains the dynamics 
(i.e. the connected time-ordered products) of the physical states of the SSB {\em extended}-AHM.
We make use here of all of the results in Appendix A concerning ${J}^{\mu}_{AHM}$.

{\bf 1) We study a certain total differential of a connected time-ordered product:}
\begin{eqnarray}
\label{ExtendedTotalDerivativeAHM}
&&\partial_{\mu} \Big< 0\vert T\Big[ { J}^{\mu}_{E-AHM}(z)  \\
&&\quad \quad \times h(x_1)...h(x_N) \pi_(y_1)...\pi(y_M)\Big]\vert 0\Big>_{\rm connected} \nonumber
\end{eqnarray}
written in terms of the {\em physical states} of the complex scalar $\phi$.
Here we have N external renormalized scalars $h=H-\HVEV$ (coordinates x, momenta p), 
and M external ($CP=-1$) renormalized pseudo-scalars ${ \pi}$ (coordinates y, momenta q).

{\bf 2) Conservation of the global $U(1)_Y$ current for the {\em physical states}:} 
Strict quantum constraints are imposed 
that force the relativistically-covariant theory of a massive transverse gauge boson 
to propagate {\it only} its true number of quantum spin $S=1$ degrees of freedom.   
Physical states and their time-ordered products, but not the BRST-invariant Lagrangian,  
obey the gauge-fixing condition ${F}_A = \partial_{\beta}{A}^{\beta}=0$ in
Lorenz gauge \cite{tHooft1971}
\begin{eqnarray}
\label{GaugeConditionExtended} 
&&\big< 0\vert T\Big[ \Big( \partial_{\beta}{A}^{\beta}(z)\Big) \nonumber \\
&&\quad \quad  \times h(x_1)...h(x_N)\pi(y_1)...\pi(y_M)\Big]\vert 0\big>_{\rm connected} \nonumber \\
&&\quad \quad =0 \,,
\end{eqnarray}
which restores conservation of the rigid/global $U(1)_Y$ extended current for physical states
\begin{eqnarray}
\label{ExtendedQuantumCurrentConservation}
&&\Big< 0\vert T\Big[ \Big( \partial_{\mu}{J}^{\mu}_{E-AHM}(z) \Big) \nonumber \\
&&\quad \quad \times h(x_1)...h(x_N) \pi(y_1)...\pi(y_M)\Big]\vert 0\Big>_{\rm connected} \nonumber \\
 &&\quad \quad=0
\end{eqnarray}
It is in this ``time-ordered-product" sense that the rigid global extended $U(1)_Y$ current ${J}^{\mu}_{E-AHM}$ is conserved, and
it is this conserved current  that generates our tower of $U(1)_Y$ extended WTI. 
These extended WTI severely constrain the dynamics of $\phi$.

{\bf 3) Vintage QFT and canonical quantization:} Equal-time commutators are imposed on the exact renormalized Beyond-AHM fields, yielding equal-time quantum commutators at space-time points $y, z$.
\begin{eqnarray}
\label{EqTimeCommE-AHM}
 \delta(z_0-y_0)\left[ {J}^0_{BeyondAHM}(z),H(y)\right] &=&0 \nonumber \\
 \delta(z_0-y_0)\left[ {J}^0_{BeyondAHM}(z),\pi(y)\right] &=&0
\end{eqnarray} 
Only certain  $U(1)_Y$  matter particles $\Phi, \psi$ obey this condition.

{\bf $\bullet$ Renormalized $\HVEV$ is defined to match the (un-extended) AHM}. 
Our extended $U(1)_Y$ WTI therefore  require that all of the new spin $S=0$ fields in $ {J}^\mu_{BeyondAHM}$ have zero vacuum expectation value (VEV):
\begin{eqnarray}
\label{E-AHMVEV}
\big<  \Phi_{BeyondAHM}\big> =0
\end{eqnarray}
Only certain  $U(1)_Y$  matter particles $\Phi$ obey this condition.

{\bf 4) Certain connected surface integrals must vanish:} 
As appropriate to our study of massless $\pi$, we again use pion pole dominance to derive 1-soft-pion theorems, and require that the {\em connected} surface integral
\begin{eqnarray}
\label{ExtendedSurfaceIntegral}
&&\lim_{k_\lambda \to 0} \int d^4z e^{ikz} \partial_{\mu} \Big< 0\vert T\Big[ \Big({J}^{\mu}_{BeyondAHM} (z)\Big)  \nonumber \\
&&\quad \quad \times h(x_1)...h(x_N) \pi_(y_1)...\pi(y_M)\Big]\vert 0\Big>_{\rm connected} \nonumber \\
&& =\int d^4z \partial_{\mu} \Big< 0\vert T\Big[ \Big({J}^{\mu}_{BeyondAHM} (z)\Big) \nonumber \\
&&\quad \quad \times h(x_1)...h(x_N) \pi_(y_1)...\pi(y_M)\Big]\vert 0\Big>_{\rm connected} \nonumber \\
&& =\int_{3-Surface\to\infty}  d^3z \quad {\widehat {z}_\mu}^{3-surface} \nonumber \\
&&\quad \quad \times \Big< 0\vert T\Big[ \Big({J}^{\mu}_{BeyondAHM} (z) \Big) \nonumber \\
&&\quad \quad \times h(x_1)...h(x_N) \pi_(y_1)...\pi(y_M)\Big]\vert 0\Big>_{\rm connected} \nonumber \\
&&\quad \quad =0
\end{eqnarray}
where we have used Stokes' theorem, and $ {\widehat {z}_\mu}^{3-surface}$ is a unit vector normal to the $3-surface$. The time-ordered-product constrains the $3-surface$ to lie on-or-inside the light-cone.

At a given point on the surface of a large enough 4-volume $\int d^4z$ (i.e. the volume of all space-time): all fields are asymptotic in-states and out-states; are properly quantized as free fields; with each field species orthogonal to the others;
and  they are evaluated at equal times, making time-ordering un-necessary at $(z^{3-surface}\to \infty)$. 

Only certain  $U(1)_Y$  {\bf massive} matter particles $\Phi, \psi$ obey this condition.

{\bf 5) Extended master equation:}
Using (\ref{ExtendedQuantumCurrentConservation},\ref{EqTimeCommE-AHM}) 
in (\ref{ExtendedTotalDerivativeAHM}) to form the right-hand-side, 
and (\ref{ExtendedSurfaceIntegral}) in (\ref{ExtendedTotalDerivativeAHM}) 
to form the left-hand-side, 
we write the {\em extended} master equation, 
which relates {\em connected} time-ordered products:
\begin{eqnarray}
\label{ExtendedMasterEquation}
&&\lim_{k_\lambda\to 0}\int d^4z e^{ikz} \nonumber \\
&&\quad \times \Big\{
-\HVEV\partial_{\mu}^z \big< 0\vert T\Big[  \big(\partial^\mu\pi(z)\big)  \nonumber \\
&&\quad \quad \quad \quad \times h(x_1)...h(x_N) \pi_(y_1)...\pi(y_M)\Big]\vert 0\big>_{\rm connected} \nonumber \\
&&\quad - \sum_{m=1}^M \quad  i\delta^4(z-y_m) \big< 0\vert T\Big[ h(z) h(x_1)...h(x_N) \nonumber \\
&&\quad \quad \quad \quad \times  \pi(y_1)...{\widehat {\pi (y_m)}}...\pi(y_M)\Big]\vert 0\big>_{\rm connected} \nonumber \\
&&\quad + \sum_{n=1}^N \quad  i\delta^4(z-x_n) \big< 0\vert T\Big[ h(x_1)...{\widehat {h(x_n)}}...h(x_N) \nonumber \\
&&\quad \quad \quad \quad \times  \pi(z){\pi}(y_1)...\pi(y_M)\big]\vert 0\big>_{\rm connected} \Big\} \nonumber \\
&&\quad =0
\end{eqnarray}
where the ``hatted" fields ${\widehat {h(x_n)}}$ and ${\widehat {\pi (y_m)}}$ are to be removed. We have also thrown away a sum of $M$ terms, proportional to $\HVEV$,
that corresponds entirely to disconnected graphs.

{\bf $\bullet$ $U(1)_Y$ Ward-Takahashi identities for the $\phi$-sector of the E-AHM:} 
The extended master equation (\ref{ExtendedMasterEquation}) 
governing the $\phi$-sector of the E-AHM, 
is idential to the master equation (\ref{MasterEquation}) 
governing the $\phi$-sector of the (un-extended) AHM. 
This proves that, for each $U(1)_Y$ WTI that is true in the AHM, 
an analogous $U(1)_Y$ WTI is true for the E-AHM.
Appendix A proved $U(1)_Y$ WTI relations among 1-$\phi$-R $\phi$-sector  T-Matrix elements $T_{N,M}$,
 as well as $U(1)_Y$ WTI relations among 1-$\phi$-I 
$\phi$-sector Green's functions $\Gamma_{N,M}$, in the spontaneously broken AHM.
Analogous $U(1)_Y$ WTI relations among 1-$\phi$-R $\phi$-sector T-Matrix elements $T_{N,M}^{\Extended}$, 
as well as analogous $U(1)_Y$ WTI relations among 1-$\phi$-I 
$\phi$-sector  Green's functions $\Gamma_{N,M}^{\Extended}$, are therefore here proved true for the spontaneously broken {\em extended}-AHM.

But there is one {\em huge} difference! 
The renormalization of  our $U(1)_Y$ WTI, 
governing $\phi$-sector $T_{N,M}^{\Extended}$ and $\phi$-sector $\Gamma_{N,M}^{\Extended}$, 
now includes the all-loop-orders contributions of 
virtual gauge bosons, $\phi$-scalars, ghosts, new Beyond-AHM scalars and new Beyond-AHM fermions: 
i.e.  $A^\mu; h, \pi; {\bar \eta}, \omega; \Phi; \psi$ respectively. 

{\bf 10) Adler self-consistency relations, but now for the E-AHM gauge theory:}
\begin{eqnarray}
\label{ExtendedAdlerSelfConsistency} 
&&\HVEV T_{N,M+1}^{\Extended}(p_1...p_N;0q_1...q_M)\nonumber \\
&& \quad \quad \times (2\pi)^4\delta^4 \Big(\sum_{n=1}^N p_n +\sum_{m=1}^M q_m \Big) \Big\vert^{p_1^2 =p_2^2...=p_N^2=m_{BEH}^2}_{q_1^2 =q_2^2...=q_M^2=0}  \nonumber \\
&& \quad \quad =0 
\end{eqnarray}
These prove the infra-red (IR) finiteness of the $\phi$-sector on-shell connected T-matrix in the E-AHM gauge theory, with massless 
$\pi$, in 
Lorenz gauge, in the 1-soft-pion limit.

{\bf 11) 1-$(h,\pi)$ Reducibility (1-$\phi$-R) and 1-$(h,\pi)$ Irreducibility (1-$\phi$-I):} With some exceptions, the extended $\phi$-sector connected amputated T-Matrix elements  $T_{N,M}^{\Extended}$ can be cut apart by cutting an internal $h$ or $\pi$ line: they are designated 1-$\phi$-R. In contrast, the extended $\phi$-sector Green's functions $\Gamma_{N,M}^{\Extended}$ are defined to be 1-$\phi$-I: ie. they cannot be cut apart by cutting an internal $h$ or $\pi$ line.
\begin{eqnarray}
\label{Extended1SPReducibility}
T_{N,M}^{\Extended} = \Gamma_{N,M}^{\Extended} + (1-\phi -R)
\end{eqnarray}

As usual, both $T_{N,M}^{\Extended}$ and $\Gamma_{N,M}^{\Extended}$ 
are 1-$(A^\mu)$-Reducible (1-$A^\mu$-R), 
i.e. they can be cut apart by cutting an internal transverse-vector $A^\mu$ gauge-particle line.
They are also 1-$\Phi$-Reducible (1-$\Phi$-R), 
i.e. they can be cut apart by cutting an internal $\Phi$ scalar line.

{\bf 12) $\phi$-sector 2-point functions, propagators and a 3-point vertex:} The 2-point functions
\begin{eqnarray}
\label{TMatrix2PointB}
T_{2,0}^{\Extended}(p,-p;)&=&\Gamma_{2,0}^{\Extended}(p,-p;)=\big[\Delta_{BEH}(p^2)\big]^{-1} \nonumber \\
T_{0,2}^{\Extended}(;q,-q)&=&\Gamma_{0,2}^{\Extended}(;q,-q)= \big[\Delta_{\pi}(q^2)\big]^{-1} \qquad \quad
\end{eqnarray}
are related to the $(1h,2\pi)$ 3-point $h\pi^2$ vertex 
\begin{eqnarray}
\label{3PointVertexB}
T_{1,2}^{\Extended}(p;q,-p-q) = \Gamma^{\Extended}_{1,2}(p;q,-p-q) \quad \quad
\end{eqnarray}
by a 1-soft-pion theorem (\ref{ExtendedGreensFWTI})
\begin{eqnarray}
\label{TMatrix2and3PointExtended}
&&\HVEV T_{1,2}^{\Extended}(q;0,-q) \nonumber \\
&&\quad \quad =\big[\Delta_{BEH}(q^2)\big]^{-1} -\big[\Delta_{\pi}(q^2)\big]^{-1} 
\end{eqnarray}

{\bf 13) The LSS theorem, in 
Lorenz-gauge E-AHM} is the $N=0,M=1$ case 
of (\ref{ExtendedAdlerSelfConsistency}):
\begin{eqnarray}
\label{TMatrixGoldstoneTheoremExtended} 
\HVEV T_{0,2}^{\Extended}(;00)&=&0 \nonumber \\
\HVEV \Gamma_{0,2}^{\Extended}(;00)&=&0 \nonumber \\
\HVEV \big[ \Delta_\pi (0) \big]^{-1} &=&0
\end{eqnarray}
proves that $\pi$ is still massless in the E-AHM, whose all-loop-orders renormalized massless-ness is protected/guarranteed by the global $U(1)_Y$ symmetry of the physical states of the E-AHM gauge theory after SSB.

{\bf 14)} $T_{N,M+1}^{\Extended;External}$ are the 1-$\phi$-R $\phi$-sector connected amputated T-Matrix elements, {\bf with one soft $\pi (q_\mu =0)$ attached to an external-leg,} as shown in Figure \ref{fig:LeeFig10}.
With the separation
\begin{eqnarray}
\label{DefineInternalTMatrixB}
&&T^{\Extended}_{N,M+1}(p_1...p_N;0q_1...q_M) \nonumber \\
&&\quad \quad =T_{N,M+1}^{\Extended;External}(p_1...p_N;0q_1...q_M) \nonumber \\
&&\quad \quad +T_{N,M+1}^{\Extended;Internal}(p_1...p_N;0q_1...q_M)  \qquad \qquad
\end{eqnarray}
we have the recursive $U(1)_Y$ T-Matrix WTI
\begin{eqnarray}
\label{InternalTMatrixExtended}
&&\HVEV T_{N,M+1}^{\Extended;Internal}(p_1...p_N;0q_1...q_M) \nonumber \\
&&\quad \quad =\sum_{m=1}^M T^{\Extended}_{N+1,M-1}(q_mp_1...p_N;q_1....{\widehat{q_m}}...q_M)  \nonumber \\
&&\quad \quad -\sum_{n=1}^N T^{\Extended}_{N-1,M+1}(p_1...{\widehat{p_n}}...p_N;p_nq_1...q_M) \quad\quad
\end{eqnarray}

{\bf 15) Recursive $U(1)_Y$ WTIs for 1-$\phi$-I $\phi$-sector connected amputated extended Green's functions $\Gamma_{N,M}^{\Extended}$}
are a solution to (\ref{InternalTMatrixExtended})
\begin{eqnarray}
\label{ExtendedGreensFWTI}
&&\HVEV \Gamma_{N,M+1}^{\Extended}(p_1...p_N;0q_1...q_M) \nonumber \\
&&\quad \quad =\sum_{m=1}^M \Gamma_{N+1,M-1}^{\Extended}(q_mp_1...p_N;q_1....{\widehat{q_m}}...q_M)  \nonumber \\
&&\quad \quad -\sum_{n=1}^N \Gamma_{N-1,M+1}^{\Extended}(p_1...{\widehat{p_n}}...p_N;p_nq_1...q_M) \quad\quad
\end{eqnarray}

{\bf 16) The LSS theorem (\ref{TMatrixGoldstoneTheoremExtended}) makes tadpoles vanish:} 
\begin{eqnarray}
\label{TadpolesExtended}
&&\big<0\vert h(x=0)\vert0\big>_{\rm connected}  \\
&& \qquad \qquad = i \Big[i\Delta_{BEH}(0)\Big]\Gamma^{\Extended}_{1,0}(0;) \nonumber
\end{eqnarray}
but the $N=0,M=1$ case of (\ref{ExtendedGreensFWTI}) reads
\begin{eqnarray}
\label{GoldstoneTadpolesExtended}
\Gamma^{\Extended}_{1,0}(0;)&=& \HVEV \Gamma^{\Extended}_{0,2}(;00) \nonumber \\
&=&0 
\end{eqnarray}
where we have used (\ref{TMatrixGoldstoneTheoremExtended}), so that tadpoles all vanish automatically, and separate tadpole renormalization is un-necessary. Since we can choose the origin of coordinates anywhere we like
\begin{eqnarray}
\label{GoldstoneTadpolesVanishExtendedB}
\big<0\vert h(x)\vert0\big>_{\rm connected} &=& 0
\end{eqnarray}
\newline
{\bf 17) Renormalized gauge-independent observable $\HVEV$}.
\begin{eqnarray}
\label{HVEVExtended}
\big<0\vert H(x)\vert0\big>_{\rm connected} &=&\big<0\vert h(x)\vert0\big>_{\rm connected} +\HVEV \nonumber \\
&=& \HVEV \nonumber \\
\partial_\mu \HVEV &=&0
\end{eqnarray}

 \end{document}